\pdfoutput=1
\documentclass[12pt]{article} %

\usepackage[utf8]{inputenc}%
\usepackage[tracking=true, letterspace=50, expansion=false]{microtype}%
\usepackage{amsmath}%
\usepackage{amsfonts}%
\usepackage{amsthm}%
\usepackage[margin=1in]{geometry} %
\usepackage{setspace}
\usepackage{natbib}
\usepackage{booktabs}
\usepackage[online, flushleft]{threeparttable}
\usepackage{multirow}
\usepackage{enumitem}%

\usepackage{tikz}

\newtheorem{theorem}{Theorem}[section]
\newtheorem{lemma}{Lemma}[section]

\newtheorem{assumption}{Assumption}[section]

\theoremstyle{definition}%
\newtheorem{remark}{Remark}[section]

\newcommand{\RMSE}{\textnormal{RMSE}}
\newcommand{\FLCI}{\textnormal{FLCI}}
\newcommand{\FLip}{\mathcal{F}_{\textnormal{Lip}}} 
\newcommand{\FRLip}[1]{\mathcal{F}_{#1,\textnormal{Lip}}} 
\newcommand{\FLipn}{\widetilde{\mathcal{F}}_{\textnormal{Lip},n}} 

\newcommand{\Rlower}[1]{\ensuremath q_{#1}}

\RequirePackage{mathtools}
\DeclarePairedDelimiter\abs{\lvert}{\rvert}
\DeclarePairedDelimiter\norm{\lVert}{\rVert}
\DeclarePairedDelimiter\1{\mathbb{I}\{}{\}}

\DeclarePairedDelimiter\hor{[}{)}

\DeclareMathOperator{\cv}{cv}

\DeclareMathOperator{\se}{se}
\newcommand{\sepate}{\se_{\ensuremath\tau}}
\DeclareMathOperator{\maxbias}{\overline{bias}}
\DeclareMathOperator{\minbias}{\underline{bias}}

\DeclareMathOperator*{\argmin}{argmin}

\DeclareMathOperator{\var}{var}
\DeclareMathOperator{\sd}{sd}

\newcommand\independent{\protect\mathpalette{\protect\independenT}{\perp}}
\def\independenT#1#2{\mathrel{\rlap{$#1#2$}\mkern2mu{#1#2}}}

\newcommand{\dimx}{p}           
\newcommand{\pscore}{e}           

\newcommand{\normx}[1]{\norm{#1}_{\mathcal{X}}} %

\usepackage{xcolor}%
\definecolor{webbrown}{rgb}{.6,0,0}%
\usepackage{hyperref} %

\usepackage{listings}
\definecolor{codegreen}{rgb}{0,0.6,0}
\definecolor{codegray}{rgb}{0.5,0.5,0.5}
\definecolor{codepurple}{rgb}{0.58,0,0.82}
\definecolor{backcolour}{rgb}{0.95,0.95,0.92}
\lstdefinestyle{mystyle}{
    commentstyle=\color{codegreen},
    keywordstyle=\ttfamily,
    stringstyle=\color{codepurple},
    basicstyle=\ttfamily\footnotesize,
    breakatwhitespace=false,
    numberstyle=\tiny\color{codegray},
    numbers=left,
    breaklines=true,
    captionpos=b,
    keepspaces=true,
    numbersep=5pt,
    showspaces=false,
    showstringspaces=false,
    showtabs=false,
    tabsize=2
}
\usepackage{cleveref}
\crefname{lstlisting}{listing}{listings}
\crefname{appsec}{appendix}{appendices}
\crefname{sappsec}{Supplemental Appendix}{Supplemental Appendices}
\crefname{appsubsec}{appendix}{appendices} %
\crefname{assumption}{assumption}{assumptions}

\usepackage[nolist]{acronym}
\begin{acronym}
  \acro{CI}{confidence interval}%
  \acro{FLCI}{fixed-length confidence interval}%
  \acro{RMSE}{root mean squared error}%
  \acro{ATE}{average treatment effect}%
  \acro{CATE}{conditional average treatment effect}%
  \acro{CATT}{conditional average treatment effect on the treated}%
  \acro{PATE}{population average treatment effect}%
  \acro{PATT}{population average treatment effect on the treated}%
  \acro{CLT}{central limit theorem}
  \acro{OLS}{ordinary least squares}
  \acro{ACI}{average coverage interval}
  \acro{CZ}{commuting zone}
  \acro{FPLIB}{flat prior limited information Bayes}
  \acro{PMT}{posterior mean trimming}
  \acro{WLS}{weighted least squares}
  \acro{DGP}{data generating process}
  \acro{NSW}{National Supported Work}
  \acro{BLP}{best linear predictor}
  \acroplural{DGP}[DGPs]{data generating processes}
\end{acronym}
\usepackage{appendix}

\hypersetup{%
  breaklinks = true,%
  colorlinks = true,%
  anchorcolor = webbrown,%
  citecolor = webbrown,%
  filecolor = webbrown,%
  linkcolor = webbrown,%
  menucolor = webbrown,%
  urlcolor= webbrown,%
  citebordercolor= 1 0 0,%
  menubordercolor=1 0 0,%
  urlbordercolor=1 0 0,%
  runbordercolor=1 0 0,%
  pdftitle=Finite-Sample Optimal Estimation and Inference on Average Treatment Effects Under Unconfoundedness,%
  pdfauthor=Tim Armstrong \& Michal Kolesár}

\title{Finite-Sample Optimal Estimation and Inference on Average Treatment
  Effects Under Unconfoundedness\thanks{We thank Xiaohong Chen, Jin Hahn, Guido
    Imbens, Luke Miratrix, Pepe Montiel Olea, Christoph Rothe, Pedro Sant'Anna,
    Andres Santos, Azeem Shaikh, Jeff Smith, and Alex Torgovitsky for
    illuminating discussions. We also thank the editor, four anonymous referees,
    and numerous seminar and conference participants for helpful comments and
    suggestions. All errors are our own. Armstrong acknowledges support by the
    National Science Foundation Grant SES-1628939. Kolesár acknowledges support
    by the Sloan Research Fellowship, and by the National Science Foundation
    Grant SES-1628878.}}
\author{Timothy B. Armstrong\thanks{email: \texttt{timothy.armstrong@yale.edu}}\\
  Yale University \and
  Michal Kolesár\thanks{email: \texttt{mkolesar@princeton.edu}}\\
  Princeton University}%
\date{\today}

\begin{document}
\maketitle

\begin{abstract}
  We consider estimation and inference on average treatment effects under
  unconfoundedness conditional on the realizations of the treatment variable and
  covariates. Given nonparametric smoothness and/or shape restrictions on the
  conditional mean of the outcome variable, we derive estimators and \acp{CI}
  that are optimal in finite samples when the regression errors are normal with
  known variance. In contrast to conventional \acp{CI}, our \acp{CI} use a larger critical
  value that explicitly takes into account the potential bias of the estimator.
  When the error distribution is unknown, feasible versions of our \acp{CI} are valid
  asymptotically, even when $\sqrt{n}$-inference is not possible due to lack of
  overlap, or low smoothness of the conditional mean. We also derive the minimum
  smoothness conditions on the conditional mean that are necessary for
  $\sqrt{n}$-inference. When the conditional mean is restricted to be Lipschitz
  with a large enough bound on the Lipschitz constant, the optimal estimator
  reduces to a matching estimator with the number of matches set to one. We
  illustrate our methods in an application to the National Supported Work
  Demonstration.
\end{abstract}

\clearpage

\section{Introduction}

To estimate the \ac{ATE} of a binary treatment in observational studies, it is
often assumed that the treatment is unconfounded given a set of pretreatment
covariates. This assumption implies that systematic differences in outcomes
between treated and control units with the same values of the covariates are
attributable to the treatment. When the covariates are continuously distributed,
it is not possible to perfectly match the treated and control units based on
their covariate values, and estimation of the \ac{ATE} requires nonparametric
regularization methods such as kernel, series or sieve estimators, or matching
estimators that allow for imperfect matches.

Many of these estimators are $\sqrt{n}$-consistent, asymptotically unbiased and
normally distributed, provided that, in addition to unconfoundedness, one also
assumes overlap of the covariate distributions of treated and untreated
subpopulations, as well as enough smoothness of either the propensity score or
the conditional mean of the outcome given the treatment and covariates
\citep[see, among
others,][]{hahn_role_1998,heckman_matching_1998,hirano_efficient_2003,chen_semiparametric_2008}.
The standard approach to constructing \acfp{CI}, which we refer to as
$\sqrt{n}$-inference, is to take any such estimator, and add and subtract its
standard deviation times the conventional 1.96 critical value (for nominal 95\%
\acp{CI}).

However, in many applications, this approach may not perform well for two
reasons. First, often the overlap is limited, which may lead to failure of
asymptotic normality and slower convergence rates
\citep[e.g.][]{KhTa10,KhNe13,busso_new_2014,rothe_robust_2017}.\footnote{To deal
  with limited overlap, one can redefine the estimand to a treatment effect for
  a subset of the population for which overlap holds, as in
  \citet{heckman_matching_1997}, \citet{galiani_water_2005},
  \citet{bailey_war_2015} or \citet{crump_dealing_2009}. However, this estimand
  is typically less policy relevant.} Second, even under perfect overlap,
asymptotic unbiasedness requires a lot of smoothness: one typically needs
continuous differentiability of the order $\dimx/2$ at minimum
\citep[e.g.][]{chen_semiparametric_2008}, and often of the order $\dimx+1$ or
higher
\citep[e.g.][]{hahn_role_1998,heckman_matching_1998,hirano_efficient_2003},
where $\dimx$ is the dimension of the covariates. Unless $\dimx$ is very small,
such assumptions are hard to evaluate, and may be much stronger than the
researcher is willing to impose. Furthermore, even if the asymptotic bias is
negligible, the actual finite-sample bias may affect coverage of standard
\acp{CI} \citep[e.g.][]{robins_toward_1997}.

In this paper, we instead treat the smoothness and/or shape restrictions on the
conditional mean of the outcome given the treatment and covariates as given and
determined by the researcher. Given these restrictions, we show how to construct
finite-sample valid \acp{CI} based on any estimator $\sum_{i=1}^{n}k_{i}Y_{i}$
that is linear in the outcomes $Y_{i}$, where the weights $\{k_{i}\}_{i=1}^{n}$
depend on the covariates and treatments for the entire
sample.\footnote{\label{fn:centrosymmetry}If the restrictions are asymmetric, we
  allow for the estimators to be affine rather than linear, taking the form
  $a+\sum_{i=1}^{n}k_{i}Y_{i}$ where the intercept $a$ may also depend on the
  covariates and treatment. See \Cref{proofs_sec_append}.} To do so, we assume
that the regression errors are normal with known variance, and we view the
treatment and covariates as fixed. The latter allows us to explicitly calculate
the worst-case finite-sample bias of the estimator under the maintained
restrictions on the conditional mean. Our \acp{CI} are constructed by simply
adding and subtracting the estimator's standard error times a critical value
that is larger than the usual $1.96$ critical value, and takes into account the
potential bias of the estimator.\footnote{The worst-case bias calculations
  require the researcher to fully specify the restrictions on the conditional
  mean, including any smoothness constants. The results in
  \Cref{sec:bounds-adaptation} imply that a priori specification of the
  smoothness constants is unavoidable, and we therefore recommend reporting
  \acp{CI} for a range of smoothness constants as a form of sensitivity
  analysis.} We then show how to choose the weights $\{k_{i}\}_{i=1}^{n}$
optimally, in order to minimize the \ac{RMSE} of the estimator, or the length of
the resulting \ac{CI}: one needs to solve a finite-sample bias-variance tradeoff
problem, which can be cast as a convex programming problem. Furthermore, we show
that, once the weights are optimized, such estimators and \acp{CI} are highly
efficient among all procedures.

To make further progress on characterizing the optimal weights, we focus on the
case where the conditional mean is assumed to satisfy a Lipschitz constraint. We
show that, for a given sample size, the optimal estimator (both for \ac{RMSE}
and \ac{CI} length) reduces to a matching estimator with a single match when the
Lipschitz constant is large enough. While the optimal estimator does not admit a
closed form in general, we develop a computationally fast algorithm that traces
out the weights as a function of the Lipschitz constant, analogous to the least
angle regression algorithm for the LASSO solution path \citep{ehjt04}.

When the assumption of normal errors and known variance is dropped, feasible
versions of our \acp{CI} are valid asymptotically, uniformly in the underlying
distribution \citep[i.e. they are honest in the sense of][]{li89honest}.
Importantly, this result obtains whether $\sqrt{n}$-inference is possible or
whether it is impossible, be it due to low regularity of the regression
function\footnote{\label{fn:smoothness_bound}We show that for
  $\sqrt{n}$-inference to be possible, one needs to assume a bound on the
  derivative of the conditional mean of order at least $\dimx/2$. If one only
  bounds derivatives of lower order, the bias will asymptotically dominate the
  variance---in contrast to, say, estimation of a conditional mean at a point,
  it is not possible to ``undersmooth''.}, limited overlap, or complete lack of
overlap; we do not even require that the \ac{ATE} be point identified. As we
further discuss in \Cref{remark:feasible_finsamp_efficiency}, our efficiency
theory is analogous to the classic theory in the linear regression model, where
finite-sample optimality results rely on the strong assumption of normal
homoskedastic errors, but asymptotic validity of \acp{CI} based on
heteroskedasticity robust standard errors obtains under weak assumptions.

The key condition underlying the asymptotic validity of our \acp{CI} is that the
estimator doesn't put too much weight $k_{i}$ on any individual observation:
this ensures that the estimator is asymptotically normal when normalized by its
standard deviation. However, since the weights solve a bias-variance tradeoff,
no single observation can receive too much weight---otherwise, in large samples,
regardless of the amount of overlap, a large decrease in variance could be
achieved at a small cost to bias. On the other hand, asymptotic normality may
fail under limited overlap for other estimators: we show that asymptotic
normality for matching estimators generally requires strong overlap.

We illustrate the results in an application to the \ac{NSW} Demonstration. We
find that our \acp{CI} are substantially different from those based on
conventional $\sqrt{n}$-asymptotic theory, with bias determining a substantial
portion of the \ac{CI} width. In line with the theoretical results above, we
also find evidence that, in contrast to the estimator we propose, the weights
for the matching estimator are too large for a normal approximation to its
sampling distribution to be reliable.

Our results rely on the key insight that, if we condition on the treatment and
covariates, the \ac{ATE} is a linear functional of a regression function. This
puts the problem in the general framework of \citet{donoho94} and \citet{CaLo04}
and allows us to apply the sharp efficiency bounds in \citet{ArKo18optimal}. The
form of the optimal estimator and \acp{CI} follows by applying the general
framework. The rest of our finite-sample results, as well as all asymptotic
results, are novel and require substantial further analysis. In particular,
solving for the optimal weights $k_{i}$ in general requires solving an
optimization problem over the space of functions in $\dimx$ variables, which
makes simple strategies, such as gridding, infeasible unless the dimension of
covariates $\dimx$ is very small. We show that under Lipschitz smoothness, the
problem can be recast so that the computational complexity depends only on $n$
and not on $p$, and our solution algorithm uses insights from
\citet{rosset_piecewise_2007} to further speed up the computation. In
independent and contemporaneous work, \citet{kallus2017} computes optimal linear
weights using a different characterization of the optimization problem.

In contrast, without conditioning on the treatment and covariates, the problem
is more difficult: while upper and lower bounds for the rate of convergence have
been developed \citep{robins_semiparametric_2009}, efficiency bounds that are
sharp in finite samples remain elusive. Whether one should condition on the
treatment and covariates when evaluating estimators and \acp{CI} is itself an
interesting question. A previous version of this paper \citep[][Section
6.4]{ArKo18ate} considers this question in the context of our empirical
application, and \citet{AbImZh14,aaiw14} give a discussion in related settings.
Since our \acp{CI} are valid unconditionally, they can be used in either
setting.\footnote{While we focus on a treatment effect that conditions on
  realized covariates in the sample, our approach can also be used to construct
  \acp{CI} for the population \ac{ATE}\@; see~\Cref{sec:pate}.}

The remainder of this paper is organized as follows. \Cref{results_sec} presents
the model and gives the main finite-sample results.
\Cref{practical_implementation_sec} considers practical implementation issues.
\Cref{sec:asymptotic-results} presents asymptotic results. \Cref{nsw_sec}
discusses an application to the \ac{NSW}. Additional results, details, and
proofs are collected in the appendices and the supplemental materials.

\section{Setup and finite-sample results}\label{results_sec}

This section sets up the model, and shows how to construct finite-sample optimal
estimators and well as finite-sample valid and optimal \acp{CI} under general
smoothness restrictions on the conditional mean of the outcome. We then
specialize the results to the case with Lipschitz smoothness. Proofs and
additional details are given in \Cref{proofs_sec_append}.

\subsection{Setup}\label{sec:setup}
We have a random sample of $n$ units. Each unit $i=1, \dotsc, n$ is
characterized by a pair of potential outcomes $Y_{i}(0)$ and $Y_{i}(1)$ under no
treatment and treatment, respectively, a covariate vector
$X_{i}\in\mathbb{R}^{\dimx}$, and a treatment indicator $D_{i}\in\{0,1\}$.
Unless stated otherwise, we condition on the realized values
$\{x_{i}, d_{i}\}_{i=1}^{n}$ of the covariates and treatment
$\{X_{i}, D_{i}\}_{i=1}^{n}$, so that probability statements are with respect to
the conditional distribution of $\{Y_{i}(0), Y_{i}(1)\}_{i=1}^{n}$. The realized
outcome is given by $Y_{i}=Y_{i}(1)d_{i}+Y_{i}(0)(1-d_{i})$.  Letting
$f(x_{i}, d_{i})$ denote the conditional mean of $Y_{i}$, we obtain a fixed
design regression model
\begin{equation}\label{fixed_design_eq}
  Y_{i}=f(x_{i}, d_{i}) + u_{i},
  \qquad\text{$\{u_{i}\}_{i=1}^n$ mutually independent with $E(u_i)=0$.}
\end{equation}
We are interested in the \acf{CATE}\footnote{\label{fn:cate-terminology}We note
  that the terminology varies in the literature. Some papers call this object
  the sample average treatment effect (SATE); other papers use the terms
  \ac{CATE} and SATE for different objects entirely.}, which, under the
assumption of unconfoundedness,
$(Y_{i}(1), Y_{i}(0))\independent D_{i}\mid X_{i}$, is given by
\begin{equation}\label{eq:cate}
  Lf=\frac{1}{n}\sum_{i=1}^{n} [f(x_i,1)-f(x_i,0)]=\frac{1}{n}\sum_{i=1}^{n}
  E[Y_{i}(1)-Y_{i}(0) \mid X_{i}=x_{i}].
\end{equation}
To obtain finite-sample results, we further assume that $u_{i}$ is normal,
\begin{equation}\label{normal_u_eq}
  u_{i} \sim N(0, \sigma^2(x_i, d_i)),
\end{equation}
and that the variance function $\sigma^2(x_i, d_i)$ is known. We relax both
assumptions in \Cref{practical_implementation_sec}.

We assume that $f$ lies in a known function class $\mathcal{F}$, which
formalizes the ``regularity'' or ``smoothness'' that we are willing to impose.
We require that $\mathcal{F}$ be convex and centrosymmetric, i.e.\ that
$f\in\mathcal{F}$ implies $-f\in\mathcal{F}$. While convexity is essential for
most of our results, centrosymmetry can be relaxed---see
\Cref{proofs_sec_append}. Our setup covers classical nonparametric function
classes, which place bounds on (possibly higher order) derivatives of $f$. As a
leading example, we place Lipschitz constraints on $f(\cdot,0)$ and
$f(\cdot,1)$:
\begin{equation}\label{eq:lipschitz-class}
  \FLip(C)=\{f\colon \abs{f(x, d)-f(\tilde x, d)}\le C
  \normx{x-\tilde x}, \, d\in \{0,1\}\},
\end{equation}
where $\normx{\cdot}$ is a norm on $x$, and $C$ denotes the Lipschitz constant,
which for simplicity we take to be the same for both $f(\cdot, 1)$ and
$f(\cdot, 0)$. As we discuss in \Cref{sec:bounds-adaptation}, the function class
$\mathcal{F}$ needs to be specified ex ante: for the Lipschitz case, for
instance, one cannot use data-driven procedures to estimate $C$, or pick the
norm $\normx{\cdot}$. We discuss how to make these choices in practice in the
context of our empirical application in \Cref{nsw_sec}.

Our goal is to construct estimators and confidence sets for the \ac{CATE}
parameter $Lf$. Letting $P_{f}$ denote the probability computed under $f$, a set
$\mathcal{C}$ is a $100\cdot (1-\alpha)\%$ confidence set for $Lf$ if
\begin{equation}\label{coverage_eq}
  \inf_{f\in\mathcal{F}} P_{f}(Lf \in\mathcal{C})\ge 1-\alpha.
\end{equation}

\subsection{Linear estimators}\label{linear_estimators_sec}
We start by showing how to construct \acp{CI} based on estimators that are linear in the
outcomes,
\begin{equation}\label{linear_estimator_eq}
  \hat L_{k}=\sum_{i=1}^n k(x_i, d_i)Y_{i}.
\end{equation}
For now, we treat the weights $k$ as given---in~\Cref{sec:optim-estim-cis}, we
will show how to choose them optimally for a general class of criteria that
include \ac{RMSE} and \ac{CI} length. In \Cref{sec:bounds-adaptation}
and \Cref{proofs_sec_append}, we show that, if we choose the weights optimally,
the resulting estimator and \acp{CI} are optimal or near optimal among all
procedures, including nonlinear ones. The class of linear estimators covers many
estimators that are popular in practice, such as series or kernel estimators, or
various matching estimators.\footnote{\label{fn:non-linear-estimators}Nonlinear
  estimators include those based on regression trees and artificial neural
  networks, as well as those using nonlinear thresholding to perform variable
  selection; see \citet{donoho_minimax_1998} for a discussion of cases where, in
  contrast to the present setting, nonlinear estimators outperform linear
  estimators.} For example, the matching estimator with $M$ matches that matches
(with replacement) on covariates takes the form $L\hat{f}_{M}$, where
$\hat{f}_{M}(x_{i}, d_{i})=Y_{i}$, and
$\hat{f}_{M}(x_{i},1-d_{i})=\sum_{j=1}^{n}W_{M, ij} Y_{j}$. Here $W_{M, ij}=1/M$
if $j$ is among the $M$ observations with treatment status $d_{j}=1-d_{i}$ that
are the closest to $i$ (using the norm $\normx{\cdot}$), and zero otherwise. For
this estimator, the weights take the form
\begin{equation}\label{match_weight_eq}
  k_{\text{match}, M}(x_i, d_i) = \frac{1}{n}(2d_{i}-1)
  \left(1+\frac{K_{M}(i)}{M}\right),
\end{equation}
where $K_{M}(i)=M\sum_{j=1}^{n}W_{M, ji}$ is the number of times observation $i$
is matched.

The estimator $\hat{L}_{k}$ is normally distributed with variance
$\sd(\hat L_{k})^2=\sum_{i=1}^{n} k(x_i, d_i)^2\sigma^2(x_i, d_i)$ and maximum
bias
\begin{equation}\label{eq:maxbias}
  \maxbias_{\mathcal{F}}(\hat{L}_{k})=\sup_{f\in\mathcal{F}} E_f(\hat L_{k}-Lf)
  =\sup_{f\in\mathcal{F}} \left[\sum_{i=1}^{n} k(x_i, d_i)f(x_i, d_i)-Lf\right].
\end{equation}
By centrosymmetry of $\mathcal{F}$, the minimum bias is given by
$-\maxbias_{\mathcal{F}}(\hat{L}_{k})$.

We form one-sided $100\cdot (1-\alpha)\%$ \acp{CI} based on $\hat L_{k}$, as
\begin{equation}\label{eq:onesided-ci}
\hor{\hat{c}, \infty}, \qquad \hat c=\hat L_{k}
  -\maxbias_{\mathcal{F}}(\hat L_{k})
  -z_{1-\alpha}\sd(\hat L_{k}),
\end{equation}
where $z_{1-\alpha}$ is the $1-\alpha$ quantile of a standard normal
distribution. Subtracting the maximum bias, in addition to subtracting
$z_{1-\alpha}\sd(\hat L_{k})$, is necessary to prevent undercoverage.

One could form a two-sided \ac{CI} centered at $\hat{L}_{k}$ by adding and
subtracting
$\maxbias_{\mathcal{F}}(\hat{L}_{k})+z_{1-\alpha/2}\sd(\hat{L}_{k})$. However,
this is conservative since the bias cannot be equal to
$\maxbias_{\mathcal{F}}(\hat{L}_{k})$ and to
$-\maxbias_{\mathcal{F}}(\hat{L}_{k})$ at once. Instead, observe that under any
$f\in\mathcal{F}$, the $z$-statistic $(\hat L_{k}-Lf)/\sd(\hat L_{k})$ is
distributed $N(t,1)$, where $t=E_{f}(\hat{L}_{k}-Lf)/\sd(\hat L_{k})$, and $t$
is bounded in absolute value by
$b=\maxbias_{\mathcal{F}}(\hat{L}_{k})/\sd(\hat L_{k})$, the ratio of the
worst-case bias to standard deviation. Thus, denoting the $1-\alpha$ quantile of
a $\abs{N(b,1)}$ distribution by $\cv_{\alpha}(b)$, a two-sided \ac{CI} can be
formed as
\begin{equation}\label{eq:flci}
  \left\{\hat L_{k}\pm \cv_\alpha(\maxbias_{\mathcal{F}}(\hat L_{k})/\sd(\hat L_{k}))
    \cdot \sd(\hat L_{k})\right\}.
\end{equation}
If $\hat{L}_{k}$ is unbiased, the critical value reduces to the conventional
critical value: $\cv_{\alpha}(0)=z_{1-\alpha/2}$. If $b>0$, it
will be larger: for $b\geq 1.5$ and $\alpha\leq 0.2$,
$\cv_{\alpha}(b)\approx b+z_{1-\alpha}$ up to three decimal
places,\footnote{\label{fn:compute_cv}The critical value $\cv_{1-\alpha}(b)$ can
  be computed as the square root of the $1-\alpha$ quantile of a non-central
  $\chi^{2}$ distribution with $1$ degree of freedom and non-centrality
  parameter $b^{2}$.}
Following \citet{donoho94}, we refer to the \ac{CI} in~\eqref{eq:flci} as a
\ac{FLCI}, since its length does not depend on the realized outcomes, only on
the known variance function $\sigma^{2}(\cdot, \cdot)$ and on $\{x_{i},
d_{i}\}_{i=1}^{n}$.

\subsection{Optimal linear estimators and \texorpdfstring{\acp{CI}}{CIs}}\label{sec:optim-estim-cis}
We now show how to choose the weights $k$ optimally. To that end, we need to
define the criteria that we wish to optimize. To evaluate estimators, we
consider their maximum \acf{RMSE},
\begin{equation}\label{eq:rmse}
  R_{\RMSE, \mathcal{F}}(\hat L_{k})
  =(\textstyle\sup_{f\in\mathcal{F}}E_{f}(\hat{L}_{k}-Lf)^{2})^{1/2}
  =(\maxbias_{\mathcal{F}}(\hat L_{k})^2+\sd(\hat L_{k})^2)^{1/2}.
\end{equation}
One-sided \acp{CI} can be compared using quantiles of excess length (see
\Cref{proofs_sec_append}). Finally, to evaluate \acp{FLCI}, we simply consider
their length,
$2\cv_{\alpha}(\maxbias_{\mathcal{F}}(\hat{L}_{k})/\sd(\hat{L}_{k}))\cdot
\sd(\hat{L}_{k})$. Since the length is fixed---it doesn't depend on the data
$\{Y_{i}\}_{i=1}^{n}$---choosing the weights
$k$ to minimize the length does not affect the coverage properties of the
resulting \ac{CI}.

Both performance criteria---\ac{FLCI} and \ac{RMSE}---depend on $k$ only through
$\maxbias_{\mathcal{F}}(\hat L_{k})$, and $\sd(\hat L_{k})$, and they are
increasing in both quantities (this is also true for performance of one-sided
\acp{CI}; see \Cref{proofs_sec_append}). Therefore, to find the optimal weights,
it suffices to first find weights that minimize the worst-case bias
$\maxbias_{\mathcal{F}}(\hat{L}_{k})$ subject to a bound on variance. We can
then vary the bound to find the optimal bias-variance tradeoff for a given
performance criterion (\ac{FLCI} or \ac{RMSE}). It follows from \citet{donoho94}
and \citet{low95} that this bias-variance frontier can be traced out by solving
a convex optimization indexed by a parameter $\delta$ that plays a role
analogous to that of a bandwidth; it can be thought of as indexing the relative
weight on the variance. Varying $\delta$ then traces out the optimal
bias-variance frontier.

For a simple statement of the Donoho-Low result, assume that the parameter space
$\mathcal{F}$, in addition to being convex and centrosymmetric, does not
restrict the value of \ac{CATE} in the sense that the function
$\iota_{\kappa}(x, d)=\kappa d$ lies in $\mathcal{F}$ for all
$\kappa\in\mathbb{R}$ (see \Cref{proofs_sec_append} for a general
statement)\footnote{\label{fn:closed-F}We also assume the regularity condition
  that if $\lambda f+\iota_{\kappa}\in\mathcal{F}$ for all $0\leq \lambda< 1$,
  then $f+\iota_{\kappa}\in \mathcal{F}$. Since $L\iota_{\kappa}=\kappa$,
  $\{\iota_{\kappa}\}_{\kappa\in\mathbb{R}}$ is the smoothest set of functions
  that span the potential values of the \ac{CATE} parameter, so this assumption
  will typically hold if the possible values of $Lf$ are unrestricted.}. Given
$\delta>0$, let $f^{*}_{\delta}$ solve
\begin{equation}\label{eq:single-class-modulus-problem}
  \max_{f\in\mathcal{F}} \, 2 Lf
  \quad\text{s.t.}\quad
  {\sum_{i=1}^n\frac{f(x_i, d_i)^2}{\sigma^2(x_i, d_i)}}\le \frac{\delta^{2}}{4}.
\end{equation}
With a slight abuse of notation, define
\begin{equation}\label{eq:optimal-weights}
  \hat{L}_{\delta}=\hat{L}_{k^{*}_{\delta}}, \qquad
  k^*_\delta(x_i, d_i)=\frac{f^*_\delta(x_i, d_i)/\sigma^2(x_i, d_i)}{
    \sum_{j=1}^{n}d_{j}f^*_\delta(x_{j}, d_{j})/\sigma^2(x_{j}, d_{j})}.
\end{equation}
Then the maximum bias of $\hat{L}_{\delta}$ occurs at $-f^{*}_{\delta}$, and the
minimum bias occurs at $f^{*}_{\delta}$, so that
\begin{equation*}
  \maxbias_{\mathcal{F}}(\hat{L}_{\delta})=
  \frac{1}{n}\sum_{i=1}^n[f^*_\delta(x_i,1)-f^*_\delta(x_i,0)]
  -\sum_{i=1}^n k^*_\delta(x_i, d_i)f^*_\delta(x_i, d_i).
\end{equation*}
Also, $\hat{L}_{\delta}$ minimizes the worst-case bias among all linear
estimators with variance bounded by
$\sd(\hat{L}_{\delta})^{2}=\delta^{2}/(2\sum_{j=1}^{n}d_{j}f^*_\delta(x_{j},
d_{j})/\sigma^2(x_{j}, d_{j}))^{2}$. Thus, the estimators
$\{\hat{L}_{\delta}\}_{\delta>0}$ trace out the optimal bias-variance frontier.

The weights leading to the shortest possible \ac{FLCI} are given by
$k^{*}_{\delta_{\FLCI}}$, where $\delta_{\FLCI}$ minimizes
$\cv_{\alpha}(\maxbias_{\mathcal{F}}(\hat{L}_{\delta})/\sd(\hat{L}_{\delta}))\cdot\sd(\hat{L}_{\delta})$
over $\delta$. Similarly, the optimal weights for estimation are given by
$k^{*}_{\delta_{\RMSE}}$, where $\delta_{\RMSE}$ minimizes
$\maxbias_{\mathcal{F}}(\hat{L}_{\delta})^{2}+\sd(\hat{L}_{\delta})^{2}$.

\subsection{Estimators and \texorpdfstring{\acp{CI}}{CIs} under Lipschitz
  smoothness}\label{lipschitz_optimal_sec}

Computing a \ac{FLCI} based on a linear estimator $\hat{L}_{k}$ with a given set
of weights $k$ requires computing the worst-case bias~\eqref{eq:maxbias}.
Computing the \ac{RMSE}-optimal estimator, and the optimal \ac{FLCI} requires
solving the optimization problem~\eqref{eq:single-class-modulus-problem}, and
then varying $\delta$ to find the optimal bias-variance tradeoff. Both
optimization problems require optimizing over the set $\mathcal{F}$, which, in
nonparametric settings, is infinite-dimensional. We now focus on the Lipschitz
class $\mathcal{F}=\FLip(C)$, and show that in this case, the solution
to~\eqref{eq:maxbias} can be found by solving a finite-dimensional linear
program. The optimization problem~\eqref{eq:single-class-modulus-problem} can be
cast as a finite-dimensional convex program. Furthermore, if the program is put
into a Lagrangian form, then the solution is a piecewise linear function of the
Lagrange multiplier, and one can trace the entire solution path
$\{\hat{L}_{\delta}\}_{\delta>0}$ using an algorithm similar to the LASSO/LAR
algorithm of \citet{ehjt04}.

We leverage the fact that in both optimization problems, we can identify
$f$ with the vector
$(f(x_1,0), \dotsc, f(x_n,0), f(x_1,1), \dotsc, \allowbreak f(x_n,1))' \in
\mathbb{R}^{2n}$, and replace the functional constraint
$f\in\mathcal{F}=\FLip(C)$ with $2n(n-1)$ linear inequality constraints
\begin{equation}\label{eq:lipschitz-constraints-finite}
  f(x_i, d)-f(x_j, d)\le C\normx{x_{i}-x_{j}}
 \quad d\in\{0,1\}, \; i, j\in \{1,\dotsc, n\}.
\end{equation}
This fact follows from the observation that both in~\eqref{eq:maxbias} and
in~\eqref{eq:single-class-modulus-problem}, the objective and constraints depend
on $f$ only through its value at these $2n$ points, and from the result that if
the Lipschitz constraints hold at these points, it is always possible to find a
function $f\in\FLip(C)$ that interpolates these points \citep[Theorem 4, which
we restate in \Cref{theorem:beliakov}]{beliakov_interpolation_2006}.

\begin{theorem}\label{k_maxbias_thm}
  Consider a linear estimator~\eqref{linear_estimator_eq} with weights $k$ that satisfy
  \begin{equation}\label{k_finite_bias_eq}
    \sum_{i=1}^n d_{i}k(x_i, d_i)=1, \quad
    \text{and}\quad \sum_{i=1}^n (1-d_{i})k(x_i, d_i)=-1.
  \end{equation}
  Then
  \begin{equation}\label{k_max_bias_eq}
    \maxbias_{\FLip(C)}(\hat L_{k})=
    \max_{f\in\mathbb{R}^{2n}}\left\{\sum_{i=1}^{n}k(x_i, d_i)f(x_i, d_i)
      -\frac{1}{n}\sum_{i=1}^{n}\left[f(x_i,1)-f(x_i,0)\right]\right\},\,
    \text{s.t.~\eqref{eq:lipschitz-constraints-finite}.}
  \end{equation}
  Furthermore, if $k(x_{i}, d_{i})\geq 1/n$ if $d_{i}=1$ and
  $k(x_{i}, d_{i})\leq -1/n$ if $d_{i}=0$, it suffices to
  impose~\eqref{eq:lipschitz-constraints-finite} for $i, j\in\{1,\dotsc, n\}$ with
  $d_{i}=1$, $d_{j}=0$ such that either $k(x_{i},1)>1/n$, or $k(x_{j},0)<-1/n$.
\end{theorem}
The assumption that $\hat{L}_{k}$ satisfies~\eqref{k_finite_bias_eq} is
necessary to prevent the bias from becoming arbitrarily large at multiples of
$f(x, d)=d$ and $f(x, d)=1-d$. \Cref{k_maxbias_thm} implies that the formulas
for one-sided \acp{CI} and two-sided \acp{FLCI} given in
\Cref{linear_estimators_sec} hold with $\maxbias_{\FLip(C)}(\hat{L}_{k})$ given
in~\eqref{k_max_bias_eq}. The last part of the theorem says that it suffices to
impose at most $2n_{0}n_{1}$ of the constraints
in~\eqref{eq:lipschitz-constraints-finite}, where $n_{d}$ is the number of
observations with $d_{i}=d$. The condition on the weights $k$ holds, for
example, for the matching estimator given in~\eqref{match_weight_eq}. Since for
the matching estimator $k(x_{i}, d_{i})=(2d_{i}-1)/n$ if observation $i$ is not
used as a match, the theorem says that one only needs to
impose~\eqref{eq:lipschitz-constraints-finite} for pairs of observations with
opposite treatment status for which one of the observations is used as a match.

For \ac{RMSE}-optimal estimators and optimal \acp{FLCI}, we have the following result:
\begin{theorem}\label{lipschitz_optimal_thm}
  Given $\delta>0$, the value of the maximizer $f^{*}_{\delta}$
  of~\eqref{eq:single-class-modulus-problem} under $\mathcal{F}=\FLip(C)$ is
  given by the solution to the convex program
  \begin{equation}\label{eq:modulus-lipchitz}
    \max_{f\in\mathbb{R}^{2n}} \, 2Lf
    \quad\text{s.t.}\quad
    {\sum_{i=1}^n\frac{f(x_i, d_i)^2}{\sigma^2(x_i, d_i)}}\le
    \frac{\delta^{2}}{4}
    \quad\text{and s.t.\ \eqref{eq:lipschitz-constraints-finite}.}
  \end{equation}
  Furthermore, if $\sigma^{2}(x, d)$ doesn't depend on $x$, it suffices to impose
  the constraints~\eqref{eq:lipschitz-constraints-finite} for
  $i, j\in\{1,\dotsc, n\}$ with $d_{i}=0$ and $d_{j}=1$, and the solution path
  $\{f^{*}_{\delta}\}_{\delta>0}$ can be computed by the piecewise linear
  algorithm given in \Cref{sec:opt_proof}.
\end{theorem}
\Cref{lipschitz_optimal_thm} reduces the infinite-dimensional
program~\eqref{eq:single-class-modulus-problem} to a quadratic optimization
problem in $\mathbb{R}^{2n}$ with one quadratic and $2n(n-1)$ linear
constraints, and a linear objective function. The computational difficulty can
be shown to be polynomial in $n$, and it does not depend on the covariate
dimension $p$. If the variance is homoskedastic for each treatment group, then
the number of linear constraints can be reduced to $2n_{0}n_{1}$, and the entire
solution path can be computed efficiently using the piecewise linear algorithm
given in \Cref{sec:opt_proof}. As a result, implementing the estimator is quite
fast: the main specification in the empirical application in \Cref{nsw_sec}
takes less than a minute on a laptop computer.

As we discuss in more detail in \Cref{sec:opt_proof}, it follows from the
algorithm that, similarly to the matching estimator (see
\cref{match_weight_eq}), the optimal estimator takes the form
$\hat{L}_{\delta}=L\hat{f}_{\delta}$, where
$\hat{f}_{\delta}(x_{i}, d_{i})=Y_{i}$, and
$\hat{f}_{\delta}(x_{i},1-d_{i})=\sum_{j=1}^{n}W_{\delta, ij}Y_{j}$, and the
weights $W_{\delta, ij}$ correspond to the Lagrange multipliers associated with
the constraints~\eqref{eq:lipschitz-constraints-finite} for $d=d_{i}$, scaled to
sum to one, $\sum_{j=1}^{n}W_{\delta, ij}=1$. The weights are zero unless
$d_{j}=1-d_{i}$ and $j$ is close to $i$ according to a matrix of ``effective
distances.'' The ``effective distance'' between $i$ and $j$ increases with the
total weight $\sum_{i=1}^{n}W_{\delta, ij}$ that we already put on $j$. Thus, we
may interpret observations $j$ with non-zero weight $W_{\delta, ij}$ as being
``matched'' to $i$. The number of matches varies across observations $i$,
increases with $\delta$, and depends on the number of units with opposite
treatment status that are close to $i$ according to the matrix of effective
distances. Observations for which there are more good matches receive relatively
more matches, since this decreases the variance of the estimator at little cost
in terms of bias. On the other hand, the weight
$k^{*}_{\delta}(x_{j}, d_{j})=\frac{1}{n}(1-2d_{j})(1+\sum_{i=1}^{n}W_{\delta,
  ij})$ on $j$ increases with the number of times it has been used as a match,
which increases the variance of the estimator. Using the ``effective distance''
matrix trades off this increase in the variance against an increase in the bias
that results from using a lower-quality match instead.

If the constant $C$ is large enough,
the increase in the bias from using more
than a single match for each $i$ is greater than any reduction in the variance
of the estimator, and the optimal estimator takes the form of a matching
estimator with a single match:
\begin{theorem}\label{match_optimality_thm}
  Suppose that $\sigma(x_{i}, d_{i})>0$ for each $i$, and suppose that each unit
  has a single closest match, so that
  $\argmin_{j\colon d_{j}\neq d_{i}}\normx{x_{i}-x_{j}}$ is a singleton for each
  $i$. Then, if $C$ is larger than a constant that depends only on
  $\sigma^{2}(x_{i}, d_{i})$ and $\{x_i, d_i\}_{i=1}^n$, the optimal estimators
  $\hat L_{\delta_{\RMSE}}$ and $\hat L_{\delta_{\FLCI}}$ are given by the
  matching estimator with $M=1$.
\end{theorem}
The single closest match condition will hold with probability one if conditional
on each treatment value, at least one of the covariates $x_{i}$ is drawn from a
continuous distribution. In contemporaneous work, \citet{kallus2017} gives a
similar result using a different method of proof. %
In the other direction, as $C\to 0$, $\hat L_{\delta_{\RMSE}}$ and
$\hat L_{\delta_{\FLCI}}$ both converge to the difference-in-means estimator
that compares the average outcomes for the treated and untreated units.

\Cref{match_optimality_thm} does not imply that one should choose a large $C$
simply to justify matching with a single match as an optimal estimator: the
chosen value of $C$ should instead represent a priori bounds on the smoothness
of $f$ formulated by the researcher. Nonetheless, if it is difficult to
formulate such bounds, a conservative choice of $C$ may be appropriate. If the
choice is conservative enough, \Cref{match_optimality_thm} will be relevant.

For the optimality result in \Cref{match_optimality_thm}, it is important that
the metric on $x$ used to define the matching estimator is the same as that used
to define the Lipschitz constraint. This formalizes the argument in
\citet{zhao_using_2004} that conditions on the regression function should be
considered when defining the metric used for matching. In the supplemental
materials, we illustrate the efficiency loss from matching with the ``wrong''
metric in the context of our empirical application presented in \Cref{nsw_sec}.

A disadvantage of imposing the Lipschitz condition directly on $f$ is that it
rules out the simple linear model for $f$, unless we impose a priori bounds on
the magnitude of the regression coefficients. In
\Cref{optimal_ci_derivation_sec_append}, we give analogs
of~\Cref{k_maxbias_thm,,lipschitz_optimal_thm,match_optimality_thm} under an
alternative specification for $\mathcal{F}$ that imposes the Lipschitz condition
on $f$ after partialling out the best linear predictor (and thus allows for
unrestricted linear response). We show that in this case, if the constant $C$ is
large enough, the optimal estimator takes the form of a regression-adjusted
matching estimator with a single match.

\subsection{Adaptation bounds and optimality among nonlinear procedures}\label{sec:bounds-adaptation}

The results in \Cref{sec:optim-estim-cis} and \Cref{lipschitz_optimal_thm} show
how to construct \ac{RMSE}-optimal linear estimators, and the shortest \ac{FLCI}
based on a linear estimator. Are these results useful, or do they overly
restrict the class of procedures?

For estimation, \Cref{theorem:donoho-minimax-efficiency} in
\Cref{sec:gener-setup-results} shows that the estimator
$\hat{L}_{\delta_{\RMSE}}$ achieving the lowest \ac{RMSE} in the class of linear
estimators is also highly efficient among all estimators: one cannot reduce the
\ac{RMSE} by more than 10.6\% by considering non-linear estimators in general,
and, in particular applications, its efficiency can be shown to be even higher.

For \acp{FLCI}, an even stronger result obtains, addressing two concerns. First,
their length is determined by the least-favorable function in $\mathcal{F}$
(that maximizes the potential bias), which may result in \acp{CI} that are ``too
long'' when $f$ turns out to be smooth. Consequently, one may prefer a
variable-length \ac{CI} that optimizes its expected length over a class of
smoother functions $\mathcal{G}\subset\mathcal{F}$ (while maintaining coverage
over all of $\mathcal{F}$), especially if this leads to substantial reduction in
expected length when $f\in\mathcal{G}$. When such a \ac{CI} also simultaneously
achieves near-optimal length over all of $\mathcal{F}$, it is referred to as
``adaptive.'' A related second concern is that implementing our \acp{CI} in
practice requires the user to explicitly specify the parameter space
$\mathcal{F}$, which in the case $\mathcal{F}=\FLip(C)$ includes the Lipschitz
constant $C$ and the norm $\normx{\cdot}$. This rules out data-driven procedures
that try to implicitly or explicitly pick $C$ or the norm using the data.

To address these concerns, we show in \Cref{sec:gener-setup-results} that
attempts to form adaptive \acp{CI} cannot substantively improve upon the
\acp{FLCI} we propose. In particular, \Cref{theorem:adaptation-theorem} gives a
sharp bound on the length of a confidence set that optimizes its expected length
at a smooth function of the form $g(x, d)=\kappa_{0}+\kappa_{1}d$, while
maintaining coverage over the original parameter space $\mathcal{F}$. The sharp
bound follows from general results in \citet{ArKo18optimal}, and it gives a
benchmark for the scope for improvement over the \ac{FLCI} centered at
$\hat{L}_{\delta_{\FLCI}}$ (the \namecref{theorem:adaptation-theorem} also gives
an analogous result for one-sided \acp{CI}). The
\namecref{theorem:adaptation-theorem} also gives a universal lower bound for
this sharp bound, which evaluates to 71.7\% when $1-\alpha=0.95$. The sharp
bound depends on the realized values of $\{x_{i}, d_{i}\}_{i=1}^{n}$ and the
variance function $\sigma^{2}(\cdot, \cdot)$, and can be explicitly computed in
a given application. We find that it typically is much higher than this lower
bound. For example, in our empirical application in \Cref{nsw_sec}, the
\ac{FLCI} efficiency is over 97\% at such smooth functions $g$ in our baseline
specification. This implies that there is very little scope for improvement over
the \ac{FLCI}\@.

Consequently, data-driven or adaptive methods for constructing \acp{CI} must
either fail to meaningfully improve over the \ac{FLCI}, or else undercover for
some $f\in\mathcal{F}$. It is thus not possible to, say, estimate the order of
differentiability of $f$, or to estimate the Lipschitz constant $C$ for the
purposes of forming a tighter \ac{CI}---the parameter space $\mathcal{F}$,
including any smoothness constants, must be specified ex ante by the researcher.
Because of this, by way of sensitivity analysis, since the \ac{CI} may
undercover if $C$ is chosen too small in the sense that $\FLip(C)$ excludes the
true $f$, we recommend reporting estimates and \acp{CI} for a range of choices
of $C$ to see how assumptions about the parameter space affect the results (the
estimates and \acp{CI} vary with $C$ because $C$ affects both the optimal tuning
parameters $\delta_{\RMSE}$ and $\delta_{\FLCI}$, and the critical value, via
the worst-case bias). We adopt this approach in the empirical application in
\Cref{nsw_sec}. This mirrors the common practice of reporting results for
different specifications of the regression function in parametric regression
problems.

The key assumption needed for these efficiency bounds is that $\mathcal{F}$ be
convex and centrosymmetric. This holds for $\FLip(C)$, and, more generally, for
parameter spaces that place bounds on derivatives of $f$. If additional
restrictions such as monotonicity are used that break either convexity or
centrosymmetry, then some degree of adaptation may be possible. While we leave
the full exploration of this question for future research, we note that the
approach in~\Cref{sec:optim-estim-cis} can still be used when the centrosymmetry
assumption is dropped. As an example, a previous version of this paper
\citep[][Section 6.4]{ArKo18ate} shows to implement our approach when
$\mathcal{F}$ imposes Lipschitz and monotonicity constraints.

\section{Practical implementation}\label{practical_implementation_sec}

We now discuss implementation of feasible versions of the estimators and
\acp{CI} given in \Cref{results_sec} when the variance function
$\sigma^{2}(x, d)$ is unknown and the errors $u_{i}$ may be non-normal. We
discuss the optimality and validity of these feasible procedures. In
\Cref{sec:pate}, we show how to use our approach to form \acp{CI} for the
\ac{PATE}.

\subsection{Baseline implementation}\label{sec:basel-impl}
As a baseline, we propose the following implementation of our
procedure:\footnote{An R package implementing this procedure, including an
  implementation of the piecewise linear algorithm, is available at
  \url{https://github.com/kolesarm/ATEHonest}.}
\begin{enumerate}
\item\label{item:initial_variance} Let $\tilde\sigma^2(x, d)$ be an initial
  (possibly incorrect) estimate or guess for $\sigma^2(x, d)$. As a default
  choice, we recommend taking $\tilde\sigma^2(x, d)=\hat\sigma^2$ where
  $\hat\sigma^2$ is an estimate of the variance computed under the assumption of
  homoskedasticity.
\item Compute the optimal weights $\{\tilde{k}^{*}_{\delta}\}_{\delta>0}$, using
  $\tilde{\sigma}^{2}(x, d)$ in place of $\sigma^{2}(x, d)$. When
  $\mathcal{F}=\FLip(C)$, and $\tilde\sigma^2(x, d)=\hat\sigma^2$ this can be
  done using the piecewise linear solution path
  $\{\tilde{f}^{*}_{\delta}\}_{\delta>0}$ in \Cref{sec:opt_proof}. Let
  $\tilde{L}_{\delta}=\sum_{i=1}^{n}\tilde{k}^{*}_{\delta}(x_{i}, d_{i})Y_{i}$
  denote the corresponding estimator,
  $\widetilde{\sd}_{\delta}^{2} = \sum_{i=1}^{n} \tilde{k}^{*}_\delta(x_{i},
  d_{i})^{2} \tilde{\sigma}^2(x_{i}, d_{i})$ denote its variance computed using
  $\tilde\sigma^2(x, d)$ as the variance function, and let
  $\maxbias_{\delta}=\maxbias_{\mathcal{F}}(\tilde{L}_{\delta})$ denote its
  worst-case bias (which doesn't depend on the variance specification).
\item\label{item:3} Compute the minimizer $\tilde{\delta}_{\RMSE}$ of
  $\maxbias_{\delta}^{2}+\widetilde{\sd}_{\delta}^{2}$. Compute the standard
  error $\se(\tilde{L}_{\tilde{\delta}_{\RMSE}})$ using the robust variance
  estimator
  \begin{equation}\label{rob_se_eq}
    \se(\tilde L_{\delta})^{2}
    =\sum_{i=1}^{n} \tilde{k}^*_\delta(x_i, d_i)^2\hat u_{i}^{2},
  \end{equation}
  where $\hat{u}^{2}_{i}$ is an estimate of $\sigma^{2}(x_{i}, d_{i})$. Report
  the estimate $\tilde{L}_{\tilde{\delta}_{\RMSE}}$, and the \ac{CI}
  \begin{equation}\label{rob_flci_eq}
    \big\{\tilde{L}_{\delta}
    \pm \cv_\alpha(\maxbias_{\delta} / \se(\tilde{L}_{\delta}))
    \se(\tilde{L}_{\delta}) \big\},
  \end{equation}
  at $\delta=\tilde{\delta}_{\FLCI}$, the minimizer of
  $\cv_{\alpha}(\maxbias_{\delta} / \widetilde{\sd}_\delta)\cdot
  \widetilde{\sd}_\delta$.
\end{enumerate}
The conditional variance, can, for instance, be estimated with the
nearest-neighbor variance estimator of \citet{AbIm06match}
$\hat{u}_{i}=J/(J+1)\cdot (Y_{i}-\hat{f}(x_i, d_i))^{2}$, where
$\hat{f}(x_{i}, d_{i})$ is the average outcome of $J$ observations (excluding
$i$) with treatment status $d_{i}$ that are closest to $i$ according to some
distance. Note that since the length of the feasible \ac{CI} in
\cref{rob_flci_eq} depends on the variance estimates, it is no longer fixed, in
contrast to the infeasible \ac{FLCI}.

In general, the point estimate $\tilde{L}_{\tilde{\delta}_{\RMSE}}$ will differ
from the estimate $\tilde{L}_{\tilde{\delta}_{\FLCI}}$ used to form the \ac{CI}.
Since reporting multiple estimates can be cumbersome, one can simply compute the
\ac{CI}~\eqref{rob_flci_eq} at $\delta=\tilde{\delta}_{\RMSE}$. The \ac{CI} will
then be based on the same estimator reported as a point estimate. While this
leads to some efficiency loss, in our main specification in the empirical
application in \Cref{nsw_sec}, we find that the resulting \ac{CI} is only 1.2\%
longer than the one that reoptimizes $\delta$ for \ac{CI} length.

\begin{remark}[Specification of $\mathcal{F}$]\label{remark:F-specification}
  In forming the \ac{CI}, we need to choose the function class $\mathcal{F}$. In
  case of the Lipschitz class~\eqref{eq:lipschitz-class}, we need to complete
  the specification by choosing the constant $C$ and the norm $\normx{\cdot}$ on
  $x$. The results discussed in \Cref{sec:bounds-adaptation} imply that it is
  not possible to do this automatically in a data-driven way. Thus,
  we recommend that these choices be made using problem-specific knowledge
  wherever possible, and that \acp{CI} be reported for a range of plausible
  values of $C$ as a form of sensitivity analysis. We illustrate this approach
  in our application in \Cref{norm_choice_section}. Note that conducting this
  sensitivity analysis comes at essentially no added computational cost, since
  the solution path $\{\tilde{f}^{*}_{\delta}\}_{\delta>0}$ only needs to be
  computed once. This is because multiplying both $\delta$ and $C$ by any
  constant scales the constraints in~\eqref{eq:modulus-lipchitz}, so that the
  solution simply scales with the given constant. In particular, letting
  $\tilde{f}^*_{\delta, C}$ denote the solution under a given $\delta$ and $C$,
  we have $\tilde{f}^*_{\delta, C}=C \tilde{f}^*_{\delta/C, 1}$.
\end{remark}
\begin{remark}[Efficiency and validity]\label{remark:feasible_finsamp_efficiency}
  How do the finite-sample optimality and validity of the infeasible estimators
  and \acp{CI} discussed in \Cref{results_sec} map into optimality and validity
  properties of the feasible procedures?

  The values $\tilde{\delta}_{\RMSE}$ and $\tilde{\delta}_{\FLCI}$ depend on the
  initial guess $\tilde{\sigma}^{2}(x, d)$. Thus, the resulting \ac{CI} in
  \cref{rob_flci_eq} will not be optimal if this guess is incorrect. However,
  because the standard error estimator~\eqref{rob_se_eq} does not use this
  initial estimate, the \ac{CI} remains asymptotically valid even if
  $\tilde{\sigma}^{2}(x, d)$ is incorrect. Furthermore, we show in
  \Cref{nonnormal_sec} that the \ac{CI} is asymptotically valid even in
  ``irregular'' settings when $\sqrt{n}$-inference is impossible, including when
  the \ac{CATE} is set identified (in which case the \ac{CI} is asymptotically
  valid for points in the identified set).

  Second, the worst-case bias calculations do not depend on the error
  distribution. Although its coverage guarantees are only asymptotic, the
  feasible \ac{CI} reflects the finite-sample impact of the covariate and
  treatment realizations (including the degree of overlap in the data) on the
  bias of the estimator through the critical value. Similarly, the bias-variance
  tradeoffs discussed in~\Cref{sec:optim-estim-cis} still go through even if the
  errors are not normal, since only the variance of the error distribution
  affects the underlying calculations. Our recommendation to assume
  homoskedasticity when setting the initial variance estimates is motivated by
  the fact that under homoskedasticity, using a constant initial variance
  function yields an estimator that has the finite-sample optimality property of
  minimizing variance among estimators with the same worst-case bias. Also, if
  the initial variance estimate is correct, then the feasible estimator
  $\tilde{L}_{\tilde{\delta}_{\RMSE}}$ will be optimal (in the sense discussed
  in \Cref{sec:optim-estim-cis}) even when the errors are
  non-normal.\footnote{\label{fn:non-linear-feasible}The result on finite-sample
    optimality among non-linear procedures discussed in
    \Cref{sec:bounds-adaptation} likewise goes through under non-normal errors,
    so long as the set of possible distributions for $u_{i}$ includes normal
    errors, and its second moment is bounded.} We can take advantage of this
  fact in large samples under homoskedasticity, when the initial variance
  estimator is consistent. At the same time, the \acp{CI} retain asymptotic
  validity under weak conditions. These optimality and validity properties
  mirror those of the \ac{OLS} estimator along with heteroskedasticity robust
  standard errors in a linear regression model: the estimator is optimal under
  homoskedasticity if one assumes normal errors, or if one restricts attention
  to linear estimators, while the \acp{CI} are asymptotically valid under
  heteroskedastic and non-normal errors.
\end{remark}

\begin{remark}[\acp{CI} based on other estimators]\label{remark:other_feasible_cis}
  Feasible \acp{CI} based on linear estimators
  $\hat{L}_{k}$~\eqref{linear_estimator_eq} can be formed as in the baseline
  implementation, using the weights $k(x_i, d_i)$ in \cref{rob_se_eq}, and
  computing the worst-case bias by solving the optimization problem in
  \Cref{k_maxbias_thm}.

  If one applies this method to form a feasible \ac{CI} based on matching
  estimators, one can determine the number of matches $M$ that leads to the
  shortest \ac{CI} (or smallest \ac{RMSE}) as in Steps 2 and 3 of the procedure,
  with $M$ playing the role of $\delta$. In our application, we compare the
  length of the resulting \acp{CI} to those of the optimal \acp{FLCI}. Although
  \Cref{match_optimality_thm} implies the matching estimator with a single match
  is suboptimal unless $C$ is large enough, we find that, in our application,
  the efficiency loss is modest.
\end{remark}

\subsection{\texorpdfstring{\ac{CI}}{CI} for the population average treatment effect}\label{sec:pate}

We now show how our approach can be adapted to construct \acp{CI} for the
\ac{PATE} based on linear estimators $\hat{L}_{k}$ of the
form~\eqref{linear_estimator_eq}. To do so, we treat the covariates and
treatment as random. Under random sampling, the \ac{PATE} is given by
$\tau=E[Y_{i}(1)-Y_{i}(0)]=E[Lf]$, where $Lf$ is the \ac{CATE} as
in~\cref{eq:cate}.

If we view $\hat{L}_{k}$ as an estimator of $\tau$, the quantity
$\maxbias_{\mathcal{F}}(\hat{L}_{k})$ now represents its worst-case bias
conditional on $\{X_{i}, D_{i}\}_{i=1}^{n}$, and is therefore random under
i.i.d.\ sampling. We thus cannot use the arguments underlying the construction
in \cref{eq:flci}. Instead, we simply add and subtract
$\maxbias_{\mathcal{F}}(\hat{L}_{k})$, in addition to adding and subtracting the
usual critical value times a standard error based on the marginal, rather than
conditional, variance of $\hat{L}_{k}$,
\begin{equation}\label{eq:pate_ci}
  \{\hat{L}_{k}\pm(\maxbias_{\mathcal{F}}(\hat L_k)+z_{1-\alpha/2}
  \sepate(\hat{L}_{k}))\}.
\end{equation}
Here $\sepate(\hat{L}_{k})^{2}=\se(\hat{L}_{k})^{2}+\se(Lf)^{2}$, where
$\se(\hat{L}_{k})^{2}$ is the conditional variance of the estimator using the
weights $k(\cdot)$ in~\cref{rob_se_eq}, and $\se(Lf)^{2}$ is a consistent
estimator of the variance of $Lf$,
$\frac{1}{n}E[(f(X_{i}, 1)-f(X_{i}, 0)-\tau)^{2}]$. For the latter, if we use
the nearest neighbor variance estimator $\hat{u}_{i}^{2}$ in~\eqref{rob_se_eq},
and the linear estimator takes the form $\hat{L}_{k}=L\hat{f}$, where
$\hat{f}(x_{i}, d_{i})=Y_{i}$ and
$\hat{f}(x_{i},1-d_{i})=\sum_{j=1}^{n}W_{ij}Y_{j}$, where
$\sum_{j=1}^{n}W_{ij}=1$ and $k_{j}=(1-2d_{j})(1+\sum_{i=1}^{n}W_{ij})/n$ (as
discussed in \Cref{linear_estimators_sec} and following
\Cref{lipschitz_optimal_thm}, this includes matching estimators, as well as the
optimal estimator), one may use the nearest neighbor estimator suggested by
\citet[Theorem 7]{AbIm06match},
$\frac{1}{n^{2}}\sum_{i=1}^{n}(\hat{f}(x_{i},1)-\hat{f}(x_{i},0)-\hat{L}_{k})^{2}-\frac{1}{n^{2}}
\sum_{i=1}^{n}(1+ \sum_{j=1}^{n}W_{ji}^{2})\hat{u}_{i}^{2}$. We note that the
optimality results discussed in \Cref{remark:feasible_finsamp_efficiency} do not
apply to the \ac{CI} in~\eqref{eq:pate_ci}. In \Cref{pate_sec_append}, we provide
formal asymptotic coverage results for this \ac{CI} and
its one-sided analog.

\section{Asymptotic results}\label{sec:asymptotic-results}

In \Cref{seb_sec}, we show that $\sqrt{n}$-inference is impossible when the
dimension of covariates is high enough relative to the order of smoothness
imposed by $\mathcal{F}$, regardless of how the \ac{CI} is formed. \Cref{nonnormal_sec} gives conditions for asymptotic
validity of the feasible \acp{CI} when the variance function is unknown and the
errors $u_{i}$ may be non-normal. \Cref{match_asym_validity_sec} discusses
conditions for asymptotic validity and optimality of \acp{CI} based on matching
estimators.

\subsection{Impossibility of \texorpdfstring{$\sqrt{n}$}{root-n}-inference under low smoothness}\label{seb_sec}

As discussed in the introduction, the standard approach to inference, which we
refer to as $\sqrt{n}$-inference, is based on estimators that are
$\sqrt{n}$-consistent and asymptotically normal, with asymptotically negligible
bias. We now show that if the dimension of the (continuously distributed)
covariates $p$ is high enough relative to the smoothness of $\mathcal{F}$, this
approach is infeasible.

To state the result, let $\Sigma(\gamma, C)$ denote the set of $\ell$-times
differentiable functions $f$ such that, for all integers
$k_1, k_2, \dotsc, k_{\dimx}$ with $\sum_{j=1}^{\dimx}k_j=\ell$,
$\abs*{\frac{\partial^{\ell}f(x)}{\partial x_1^{k_1}\cdots \partial
    x_{\dimx}^{k_{\dimx}}} -\frac{\partial^\ell f(x')}{\partial x_1^{k_1}\cdots
    \partial x_{\dimx}^{k_{\dimx}}}} \le C\norm{x-x'}^{\gamma-\ell}$, where
$\ell$ is the greatest integer strictly less than $\gamma$ and
$\norm{x}^{2}=\sum_{j=1}^{\dimx}x_{j}^{2}$. Note that $f\in\FLip(C)$ is
equivalent to $f(\cdot,1), f(\cdot,0)\in\Sigma(1,C)$.
\begin{theorem}\label{sate_rate_thm}
  Let $\{X_i, D_i\}$ be i.i.d.\ with $X_i\in\mathbb{R}^{\dimx}$ and
  $D_i\in\{0,1\}$. Suppose that the Gaussian regression
  model~\eqref{fixed_design_eq} and~\eqref{normal_u_eq} holds conditional on the
  realizations of the treatment and covariates. Suppose that the marginal
  probability that $D_i=1$ is not equal to zero or one and that $X_i$ has a
  bounded density conditional on $D_i$. Given $\gamma, C$, let
  $\hor{\hat c_n, \infty}$ be a sequence of \acp{CI} for $Lf$ with asymptotic
  coverage at least $1-\alpha$ under $\Sigma(\gamma, C)$ conditional on
  $\{X_i, D_i\}_{i=1}^n$:
  \begin{equation*}
    \liminf_{n\to\infty} \inf_{f(\cdot,0), f(\cdot,1)\in \Sigma(\gamma, C)}
    P_{f}(Lf
    \in \hor{\hat c_n, \infty}\mid \{X_i, D_i\}_{i=1}^{n})
    \ge 1-\alpha
  \end{equation*}
  almost surely. Then, under the zero function $f(x, d)=0$, $\hat c_n$ cannot
  converge to the \ac{CATE} (which is $0$ in this case) more quickly than
  $n^{-\gamma/\dimx}$: there exists $\eta>0$ such that
  \begin{equation*}
    \liminf_{n\to\infty}
    P_0\left(\hat c_n\le -\eta n^{-\gamma/\dimx}
    \mid \{X_i, D_i\}_{i=1}^n\right)
    \ge 1-\alpha\quad\text{almost surely.}
  \end{equation*}
\end{theorem}

The theorem shows that the excess length of a \ac{CI} with conditional coverage
in the class with $f(\cdot,0), f(\cdot,1)\in\Sigma(\gamma, C)$ must be of order
at least $n^{-\gamma/\dimx}$, even at the ``smooth'' function $f(x, d)=0$. The
Lipschitz case corresponds to $\gamma=1$, so that $\sqrt{n}$-inference is
possible only when $\dimx\le 2$. While \Cref{sate_rate_thm} considers a setting
with normal errors, the same bound applies if the normality assumption is
dropped (so long as the class of possible distributions for $u_{i}$ includes the
normal distribution), since including other distributions only makes the problem
more difficult. \Cref{sate_rate_thm} requires coverage conditional on the
realizations of the covariates and treatment. For unconditional coverage, the
results of \citet{robins_semiparametric_2009} imply that if
$\pscore\in\Sigma(\gamma_{\pscore}, C)$, where
$\pscore(x)=P(D_{i}=1\mid X_{i}=x)$ denotes the propensity score, and if
$f(\cdot,0), f(\cdot,1)\in\Sigma(\gamma, C)$, then $\sqrt{n}$-inference is
impossible unless $\gamma_{\pscore}+\gamma\ge \dimx/2$. Thus, conditioning
effectively takes away any role of smoothness of the propensity score.

\subsection{Asymptotic validity of feasible \texorpdfstring{\acp{CI}}{CIs}}\label{nonnormal_sec}

The following theorem gives sufficient conditions for the asymptotic validity of
the feasible \acp{CI} given in \Cref{sec:basel-impl} based on the estimator
$\tilde L_{\delta}$ when $\mathcal{F}=\FLip(C)$. To allow us to capture cases in
which the finite-sample bias of the estimator is non-negligible even though the
asymptotic bias is negligible under standard asymptotics where the parameter
space is fixed with the sample size, we allow $C=C_{n}\to\infty$ as
$n\to\infty$. For concreteness, we restrict attention to standard errors based
on nearest neighbor estimates, and for simplicity, we focus on the case where
the preliminary variance estimate $\tilde{\sigma}^{2}(x, d)$ is non-random,
leaving the extension to random $\tilde{\sigma}(x, d)$ to future research.

\begin{theorem}\label{asymptotic_coverage_theorem}
  Consider the model~\eqref{fixed_design_eq}. Suppose that (a)
  $\mathcal{F}=\FLip(C_{n})$, $1/K\le Eu_i^2\le K$ and
  $E\abs{u_{i}}^{2+1/K}\le K$ for some constant $K$, and that the variance
  function $\sigma^{2}(x, d)$ is uniformly continuous in $x$ for $d\in\{0,1\}$;
  and (b)
  \begin{equation}\label{x_density_condition}
    \text{for all $\eta>0$}\;
    \min_{1\le i\le n} \#\{j\in \{1, \dotsc, n\}\colon
    \normx{x_{j}-x_{i}}\le \eta/C_n, \, d_i=d_j\}\to\infty.
  \end{equation}
  Let $\mathcal{C}$ be the \ac{CI} in~\cref{rob_flci_eq} based on the feasible
  estimator $\tilde L_\delta$, with $\delta$ fixed and $\tilde \sigma^2(x, d)$ a
  nonrandom function bounded away from zero and infinity. Suppose the estimator
  $\hat{u}_{i}^{2}$ in~\eqref{rob_se_eq} is the nearest-neighbor variance
  estimator based on a fixed number of nearest neighbors $J$. Then
  $\liminf_{n\to\infty}\inf_{f\in\FLip(C_n)}P_f(Lf\in \mathcal{C})\ge 1-\alpha$.
\end{theorem}
The main regularity condition, \cref{x_density_condition}, only requires that
the covariate distribution for the treated and control units is well-behaved: we
do not require overlap between these two distributions. If $C_n=C$ does not
change with $n$, a bounded support condition suffices:

\begin{lemma}\label{x_density_sufficient_condition_lemma}
  Suppose that $(X_{i}, D_{i})$ is drawn i.i.d.\ from a distribution where $X_{i}$ has
  bounded support and $0<P(D_{i}=1)<1$, and that $C_n=C$ is fixed.
  Then~\eqref{x_density_condition} holds almost surely.
\end{lemma}

Thus, by \Cref{asymptotic_coverage_theorem}, feasible \acp{CI} are
asymptotically valid in settings in which $\sqrt{n}$-inference is impossible,
including in settings in which the covariate dimension $\dimx$ is high enough so
that \Cref{sate_rate_thm} applies, and settings with imperfect overlap \citep[as
in][]{KhTa10} including set identification due to complete lack of overlap. The
estimator $\tilde{L}_{\delta}$ remains asymptotically normal, when normalized by
its standard deviation: the ``irregular'' nature of the setting only shows up
through non-negligible asymptotic bias, which is captured by the critical value
$\cv_{\alpha}$ when constructing the \ac{CI}, and through a slower rate of
convergence of the estimator (so the impossibility result of
\Cref{sate_rate_thm} is not contradicted).

\begin{remark}[Lindeberg weights]\label{remark:lindeberg-weights}
  The key to establishing \Cref{asymptotic_coverage_theorem} is showing that the
  estimator, when normalized by its standard deviation, converges to a normal
  distribution. This follows from a \ac{CLT}, provided that the Lindeberg
  condition holds. This in turn requires that the estimator doesn't put too much
  weight $\tilde{k}^{*}_{\delta}(x_{j}, d_{j})$ on any individual observation in
  the sense that for $k=\tilde{k}^{*}_{\delta}$, as $n\to\infty$,
  \begin{equation}\label{lindeberg_eq}
    \operatorname{Lind}(k)=
    \frac{\max_{1\le j\le n} k(x_{j}, d_{j})^2}{\sum_{i=1}^n k(x_i,d_i)^2}
    \to 0.
  \end{equation}
  To give intuition for why~\eqref{lindeberg_eq} indeed holds, recall that, as
  discussed below~\Cref{lipschitz_optimal_thm}, putting more weight on any
  individual observation $j$ (by using it often as a match) increases the
  variance of the estimator. Under condition~\eqref{x_density_condition}, there
  are other observations that are almost as good of a match as $j$ (since there
  are within distance $\eta/C_{n}$ to $j$), so it can't be optimal to place too
  much weight on $j$: using these other observations as a match instead of $j$
  would lower the variance of the estimator at little cost to bias.

  One may nonetheless be concerned that in finite-samples, the \ac{CLT}
  approximation is not accurate. This can be assessed directly by computing
  $\operatorname{Lind}(\tilde{k}^{*}_{\delta})$ and checking whether it is close to $0$.
  We do this in our application in
  \Cref{nsw_sec}.\footnote{\label{fn:lindeberg_enforcement}One can also directly
    ensure that $\operatorname{Lind}(\tilde{k}^{*}_{\delta})$ is small by only
    optimizing the \ac{RMSE} or \ac{CI} length in step~\ref{item:3} of the
    baseline implementation in \Cref{sec:basel-impl} over values of $\delta$
    large enough so that $\operatorname{Lind}(\tilde{k}^{*}_{\delta})$ is below a
    pre-specified cutoff. This is analogous to the suggestion by \citet{NoRo19}
    to only consider large enough bandwidth values in a regression discontinuity
    setting so that the resulting Lindeberg weights are small.}
\end{remark}

\subsection{Asymptotic properties of \texorpdfstring{\acp{CI}}{CIs} based on matching estimators}\label{match_asym_validity_sec}

For feasible \acp{CI} based on matching estimators, we obtain the following
result:

\begin{theorem}\label{match_asymptotic_coverage_theorem}
  Suppose that the conditions of \Cref{asymptotic_coverage_theorem} hold. Let
  $\mathcal{X}$ be a set containing $\{x_{i}\}_{i=1}^{n}$. Let
  $\overline{G}\colon\mathbb{R}^{+} \to\mathbb{R}^+$ and
  $\underline{G}\colon\mathbb{R}^{+} \to\mathbb{R}^+$ be functions with
  $\lim_{t\to 0}\frac{\overline{G}(\underline{G}^{-1}(t))^{2}}{t/\log t^{-1}}=0$. Suppose
  that, for any sequence $a_n$ with $n \underline G(a_n)/\log n\to \infty$, we
  have
  \begin{equation}\label{overlap_event_eq}
    \underline G(a_n)\le
    \frac{\#\{i\colon \normx{x_i-x}\le a_n, \, d_i=d\}}{n}
    \le \overline G(a_n)
    \quad\text{all}\; x\in\mathcal{X}, \, d\in\{0,1\}
  \end{equation}
  for large enough $n$. Let $\mathcal{C}$ be the \acp{CI} in
  \Cref{remark:other_feasible_cis} based on the matching estimator with a fixed
  number of matches $M$, and $\mathcal{F}=\FLip(C_n)$. Then
  $\liminf_{n\to\infty}\inf_{f\in\FLip(C_n)}P_f(Lf\in \mathcal{C})\ge 1-\alpha$.
\end{theorem}

\Cref{match_asymptotic_coverage_theorem} is related to results of
\citet{AbIm06match} on asymptotic properties of matching estimators with a fixed
number of matches. \citet{AbIm06match} note that, when $\dimx$ is large enough,
the bias term will dominate, so that conventional \acp{CI} based on matching
estimators will not be valid. In contrast, the \acp{CI} in
\Cref{match_asymptotic_coverage_theorem} remain valid even when $\dimx$ is
large, since they are widened to take into account the potential bias of the
estimator. Alternatively, one can attempt to restore asymptotic coverage by
subtracting an estimate of the bias based on higher-order smoothness
assumptions. While this can lead to asymptotic validity when additional
smoothness is available \citep{abadie_bias-corrected_2011}, it follows from
\Cref{sate_rate_thm} that such an approach will lead to asymptotic undercoverage
under some sequence of regression functions in the Lipschitz class $\FLip(C)$.

Relative to \Cref{asymptotic_coverage_theorem},
\Cref{match_asymptotic_coverage_theorem} requires the additional
condition~\eqref{overlap_event_eq}. This condition holds almost surely if
$(X_{i}, D_{i})$ are drawn i.i.d.\ from a distribution where $\underline G(a)$
and $\overline G(a)$ are lower and upper bounds (up to constants) for
$P(\normx{X_{i}-x} \le a, \, D_{i}=d)$ for $x$ on the support of
$X_{i}$ \citep[Theorem 37, p.~34]{pollard_convergence_1984}. The condition
$\lim_{t\to 0}\overline G(\underline G^{-1}(t))^2/[t/\log t^{-1}]=0$ can thus be
interpreted as an overlap condition. In particular, if the density of $X_{i}$ is
bounded away from zero and infinity on a sufficiently regular support, then the
condition holds if the propensity score $P(D_{i}=1\mid X_{i})$ is bounded away
from zero and one.

In the supplemental materials, we strengthen the conclusions of
\Cref{match_asymptotic_coverage_theorem} and give an asymptotic analog of
\Cref{match_optimality_thm} showing that, if there is sufficient overlap and
$\dimx> 2$, matching with $M=1$ matches is asymptotically efficient under the
\ac{RMSE} and \ac{FLCI} criteria. Thus, while it may at first appear that, as
argued in \citet{AbIm06match}, the matching estimator is inefficient due to its
slower than $\sqrt{n}$ rate of convergence, by \Cref{sate_rate_thm}, the
$\sqrt{n}$-rate is not feasible in this setting: the matching estimator in fact
achieves the fastest possible rate, and, when $M=1$, the constant is also
asymptotically optimal.

On the other hand, if there is not sufficient overlap, asymptotic normality may
fail for the matching estimator. As an extreme example, suppose that $\dimx=1$
and that $x_{j}<x_{i}$ for all observations where $d_{j}=0$ and $d_{i}=1$. Then
each untreated observation will be matched to the same treated observation, the
one with the smallest value of $x_{i}$ among treated observations. Consequently,
the Lindeberg weight defined in \cref{lindeberg_eq} will be bounded away from
zero for this observation, and the \ac{CLT} will fail. In contrast, by
\Cref{x_density_sufficient_condition_lemma}, the estimator $\tilde{L}_{\delta}$
will be asymptotically normal when scaled by its standard deviation even when
there is no overlap between the distribution of $x_{i}$ for treated and
untreated observations.

\citet{rothe_robust_2017} argues that, in settings with limited overlap,
estimators of the \ac{CATE} may put a large amount of weight on a small number
of observations. As a result, standard approaches to inference that rely on
normal asymptotic approximations to the distribution of the $t$-statistic will
be inaccurate in finite samples. Our results shed light on when such concerns
are relevant. The above example shows that such concerns may indeed
persist---even in large samples---if one uses a matching estimator with a fixed
number of matches. Similarly to the discussion
in~\Cref{remark:lindeberg-weights}, in finite samples, one can assess these
concerns directly by computing the maximum Lindeberg weight
$\operatorname{Lind}(k_{\text{match}, M})$. Furthermore, it follows from the
proof of \Cref{match_asymptotic_coverage_theorem} that when $\dimx>2$, bias will
dominate variance asymptotically even if one attempts to ``undersmooth'' by
using a matching estimator with a single match. In such settings, it is
important to widen the \acp{CI} to take the bias into account, in addition to
accounting for the potential inaccuracy of the normal asymptotic approximation.

\section{Empirical application}\label{nsw_sec}

We now illustrate our methods with an application to the \acf{NSW}
demonstration. We use the same dataset as \citet{dw99} and
\citet{abadie_bias-corrected_2011}.\footnote{We use the data from Rajeev
  Dehejia's website, \url{http://users.nber.org/~rdehejia/nswdata2.html}.} In
particular, the treated sample corresponds to 185 men in the \ac{NSW}
experimental sample with non-missing prior earnings data who were randomly
assigned to receive job training after December 1975, and completed it by
January 1978; the sample with $d_i=0$ is a non-experimental sample of $2,490$
men taken from the PSID\@. The outcome $Y_{i}$ corresponds to earnings in 1978
in thousands of dollars, and the covariate vector $x_i$ contains the variables:
age, education, indicators for black and Hispanic, indicator for marriage,
earnings in 1974, earnings in 1975, and employment indicators for 1974 and
1975.\footnote{Following \citet{abadie_bias-corrected_2011}, the no-degree
  indicator variable is dropped, and the employment indicators are defined as an
  indicator for nonzero earnings.} We assume that the unconfoundedness
assumption holds given this covariate vector.

We are interested in the \ac{CATT},
\begin{equation*}
  \text{CATT}(f)
  =\frac{\sum_{i=1}^{n}d_{i}(f(x_i,1)-f(x_i,0))}{\sum_{i=1}^n d_i}
  =\frac{\sum_{i=1}^{n}d_{i}E[Y_{i}(1)-Y_{i}(0) \mid X_{i}=x_{i}].}{\sum_{i=1}^{n}d_{i}}
\end{equation*}
The analysis in \Cref{results_sec} goes through essentially unchanged with this
definition of $Lf$ (see \Cref{proofs_sec_append}).

\subsection{Implementation details}\label{norm_choice_section}

To construct feasible versions of our estimators and \acp{CI}, we follow the
baseline implementation in \Cref{sec:basel-impl}. Implementing the procedure
requires us to fix the norm $\normx{\cdot}$ and the smoothness
constant $C$ in the definition~\eqref{eq:lipschitz-class}. We consider weighted
$\ell_{q}$ norms of the form
\begin{equation}\label{eq:norm-nsw}
  \norm{x}_{A, q}=\left(\textstyle\sum_{j=1}^{\dimx}\abs{A_{jj}x_{j}}^{q} \right)^{1/q},
\end{equation}
where $A$ is a diagonal matrix. Let us now describe our main specification; we
discuss other specifications in the supplemental materials. To make the
restrictions on $f$ implied by the choice of norm and $C$ interpretable, in our
main specification, we set $C=1$, and $A=A_{\text{main}}$, with the diagonal
elements of $A_{\text{main}}$ given in~\Cref{Ane_table}. The $j$th diagonal
element $A_{jj}$ then gives the a priori bound on the derivative of the
regression function with respect to $x_{j}$ (i.e.\ the partial effect of
increasing $x_{j}$ by one unit). We set $q=1$, so that the cumulative effect of
changing multiple elements of $x_{j}$ by one unit is bounded by the sum of the
corresponding elements $A_{jj}$.

\begin{table}[tp]
  \centering
  \begin{threeparttable}
  \caption{Diagonal elements of the weight matrix $A$ in definition of the
    norm~\eqref{eq:norm-nsw} for the main specification,
    $A_{\text{main}}$.}\label{Ane_table}
  \begin{tabular}{@{}lrrrrrcccc@{}}
    &&&&&& \multicolumn{2}{c}{Earnings}
    & \multicolumn{2}{c@{}}{Employed}\\
    \cmidrule(rl){7-8}\cmidrule(l){9-10}
    & Age & Educ. & Black & Hispanic & Married& 1974 & 1975 & 1974 & 1975 \\
    \midrule
    $A_{\text{main}}$ &0.15& 0.60& 2.50& 2.50& 2.50& 0.50&0.50&0.10& 0.10\\
  \end{tabular}
\end{threeparttable}
\end{table}

The elements of $A_{\text{main}}$ are chosen to give restrictions on the
conditional mean $f$ that are plausible when $C=1$; we report results for a
range of choices of $C$ as a form of sensitivity analysis. While the outcome is
measured in levels, it is easier to interpret the bounds in terms of percentage
increase in expected earnings. As a benchmark, consider deviations from expected
earnings when the expected earnings are \$10,000, that is $f(x_{i}, d_{i})=10$.
Since the average earnings for the $d_{i}=1$ sample are \$6,400, with 78\% of
the treated sample reporting income below 10 thousand dollars, the implied
percentage bounds for most people in the treated sample will be even more
conservative than this benchmark. We set the coefficients on black, Hispanic,
and married to $2.5$, implying that the wage gap due to race and marriage status
is at most 25\% at this benchmark. We set the coefficients on 1974 and 1975
earnings so that increasing earnings in each of the these years by $x$ units
leads to at most $(0.5+0.5)x$ increase in 1978 earnings, that is at most a
one-to-one increase. Including the employment indicators allows for a small
discontinuous jump in addition for people with zero previous years' earnings.
The implied bounds for the effect of age and education on expected earnings at
10 thousand dollars are 1.5\% and 6\%, respectively, in line with the 1980
census data.

With this choice of norm,
$\normx{x}=\norm{x}_{A_{\text{main}},
  1}=\sum_{j=1}^{\dimx}\abs{A_{\text{main}, jj}x_{j}}$, we follow the baseline
implementation in \Cref{sec:basel-impl}. In step~\ref{item:initial_variance}, we
use the homoskedastic estimate $\tilde \sigma^2(x, d)=\hat{\sigma}^{2}$. Here
$\hat{\sigma}^{2}=\frac{1}{n}\sum_{i=1}^n \hat{u}_{i}^2$, where
$\hat{u}^{2}_{i}$ is the nearest-neighbor variance estimator with $J=2$
neighbors, using Mahalanobis distance (using the metric
$\norm{\cdot}_{A_{\text{main}},1}$ leads to very similar results). We compare
our estimators and \acp{CI} to those based on matching estimators: here we again
use the norm $\norm{\cdot}_{A_{\text{main}}, 1}$ to define distance. While all
reported confidence intervals and standard errors use the
heteroskedasticity-robust formula~\eqref{rob_se_eq} (using the nearest-neighbor
estimate $\hat{u}_{i}^{2}$ above), when making efficiency comparisons, we use
homoskedastic standard errors, so that \ac{RMSE} and \ac{CI} length are the same
as those used to optimize the choice of the smoothing parameter $\delta$, or the
number of matches $M$.

\subsection{Results}\label{sec:results}

\begin{table}[t]
  \centering
  \begin{threeparttable}
  \caption{Results for \ac{NSW} application, main specification with $q=1$,
    $A=A_{\text{main}}$, and $C=1$.}\label{nsw_results_table_Amain_C1}
  \begin{tabular*}{0.99\linewidth}{@{\extracolsep{\fill}}@{}l@{}lcc cc@{}}
    & & \multicolumn{2}{c}{Feasible estimator $\tilde{L}_{\delta}$}
    & \multicolumn{2}{c}{Matching estimator}\\
    \cmidrule(rl){3-4} \cmidrule(rl){5-6}
    & & (1) & (2) & (3) & (4)\\
    \phantom{a}& Criterion & RMSE & CI length &RMSE & CI length \\
    \midrule
    \multicolumn{3}{@{}l}{Panel A\@: Inference on the \ac{CATT}}\\
    \cmidrule(r){1-3}
    &$\delta$ & 1.82 & 3.30 & &\\
    &$M$ & & & 1 & 18\\
    &Estimate & 0.96 & 0.94 & 1.39 & 1.26\\
    &Worst-case bias & 1.63 & 1.78 & 1.48 & 2.21\\
    &Rob.\ std.\ error & 1.01 & 0.94 & 1.09 & 0.89\\
    &Critical value ($\cv_{0.05}$) & 3.27 & 3.55 & 3.01 & 4.13\\
    &95\% conf.\ interval & $(-2.33,4.25)$ & $(-2.38,4.27)$ & $(-1.88, 4.66)$ & $(2.41, 4.93)$\\
    \multicolumn{3}{@{}l}{Panel B\@: Inference on the \acs{PATT}}\\
    \cmidrule(r){1-3}
    &Rob.\ marginal std.\ error & 1.06 & 1.00 & 1.14 & 0.94\\
    &95\% conf.\ interval & $(-2.75,4.67)$ & $(-2.80,4.69)$ & $(-2.32,5.11)$ & $(-2.78,5.30)$\\[1ex]
    &$\operatorname{Lind}(k)$ & 0.073 & 0.062 & 0.192 & 0.062\\
  \end{tabular*}
    \begin{tablenotes}
    \item\emph{Notes:} In each column, the results in both panels are based on
      an estimator with the smoothing parameter ($\delta$ or $M$) chosen to
      optimize the criterion listed under ``Criterion''.
      $\operatorname{Lind}(k)$ corresponds to the maximum Lindeberg weight given
      in~\cref{lindeberg_eq}.
    \end{tablenotes}
\end{threeparttable}
\end{table}

\Cref{nsw_results_table_Amain_C1} reports the point estimates and \acp{CI} at
values of $\delta$ and $M$ that optimize the \ac{RMSE} and \ac{CI} length
criteria. There are three aspects of the results worth highlighting. First, in
all cases, the worst-case bias is non-negligible relative to the standard error.
This is consistent with \Cref{sate_rate_thm} on impossibility of
$\sqrt{n}$-inference: under the Lipschitz smoothness assumption it is not
possible to construct asymptotically unbiased estimators in this application
given that the dimension of continuously distributed covariates is $4$. Our
\acp{CI} reflect this by explicitly taking the bias into account. Second, in
line with the predictions of \Cref{asymptotic_coverage_theorem}, the maximum
Lindeberg weights~\eqref{lindeberg_eq} are low for the feasible estimators
$\tilde{L}_{\delta}$ in columns (1) and (2). However, due to imperfect overlap,
for the matching estimator with $M=1$ match in column (3), it is considerably
higher, and above the 0.075 cutoff suggested in \citet{NoRo19}. Thus, its
distribution may not be well-approximated by a normal distribution, in line with
the discussion in \Cref{match_asym_validity_sec}. Third, in Panel B, the table
also reports \acp{CI} for the \ac{PATT}, constructed using the
formula~\eqref{eq:pate_ci}, using the nearest neighbor variance estimator to
estimate the marginal variance. These \acp{CI} are only slightly longer than
those for the \ac{CATT}: depending on the column, the increase in length is
10.3--13.6\%.

\begin{figure}[t]
  \centering
  \input{./nsw_opt_het3.tex}
  \caption{Feasible estimator and \acp{CI} for \ac{CATT} and \ac{PATT} in
    \ac{NSW} application as a function of the Lipschitz constant $C$. Dashed
    line corresponds to point estimator $\tilde{L}_{\delta_{\RMSE}}$, shaded
    region denotes the estimator $\pm$ its worst-case bias. Solid lines give
    95\% robust \acp{CI} for the \ac{CATT} based on the estimator
    $\tilde{L}_{\delta_{\FLCI}}$, and dotted lines give \acp{CI} for the
    \ac{PATT}.}\label{plot_C}
\end{figure}

To examine sensitivity of the results to the specification of the parameter
space $\mathcal{F}$, \Cref{plot_C} plots the estimator $\tilde{L}_{\delta}$ at
the \ac{RMSE}-optimal choice of $\delta$, as well as \ac{CI} for the \ac{CATT}
and \ac{PATT} when the smoothing parameter $\delta$ is chosen to optimize
\ac{CI} length. For very small values of $C$---smaller than $0.1$---the
Lipschitz assumption implies that selection on pretreatment variables does not
lead to substantial bias, and the estimator and \acp{CI} incorporate this by
tending toward the raw difference in means between treated and untreated
individuals, which in this data set is negative. For $C\geq 0.2$, the point
estimate is positive and remarkably stable as a function of $C$, ranging between
$0.94$ and $1.14$, which suggests that the estimator and \acp{CI} are accounting
for the possibility of selection bias by controlling for observables. The
two-sided \acp{CI} become wider as $C$ increases, which, as can be seen from the
figure, is due to greater potential bias resulting from a less restrictive
parameter space.

According to \Cref{match_optimality_thm}, matching with $M=1$ is efficient when
$C$ is ``large enough''. In our application, for $C\geq 2.8$, the efficiency of
the matching estimator is at least 95\% for both \ac{RMSE} and \ac{CI} length.
Matching with $M=1$ leads to a modest efficiency loss in our main specification,
where $C=1$: its efficiency is $90.4\%$ for \ac{RMSE}, and $86.0\%$ for the
construction of two-sided \acp{CI}. However, inference results based on the
matching estimator should be taken with a grain of caution due to the concerns
with the accuracy of the \ac{CLT} approximation discussed above.

\subsection{Comparison with experimental estimates}\label{comparison_with_experimental_section}

The present analysis follows, among other, \citet{LaLonde1986}, \citet{dw99},
\citet{smith_reconciling_2001}, \citet{smith_does_2005} and
\citet{abadie_bias-corrected_2011} in using a non-experimental sample to
estimate treatment effects of the \ac{NSW} program. A major question in this
literature has been whether the non-experimental sample can be used to obtain
results that are in line with the estimates based on the original experimental
sample of individuals who were randomized out of the \ac{NSW} program. In the
experimental sample, the difference in means between the outcome for the treated
and untreated individuals is $1.79$. Treating this estimator as an estimator of
the \ac{CATT}, the (unconditional) robust standard error is $0.64$; treating it
as an estimator of the \ac{PATT} (which also coincides with the \ac{PATE}), it
is $0.67$.

The estimates in columns (1) and (2) of \Cref{nsw_results_table_Amain_C1} are
slightly lower, although the difference between them and the experimental
estimate is much smaller than the worst-case bias. Consequently, all the
difference between the estimates can be explained by the bias alone. The large
value of the worst-case bias also suggests that the goal of recovering the
experimental estimates using the current non-experimental dataset is too
ambitious, unless one is willing to impose substantially stronger smoothness
assumptions. Furthermore, differences between the estimates reported here and
the experimental estimate may also arise from (1) failure of the selection on
observables assumption; and (2) the sampling error in the experimental and
non-experimental estimates.

\begin{appendices}
  \crefalias{section}{appsec}%
  \crefalias{subsection}{appsubsec}%
  \allowdisplaybreaks%

  \section{Finite-sample results: proofs and additional details}\label{proofs_sec_append}

  This appendix contains proofs and derivations in \Cref{results_sec}, as well
  as additional results. \Cref{sec:gener-setup-results} maps a generalization of
  the setup in \Cref{sec:setup} to the framework of \citet{donoho94} and
  \citet{ArKo18optimal}. Specializing their general results to the current
  setting yields the formulas for optimal estimators and \acp{CI} given in
  \Cref{sec:optim-estim-cis}, and the efficiency bounds discussed in
  \Cref{sec:bounds-adaptation}. \Cref{optimal_ci_derivation_sec_append}
  considers the case with Lipschitz constraints, and when the Lipschitz
  constraints are imposed after partialling out the \ac{BLP}. It also gives
  proofs for \Cref{k_maxbias_thm}, \Cref{match_optimality_thm}, and the first
  part of \Cref{lipschitz_optimal_thm}. \Cref{sec:opt_proof} proves the second
  part of \Cref{lipschitz_optimal_thm}.

  \subsection{General setup and results}\label{sec:gener-setup-results}

  We generalize the setup in \Cref{sec:setup} by letting the parameter of
  interest be a weighted \ac{CATE} of the form
  \begin{equation*}
    Lf=\sum_{i=1}^{n} w_{i}(f(x_i,1)-f(x_i,0)),
  \end{equation*}
  where $\{w_i\}_{i=1}^n$ is a set of known weights that sum to one,
  $\sum_{i=1}^{n} w_{i}=1$. Setting $w_i=1/n$ gives the \ac{CATE}, while setting
  $w_i=d_i/n_{1}$, gives the \acf{CATT}. Here $n_{d}=\sum_{j=1}^n \1{d_{j}=d}$
  gives the number of observations with treatment status equal to $d$. We retain
  the assumption that $\mathcal{F}$ is convex, but drop the centrosymmetry
  assumption. To handle non-centrosymmetric cases, we allow for affine
  estimators that recenter by some constant $a$,
  \begin{equation*}
    \hat{L}_{k, a}=a+\sum_{i=1}^{n}k(x_{i}, d_{i})Y_{i},
  \end{equation*}
  with the notational convention $\hat{L}_{k}=\hat{L}_{k,0}$. Define the maximum
  and minimum bias
  \begin{equation*}
    \maxbias_{\mathcal{F}}(\hat{L}_{k, a})=\sup_{f\in\mathcal{F}} E_f(\hat{L}_{k, a}-Lf),
    \qquad \minbias_{\mathcal{F}}(\hat{L}_{k, a})
    =\inf_{f\in\mathcal{F}} E_f(\hat{L}_{k, a}-Lf).
  \end{equation*}
  While the centering constant has no effect on one-sided \acp{CI}, centering by
  $a^{*}=-(\maxbias_{\mathcal{F}}(\hat{L}_{k, 0})
  +\minbias_{\mathcal{F}}(\hat{L}_{k, 0}))/2$, so that
  $\maxbias_{\mathcal{F}}(\hat{L}_{k, a^{*}})+\minbias_{\mathcal{F}}(\hat{L}_{k,
    a^{*}})=0$ reduces the estimator's \ac{RMSE} and the length of the resulting
  \ac{FLCI} (note this yields $a^{*}=0$ under centrosymmetry). To simplify the
  results below, we assume that the estimator is recentered in this way.

  One-sided \acp{CI} and \acp{FLCI} based on $\hat{L}_{k, a^{*}}$ can be formed
  as in \cref{eq:onesided-ci,eq:flci}, with $\hat{L}_{k, a^{*}}$ in place of
  $\hat{L}_{k}$, with its \ac{RMSE} given by~\cref{eq:rmse}. For comparisons of
  one-sided \acp{CI}, we focus on quantiles of excess length. Given a subset
  $\mathcal{G}\subseteq\mathcal{F}$, define the worst-case $\beta$th quantile of
  excess length over $\mathcal{G}$ of a \ac{CI} $\hor{c, \infty}$,
  $q_\beta(\hat c, \mathcal{G})=\sup_{g\in\mathcal{G}} q_{g, \beta}(Lg-\hat c)$
  where $Lg-\hat{c}$ is the excess length of the \ac{CI} $\hor{\hat c, \infty}$,
  and $q_{g, \beta}(\cdot)$ denotes the $\beta$th quantile under the function
  $g$. For a one-sided \ac{CI} based on $\hat{L}_{k, a^{*}}$,
  \begin{equation*}
    q_{\beta}(\hat c, \mathcal{G})
    =\maxbias_{\mathcal{F}}(\hat L_{k,
      a^{*}})-\minbias_{\mathcal{G}}(\hat{L}_{k, a*})
    +\sd(\hat L_{k, a^{*}})(z_{1-\alpha}+z_\beta).
  \end{equation*}
  This follows from the fact that the worst-case $\beta$th quantile of excess
  length over $\mathcal{G}$ is attained when the estimate is biased downward as
  much as possible. Taking $\mathcal{G}=\mathcal{F}$, a \ac{CI} that optimizes
  $q_\beta(\hat c, \mathcal{F})$ is minimax. Taking $\mathcal{G}$ to correspond
  to a smaller set of smoother functions amounts to ``directing power'' at such
  smooth functions.

  For constructing optimal estimators and \acp{CI}, observe that our setting is
  a fixed design regression model with normal errors and known variance, with
  the parameter of interest given by a linear functional of the regression
  function. Therefore, our setting falls into the framework of \citet{donoho94}
  and \citet{ArKo18optimal}, and we can specialize their general efficiency
  bounds and the construction of optimal affine estimators and \acp{CI} to the
  current setting.\footnote{In particular, in the notation of
    \citet{ArKo18optimal},
    $Y=(Y_1/\sigma(x_1,d_1), \dotsc, Y_n/\sigma(x_n, d_n))$,
    $\mathcal{Y}=\mathbb{R}^n$, and
    $Kf=(f(x_1,d_1)/\sigma(x_1,d_1), \dotsc, \allowbreak f(x_n, d_n)/\sigma(x_n,
    d_n))$. \citet{donoho94} denotes the outcome vector $Y$ by $\mathbf{y}$, and
    uses $\mathbf{x}$ and $\mathbf{X}$ in place of $f$ and $\mathcal{F}$.} To
  state these results, define the (single-class) modulus of continuity of $L$
  (see p.~244 in \citealp{donoho94}, and Section 3.2 in \citealp{ArKo18optimal})
  \begin{equation}\label{eq:modulus-problem-no-centrosymmetry}
    \omega(\delta)=\sup_{f, g\in\mathcal{F}}\left\{Lg-Lf\colon
      \sum_{i=1}^{n}\frac{(f(x_{i}, d_{i})-g(x_{i}, d_{i}))^{2}}{\sigma^{2}(x_{i}, d_{i})}\leq
      \delta^{2}\right\},
  \end{equation}
  and let $f^{*}_{\delta}$ and $g^{*}_{\delta}$ a pair of functions that attain
  the supremum (assuming the supremum is attained). When $\mathcal{F}$ is
  centrosymmetric, then $f^{*}_{\delta}=-g^{*}_{\delta}$, and the modulus
  problem reduces to the optimization
  problem~\eqref{eq:single-class-modulus-problem} in the main text (in the main
  text, the notation $f^*_\delta$ is used for the function denoted $g^*_\delta$
  in this appendix). Let $\omega'(\delta)$ denote an (arbitrary) element of the
  superdifferential at $\delta$ (the set of scalars $a$ such that
  $\omega(\delta)+a(d-\delta)\geq \omega(d)$ for all $d\geq 0$; the set is
  non-empty since the modulus can be shown to be concave). Define
  $\hat{L}_{\delta}=\hat{L}_{k^{*}_{\delta}, a^{*}_{\delta}}$, where
  \begin{equation*}
    k^{*}_{\delta}(x_{i},
    d_{i})=\frac{\omega'(\delta)}{\delta}\frac{g^{*}_{\delta, i}
      -f_{\delta, i}^{*}}{\sigma^{2}(x_{i}, d_{i})}, \quad
    a^{*}_{\delta}=\frac{L(f^{*}_{\delta}+g^{*}_{\delta})-\sum_{i=1}^{n}k^{*}_{\delta}(x_{i}, d_{i})
      (f^{*}_{\delta, i}+g^{*}_{\delta, i})}{2},
  \end{equation*}
  with $g^{*}_{\delta, i}=g_{\delta}^{*}(x_{i}, d_{i})$ and
  $f^{*}_{\delta, i}=f_{\delta}^{*}(x_{i}, d_{i})$. If the class $\mathcal{F}$ is
  translation invariant in the sense that $f\in\mathcal{F}$ implies
  $f+\iota_{\kappa}\in\mathcal{F}$\footnote{In the main text, we assume that
    $\{\iota_{\kappa}\}_{\kappa\in\mathbb{R}}\subset \mathcal{F}$. By convexity,
    for any $\lambda<1$,
    $\lambda f+(1-\lambda)\iota_{\kappa}=\lambda
    f+\iota_{(1-\lambda)\kappa}\in\mathcal{F}$, which implies that for all
    $\lambda<1$ and $\kappa\in\mathbb{R}$,
    $\lambda f+\iota_{\kappa}\in\mathcal{F}$. This, under the assumption
    in~\cref{fn:closed-F}, implies translation invariance.}, then by Lemma D.1
  in \citet{ArKo18optimal}, the modulus is differentiable, with the derivative
  $\omega'(\delta)=\delta/\sum_{i=1}^{n}d_{i}(g^{*}_{\delta,
    i}-f^{*}_{\delta, i})/\sigma^{2}(x_{i}, d_{i})$. The formula for
  $\hat{L}_{\delta}$ in the main text follows from this result combined with
  fact that, under centrosymmetry, $f^{*}_{\delta}=-g^{*}_{\delta}$. By Lemma
  A.1 in \citet{ArKo18optimal}, the maximum and minimum bias of
  $\hat{L}_{\delta}$ is attained at $g^*_\delta$ and $f^*_\delta$, respectively,
  which yields
  $\maxbias_{\mathcal{F}}(\hat{L}_{\delta})=
  -\minbias_{\mathcal{F}}(\hat{L}_{\delta})=\frac{1}{2}(\omega(\delta)-\delta
  \omega'(\delta))$. Note that $\sd(\hat{L}_{\delta})=\omega'(\delta)$.

  Corollary 3.1 in \citet{ArKo18optimal}, and the results in \citet{donoho94}
  then yield the following result:
  \begin{theorem}\label{theorem:Ldelta-optimality}
    Let $\mathcal{F}$ be convex, and fix $\alpha>0$. (i) Suppose that
    $f^{*}_{\delta}$ and $g^{*}_{\delta}$ attain the supremum
    in~\eqref{eq:modulus-problem-no-centrosymmetry} with
    $\sum_{i=1}^{n}\frac{(f(x_{i}, d_{i})-g(x_{i},
      d_{i}))^{2}}{\sigma^{2}(x_{i}, d_{i})}=\delta^{2}$, and let
    $\hat{c}_{\delta}^{*}=\hat{L}_{\delta}-\maxbias_{\mathcal{F}}(\hat{L}_{\delta})-z_{1-\alpha}\sd(\hat{L}_{\delta})$.
    Then $\hor{\hat{c}_{\delta}^{*}, \infty}$ is a $1-\alpha$ \ac{CI} over
    $\mathcal{F}$, and it minimaxes the $\beta$th quantile of excess length
    among all $1-\alpha$ \acp{CI} for $Lf$, where
    $\beta=\Phi(\delta-z_{1-\alpha})$, and $\Phi$ denotes the standard normal
    cdf. (ii) Let $\delta_{\FLCI}$ be the minimizer of
    $\cv_\alpha\left(\omega(\delta)/2\omega'(\delta)-\delta/2\right)\omega'(\delta)$
    over $\delta$, and suppose that $f^{*}_{\delta_{\FLCI}}$ and
    $g^{*}_{\delta_{\FLCI}}$ attain the supremum
    in~\eqref{eq:modulus-problem-no-centrosymmetry} at $\delta=\delta_{\FLCI}$.
    Then the shortest $1-\alpha$ \ac{FLCI} among all \acp{FLCI} centered at
    affine estimators is given by
    \begin{equation*}
      \left\{\hat L_{\delta_{\FLCI}} \pm
        \cv_\alpha(\maxbias_{\delta_{\FLCI}}/\sd(\hat L_{\delta_{\FLCI}}))
        \sd(\hat L_{\delta_{\FLCI}})
      \right\}.
    \end{equation*}
    (iii) Let $\delta_{\RMSE}$ minimize
    $\frac{1}{4}(\omega(\delta)-\delta\omega'(\delta))^{2}+\omega'(\delta)^{2}$
    over $\delta$, and suppose that $f^{*}_{\delta_{\FLCI}}$ and
    $g^{*}_{\delta_{\FLCI}}$ attain the supremum
    in~\eqref{eq:modulus-problem-no-centrosymmetry} at $\delta=\delta_{\RMSE}$.
    Then the estimator $\hat{L}_{\delta_{\RMSE}}$ minimaxes \ac{RMSE} among all
    affine estimators.
  \end{theorem}
  The theorem shows that a one-sided \ac{CI} based on $\hat{L}_\delta$ is
  minimax optimal for $\beta$-quantile of excess length if
  $\delta=z_{\beta}+z_{1-\alpha}$. Therefore, restricting attention to affine
  estimators does not result in any loss of efficiency if the criterion is
  $q_{\beta}(\cdot, \mathcal{F})$.

  If the criterion is \ac{RMSE}, \Cref{theorem:Ldelta-optimality} only gives
  minimax optimality in the class of affine estimators. However,
  \citet{donoho94} shows that one cannot substantially reduce the maximum risk
  by considering non-linear estimators. To state the result, let
  $\rho_{A}(\tau)=\tau/\sqrt{1+\tau}$ denote the minimax \ac{RMSE} among affine
  estimators of $\theta$ in the bounded normal mean model in which we observe a
  single draw from the $N(\theta,1)$ distribution, and $\theta\in[-\tau, \tau]$,
  and let $\rho_{N}(\tau)$ denote the minimax \ac{RMSE} among all estimators
  (affine or non-linear). \citet{DoLiMG90} give bounds on $\rho_{N}(\tau)$, and
  show that $\sup_{\tau>0}\rho_{A}(\tau)/\rho_{N}(\tau)\leq \sqrt{5/4}$, which
  is known as the Ibragimov-Hasminskii constant.
  \begin{theorem}[\citealp{donoho94}]\label{theorem:donoho-minimax-efficiency}
    Let $\mathcal{F}$ be convex. The minimax \ac{RMSE} among affine estimators risk
    equals
    $R_{\RMSE, A}^{*}(\mathcal{F})=\sup_{\delta>0}\frac{\omega(\delta)}{\delta}\rho_{A}(\delta/2)$.
    The minimax \ac{RMSE} among all estimators is bounded below by
    $\sup_{\delta>0}\frac{\omega(\delta)}{\delta}\rho_{N}(\delta/2)\geq
    \sqrt{4/5}\sup_{\delta>0}\frac{\omega(\delta)}{\delta}\rho_{A}(\delta/2)=\sqrt{4/5}R_{\RMSE, A}^{*}(\mathcal{F})$.
  \end{theorem}
  The theorem shows that the minimax efficiency of $\hat{L}_{\delta_{\RMSE}}$
  among all estimators is at least $\sqrt{4/5}=89.4\%$. In particular
  applications, the efficiency can be shown to be even higher by lower bounding
  $\sup_{\delta>0}\frac{\omega(\delta)}{\delta}\rho_{N}(\delta/2)$ directly,
  rather than using the Ibragimov-Hasminskii constant. The arguments in
  \citet{donoho94} also imply $R_{\RMSE, A}^{*}(\mathcal{F})$ can be equivalently
  computed as
  $R_{\RMSE, A}^{*}(\mathcal{F})=\inf_{\delta>0}
  \frac{1}{2}\sqrt{(\omega(\delta)-\delta\omega'(\delta))^{2}+\omega'(\delta)^{2}}=\inf_{\delta>0}
  \sup_{f\in\mathcal{F}}(E(\hat{L}_{\delta}-Lf)^{2})^{1/2}$, as implied by
  \Cref{theorem:Ldelta-optimality}.

  The one-dimensional subfamily argument used in \citet{donoho94} to derive
  \Cref{theorem:donoho-minimax-efficiency} could also be used to obtain the
  minimax efficiency of the \ac{FLCI} based on $\hat{L}_{\delta_{\FLCI}}$ among
  all \acp{CI} when the criterion is expected length. However, when the
  parameter space $\mathcal{F}$ is centrosymmetric, we can obtain a stronger
  result:
  \begin{theorem}\label{theorem:adaptation-theorem}
    Let $\mathcal{F}$ be convex and centrosymmetric, and fix $g\in\mathcal{F}$
    such that $f-g\in\mathcal{F}$ for all $f\in\mathcal{F}$. (i) Suppose
    $-f^{*}_{\delta}$ and $f^{*}_{\delta}$ attain the supremum
    in~\eqref{eq:modulus-problem-no-centrosymmetry} with
    $\sum_{i=1}^{n}\frac{(f(x_{i}, d_{i})-g(x_{i},
      d_{i}))^{2}}{\sigma^{2}(x_{i}, d_{i})}=\delta^{2}$, and
    $\delta=z_{\beta}+z_{1-\alpha}$. Define $\hat{c}_{\delta}^{*}$ as in
    \Cref{theorem:Ldelta-optimality}. Then the efficiency of
    $\hat{c}_{\delta}^{*}$ under the criterion $q_{\beta}(\cdot, \{g\})$, is
    given by
    \begin{equation*}
      \frac{\inf_{\{\hat{c}\colon \text{$\hor{\hat{c}, \infty}$ satisfies~\eqref{coverage_eq}}\}}
        \Rlower{\beta}(\hat{c}, \{g\})}{
        \Rlower{\beta}(\hat{c}^{*}_{\delta},
        \{g\})}=\frac{\omega(2\delta)}{\omega(\delta)+\delta \omega'(\delta)}\geq\frac{1}{2}.
    \end{equation*}
    (ii) Suppose the minimizer $f_{{L_0}}$ of
    $\sum_{i=1}^{n}\frac{(f(x_{i}, d_{i})-g(x_{i},
      d_{i}))^{2}}{\sigma^{2}(x_{i}, d_{i})}$ subject to $Lf={L_0}$ and
    $f\in\mathcal{F}$ exists for all ${L_0}\in\mathbb{R}$. Then the efficiency
    of the \ac{FLCI} around $\hat{L}_{\delta_{\FLCI}}$ at $g$ relative to all
    confidence sets, $\frac{\inf_{\{\mathcal{C}\colon \text{$\mathcal{C}$
            satisfies~\eqref{coverage_eq}}\}} E_{g}\lambda(\mathcal{C})}{
        \cv_\alpha(\maxbias_{\delta_{\FLCI}}/\sd(\hat L_{\delta_{\FLCI}}))
        \sd(\hat L_{\delta_{\FLCI}})}$, is given by
    \begin{equation}\label{eq:finite-sample-flci-efficiency}
      \frac{(1-\alpha)E\left[\omega(2(z_{1-\alpha}-Z))\mid Z\leq z_{1-\alpha}
        \right]}{2\cv_{\alpha}\left(\frac{\omega(\delta_{\FLCI})}{2
            \omega'(\delta_{\FLCI})} -\frac{\delta_{\FLCI}}{2} \right)
        \cdot \omega'(\delta_{\FLCI})}
      \geq \frac{z_{1-\alpha}(1-\alpha)-\tilde{z}_{\alpha}
        \Phi(\tilde{z}_{\alpha}) +\phi(z_{1-\alpha})-\phi(\tilde{z}_{\alpha})
      }{z_{1-\alpha/2}},
    \end{equation}
    where $\lambda(\mathcal{C})$ denotes the Lebesgue measure of a confidence
    set $\mathcal{C}$, $Z$ is a standard normal random variable, $\Phi(z)$ and $\phi(z)$
    denote the standard normal distribution and density, and $\tilde{z}_{\alpha}=z_{1-\alpha}-z_{1-\alpha/2}$.
  \end{theorem}
  \begin{proof}
    Both parts of the theorem, except for the lower bound
    in~\eqref{eq:finite-sample-flci-efficiency}, follow from Corollary 3.2 and
    Corollary 3.3 in \citet{ArKo18optimal}. The lower bound follows from
    Theorem C.7 in \citet{ArKo20sensitivity}.
  \end{proof}
  The theorem gives sharp efficiency bounds for one-sided \acp{CI} as well as
  \acp{FLCI} relative to \acp{CI} that direct all power at a particular function
  $g$. The condition on $g$ is satisfied if $g$ is smooth enough relative to
  $\mathcal{F}$. For example, if $\mathcal{F}=\FLip(C)$, it holds if $g$ is
  piecewise constant, $g(x, d)=\kappa_{0}+\kappa_{1}d$ for some
  $\kappa_{0}, \kappa_{1}\in\mathbb{R}$. The theorem also gives lower bounds for
  these efficiencies---for one-sided \acp{CI}, the theorem implies that the
  $\beta$-quantile excess of length of the \ac{CI}
  $\hor{\hat{c}^{*}_{\delta}, \infty}$ at $g$ cannot be reduced by more than
  50\%. For $95\%$ \acp{FLCI}, the efficiency lower bound
  in~\eqref{eq:finite-sample-flci-efficiency} evaluates to 71.7\%.
  In a particular application, sharp lower bounds can be computed directly by
  computing the modulus; as we show in \Cref{nsw_sec}, this typically yields
  much higher efficiencies.

  \subsection{Estimators and \texorpdfstring{\acp{CI}}{CIs} under Lipschitz
    smoothness}\label{optimal_ci_derivation_sec_append}

  We now specialize the results from~\Cref{sec:gener-setup-results} to the case
  with Lipschitz smoothness, $\mathcal{F}=\FLip(C)$, as well as versions of
  these classes that impose monotonicity conditions, and versions that impose
  the Lipschitz smoothness after partialling out the \acf{BLP}. We focus on the
  \ac{CATT} and \ac{CATE} estimands by requiring the weights $w_{i}$ in the
  definition of $Lf$ to depend only on $d_{i}$: we assume that $w_{i}=w(d_{i})$,
  with $w(1), w(0)\geq 0$ and $w(1)n_{1}+w(0)n_{0}=1$. For \ac{CATT},
  $w(1)=1/n_{1}$ while $w(0)=0$, and for \ac{CATE}, $w(1)=w(0)=1/n$.

  We begin by defining a version of the Lipschitz class that imposes the
  Lipschitz condition after partialling out the \ac{BLP}. Let $z$ be a subset of
  the covariates $x$ that includes the intercept. The regression coefficients
  $\beta_{0}, \beta_{1}$ in a weighted least squares regression of $Y_{i}$ onto
  $(1-d_{i})z_{i}$ and onto $d_{i}z_{i}$, weighed by the precision weights
  $\sigma^{-2}(x_{i}, d_{i})$, are given by\footnote{See \citet{AbImZh14} for a
    comparison of this conditional definition of the \ac{BLP} with the
    unconditional \ac{BLP},
    $\beta_{d}=\argmin_{\beta}E\sigma^{-2}(X_{i}, D_{i})
    (Y_{i}(d)-Z_{i}'\beta)^{2}$}
  $\beta_{d}=\argmin_{\beta} \sum_{i=1}^{n}\sigma^{-2}(x_{i},
  d_{i})\1{d_{i}=d}(E[Y_{i}(d)\mid Z_{i}=z_{i}]-z_{i}'\beta)^{2}$. The part of
  $f$ left over after partialling out the \ac{BLP} is given by
  $g(x, d)=f(x, d)-(1-d)z\beta_{0}-dz\beta_{1}$, and it satisfies the
  orthogonality restriction
  \begin{equation}\label{eq:lip-blp-eq}
    \sum_{i=1}^{n}\1{d_{i}=d}\frac{z_{i}g(x_{i}, d)}{\sigma^{2}(x_{i}, d_{i})}=0,\quad d\in\{0,1\}.
  \end{equation}
  Imposing a Lipschitz smoothness condition on $g$ then leads to the class
  \begin{equation}\label{eq:lip-blp}
    \FRLip{z}(C) =\left\{z'\beta_{d}+g(x, d)\colon g\in\FLip(C),\;
      \text{$g$ satisfies~\eqref{eq:lip-blp-eq}}, \;\beta_{0}, \beta_{1}\in\mathbb{R}^{\dimx}\right\}.
  \end{equation}
  If $z$ comprises just the intercept, then the constraint~\eqref{eq:lip-blp-eq}
  simply normalizes $f(\cdot, d)$ to have empirical mean equal to zero,
  $g(x, d)=f(x, d)-\sum_{i}\{d_{i}=d\}f(x_{i}, d)/\sigma^{2}(x_{i}, d_{i})$. Since
  this is just a normalization, $\FRLip{1}(C)=\FLip(C)$. Results for the class
  $\FRLip{z}(C)$ thus directly imply results for the class $\FLip(C)$ that the
  main text focuses on. On the other hand, if $z=x$, we can think of the space
  $\FRLip{x}(C)$ as formalizing the notion that $f$ is ``approximately linear''
  by requiring that the residual is small in the sense that it lies in
  $\FLip(C)$, with $\FRLip{x}(0)$ corresponding to the linear model. In
  practice, when the variance function is unknown, one may wish to impose this
  class under the assumption of homoskedasticity, so that we partial out the
  \ac{OLS}, rather than the weighted least squares estimand.

  To see how we can replace the functional constraint $g\in\FLip(C)$ with
  inequality constraints, let $\FLipn(C)$ denote the set of functions
  $f\colon \{x_1, \dotsc, x_n\}\times \{0,1\}\to \mathbb{R}$ such that
  $\abs{f(x, d)-f(\tilde x, d)}\le C\normx{x-\tilde x}$ for all
  $x, \tilde x\in \{x_1, \dotsc, x_n\}$ and each $d\in\{0,1\}$. That is,
  $\FLipn(C)$ denotes the class of functions that satisfy the Lipschitz
  conditions when restricted to the domain $\{x_1, \dotsc, x_n\}\times \{0,1\}$.
  Clearly, $f\in\FLip(C)$ implies $f\in\FLipn(C)$. In the other direction, we
  have the following result:
  \begin{lemma}{\citep[][Theorem
      4]{beliakov_interpolation_2006}}\label{theorem:beliakov}
    For any function $f\colon \{x_1, \dotsc, x_n\}\times \{0,1\}\to \mathbb{R}$,
    we have $f\in\FLipn(C)$ if and only if there
    exists a function $h\in\FLip(C)$ such that
    $f(x, d)=h(x, d)$ for all $(x, d)\in \{x_1, \dotsc, x_n\}\times \{0,1\}$.
  \end{lemma}
  As a direct consequence, we obtain:
  \begin{lemma}\label{theorem:linear-blp}
    Consider a linear estimator $\hat{L}_{k}=\sum_{i}k_{i}Y_{i}$ with weights
    $k$ that satisfy
    $\sum_{i}\1{d_{i}=d}z_{i}k_{i}=(2d-1)[w(d)\sum_{i}\1{d_{i}=d}z_{i}
    +w(1-d)\sum_{i}\1{d_{i}=1-d}z_{i}]$. The worst-case bias of this estimator
    over $\FRLip{z}(C)$ is given by the value of
    \begin{equation*}
      \max_{g\in\mathbb{R}^{2n}}\left\{\sum_{i=1}^{n}k_{i}g(x_{i}, d_{i})-
        \sum_{i=1}^{n}w(d_{i})[g(x_{i},1)-g(x_{i},0)]\right\},
    \end{equation*}
    where the maximum is taken subject to~\eqref{eq:lip-blp-eq} and subject to
    \begin{equation}\label{eq:lip-blp-n}
      g(x_{i}, d)-g(x_{j}, d)\leq C\normx{x_{i}-x_{j}}, \quad d\in\{0,1\}\quad i, j\in\{1,\dotsc, n\}.
    \end{equation}
  \end{lemma}
  The constraints on the weights imply that the values of $\beta_{0}$ and
  $\beta_{1}$ do not affect the bias. Specializing to the case with $z=1$ yields
  the first part of \Cref{k_maxbias_thm} (using the observation above that in
  this case the constraint~\eqref{eq:lip-blp-eq} is just a normalization that
  under the conditions on the weights does not affect the bias, so not imposing
  it doesn't affect the worst-case bias).

  To show the second part of \Cref{k_maxbias_thm}, we use the following lemma,
  the proof of which is deferred to the supplemental materials.
  \begin{lemma}\label{lemma:linear-constraints}
    Fix $d\in\{0,1\}$, and consider a vector $(g(x_{1}, d), \dotsc, g(x_{n}, d))'$.
    (i) Suppose~\eqref{eq:lip-blp-n} holds for all $i, j$ with $d_{i}=d_{j}=d$,
    and also for all $i, j$ with $d_{j}=1-d_{i}=d$. Suppose further that for each
    $i$ with $d_{i}=1-d$, there exists a $j$ with $d_{j}=d$ such
    that~\eqref{eq:lip-blp-n} holds with equality. Then~\eqref{eq:lip-blp-n}
    holds for all $i, j$. (ii) Suppose~\eqref{eq:lip-blp-n} holds for all $i, j$
    with $d_{j}=1-d_{i}=d$. Suppose further that for each $i$ with $d_{i}=1-d$,
    there exists a $j$ with $d_{j}=d$ such that~\eqref{eq:lip-blp-n} holds with
    equality. Suppose also that for each $j$ with $d_{j}=d$, there exists an $i$
    with $d_{i}=1-d$ such that~\eqref{eq:lip-blp-n} holds with equality.
    Then~\eqref{eq:lip-blp-n} holds for all $i, j$.
  \end{lemma}
  We now show that if~\eqref{eq:lipschitz-constraints-finite} holds for all
  $i, j$ with $d_{i}=1$, $d_{j}=0$ and $k_{i}=k(x_{i},1)>w(1)$, then it holds for
  $d=1$ and all $i, j$. The argument for $d=0$ is analogous and omitted. Observe
  that if $k_{i}=w(1)$ for some $i$ with $d_{i}=1$, we can set
  $f(x_{i},1)=\min_{j\colon d_{j}=0}\{f(x_{j},1)+C\normx{x_{i}-x_{j}}\}$ without
  affecting the bias, so that we may suppose
  that~\eqref{eq:lipschitz-constraints-finite} holds for all $i, j$ with
  $d_{i}=1-d_{j}=1$. \Cref{k_maxbias_thm} then follows by observing that
  $g=-(f(x_{1},1), \dotsc, f(x_{n},1))$ must satisfy the assumptions of part (ii)
  of \Cref{lemma:linear-constraints}, otherwise we could increase the bias by
  increasing $f(x_{i},1)$ (if $d_{i}=1$) or decreasing $f(x_{j},1)$ (if
  $d_{j}=0$).

  For the optimal estimator, \Cref{theorem:beliakov} implies the following
  result:
  \begin{theorem}\label{theorem:blp-lipschitz_optimal}
    Given $\delta>0$, the value of the maximizer
    $f^{*}_{\delta}(x_{i}, d_{i})=z_{i}'\beta^{*}_{d_{i}, \delta}+g^{*}_{\delta}(x_{i}, d_{i})$
    of~\cref{eq:single-class-modulus-problem} is given by the solution to the
    convex program
    \begin{equation}\label{eq:modulus-lipchitz-blp}
      \max_{g\in\mathbb{R}^{2n}, \beta_{0}, \beta_{1}\in\mathbb{R}^{\dim(z_{i})}} \, 2Lf
      \quad\text{s.t.}\quad
      {\sum_{i=1}^n\frac{g(x_i, d_i)^2+(z_{i}'\beta_{d_{i}})^{2}}{
          \sigma^2(x_i, d_i)}}\le
      \frac{\delta^{2}}{4}
      \quad\text{and s.t.~\eqref{eq:lip-blp-n} and~\eqref{eq:lip-blp-eq}.}
    \end{equation}
  \end{theorem}
  Specializing to the case with $z=1$ gives the first part of
  \Cref{lipschitz_optimal_thm}. The proof for the second part is deferred
  to~\Cref{sec:opt_proof}.

  Next, we consider the form of the optimal estimator when $\delta/C$ is small.
  To that end, let us first define the regression-adjusted matching estimator.
  The linear regression estimator imputes the counterfactual outcome
  $Y_{i}(1-d_{i})$ as $z_{i}'\hat{\beta}_{1-d_{i}}$, where
  $\hat{\beta}_{d}=(Z_{d}'\Sigma_{d}^{-1}Z_{d})^{-1}Z_{d}'\Sigma_{d}^{-1}Y_{d}$.
  Here $Y_{d}$ and $Z_{d}$ are the subsets of the outcome vector and design
  matrix of the covariates $z_{i}$ corresponding to units with $d_{i}=d$, and
  $\Sigma_{d}$ is a diagonal matrix of dimension $n_{d}\times n_{d}$ with
  elements $\sigma^{2}(x_{i}, d_{i})$ corresponding to observations with
  $d_{i}=d$ on the diagonal. Recall from \Cref{linear_estimators_sec} that in
  contrast, the matching estimator with $M$ matches uses the imputation
  $\sum_{j=1}^{n}W_{M, ij}Y_{j}$, where $W_{M, ij}=1/M$ if $j$ is among the $M$
  observations with treatment status $d_{j}=1-d_{i}$ that are closest to $i$,
  and zero otherwise. The regression-adjusted matching estimator \citep{rubin79}
  combines these approaches, imputing the counterfactual outcome as
  $\sum_{j=1}^{n}W_{M, ij}(Y_{j}-z_{j}'\hat{\beta}_{1-d_{i}})+z_{i}'\hat{\beta}_{1-d_{i}}
  =\sum_{j=1}^{n}W_{M, ij}(Y_{j}+z_{i}'\hat{\beta}_{1-d_{i}}-z_{j}'\hat{\beta}_{1-d_{i}})$.
  Relative to the matching estimator, we adjust the matching imputation for the
  difference in covariate values. The estimator thus takes the form
  \begin{equation*}
    \hat{L}_{k}= \sum_{i}(2d_{i}-1)w(d_{i})\Big[Y_{i}-z_{i}'\hat{\beta}_{1-d_{i}}-
      \sum_{j}W_{M, ij}(Y_{j}-z_{j}'\hat{\beta}_{d_{i}})\Big].
  \end{equation*}
  The next result shows that $\hat{L}_\delta$ is given by regression adjusted
  matching with $M=1$ once $C/\delta$ is large enough.
  \Cref{match_optimality_thm} follows by setting $z=1$ and the fact, shown in
  the supplemental materials, that $\delta_{\FLCI}$ and $\delta_{\RMSE}$ do not
  increase without bound as $C$ increases.
  \begin{theorem}\label{theorem:reg-adjusted-matching}
    Suppose that $\sigma(x_{i}, d_{i})>0$ for each $i$, and suppose that each
    unit has a single closest match, so that
    $\argmin_{j\colon d_{i}\neq d_{j}}\normx{x_{i}-x_{j}}$ is a singleton for
    each $i$. There exists a constant $K$ depending on $\sigma^{2}(x_{i}, d_{i})$
    and $\{x_{i}, d_{i}\}_{i=1}^{n}$, such that, if $C/\delta>K$, the optimal
    estimator $\hat{L}_{\delta}$ is given by the regression adjusted matching
    estimator with $M=1$.
  \end{theorem}
\begin{proof}[Proof of~\Cref{theorem:reg-adjusted-matching}]
  For notational convenience, let $m$ and $r$ denote the vectors of length $n$
  with elements $m_{i}=(2d_{i}-1)g(x_{i}, d_{i})$, and let
  $r_{i}=(1-2d_{i})g(x_{i},1-d_{i})$. Abusing notation, let $m_{d}$ and $r_{d}$
  denote the subvectors of $m$ and $r$ corresponding to units with treatment
  status $d$. Finally, let $\iota_{d}$ denote the vector of ones with length
  $n_{d}$. With this notation,
  \begin{equation*}
    Lf=
    w(1)\iota_{1}'(m_{1}+r_{1})
    +w(0)\iota_{0}'(r_{0}+m_{0})
    +w(1)\iota_{1}'Z_{1}(\beta_{1}-\beta_{0})
    +w(0)\iota_{0}'Z_{0}(\beta_{1}-\beta_{0}),
  \end{equation*}
  and~\eqref{eq:modulus-lipchitz-blp} can be written as
  \begin{equation}\label{eq:constraint-1-blp}
    \max_{\beta_{0}, \beta_{1}, m\in\mathbb{R}^{n}, r\in\mathbb{R}^{n}}2Lf
    \quad\text{s.t.}\quad
    {\sum_{i=1}^n\frac{m_{i}^{2}+(z_{i}'\beta_{d_{i}})^{2}}{
        \sigma^2(x_i, d_i)}}\le
    \frac{\delta^{2}}{4}
  \end{equation}
  subject to
  \begin{equation}
    Z_{d}'\Sigma_{d}^{-1}m_{d}=0,\quad \text{for $d\in\{0,1\}$}\label{eq:constraint-eq-blp}
  \end{equation}
  and subject to~\eqref{eq:lip-blp-n}. Now, note that if $w(d)=0$, setting
  $r_{d}=0$ and $m_{d}=0$ is optimal and the constraints~\eqref{eq:lip-blp-n}
  hold trivially. If $w(d)>0$, observe that
  \begin{equation}\label{eq:constraint-3-blp}
    r_{i}-m_{j}\le C \normx{x_{i}-x_{j}}\quad
    \text{for all $i, j$ with $d_{i}=1-d_{j}=d$},
  \end{equation}
  holds with equality for each $i$ for at least one $j$, otherwise we could
  increase the value of the objective function. Therefore, by
  \Cref{lemma:linear-constraints}, to maximize~\eqref{eq:modulus-lipchitz-blp},
  we can replace the constraints in~\eqref{eq:lip-blp-n}
  with~\eqref{eq:constraint-3-blp} and
  \begin{equation}\label{eq:constraint-2-blp}
    m_{i}-m_{j}\le C \normx{x_{i}-x_{j}}\quad
    \text{for all $i, j$ with $d_{i}=d_{j}=d$}.
  \end{equation}
  Next, the assumption that each observation has a unique closest match implies
  that the values of $x_{i}$ and $x_{j}$ for $d_{i}=d_{j}$ are distinct, so
  that, for $\delta/C$ small enough and hence $\sum_{i}m_{i}^{2}$ small enough,
  the constraint~\eqref{eq:constraint-1-blp}
  implies~\eqref{eq:constraint-2-blp}. For $\delta/C$ small enough, it thus
  suffices to maximize~\eqref{eq:constraint-1-blp} subject
  to~\eqref{eq:constraint-eq-blp} and~\eqref{eq:constraint-3-blp}. This is a
  convex optimization problem and constraint qualification holds since $m=0$
  satisfies Slater's condition. Thus, the solution (or set of solutions) is the
  same as the solution to the Lagrangian,
  \begin{multline*}
    -Lf+\frac{\lambda}{2}\left(
      \sum_{d=0}^{1}(m_{d}'\Sigma^{-1}_{d}m_{d}+\beta_{d}'Z_{d}'\Sigma_{d}^{-1}Z_{d}\beta_{d})
      - \frac{\delta^{2}}{4}\right)+\nu_{0}'Z_{0}'\Sigma_{0}^{-1}m_{0}
    +\nu_{1}Z_{1}'\Sigma_{1}^{-1}m_{1}+\\
    \sum_{i, j\colon d_{i}=1-d_{j}=1}[
    \Lambda^{0}_{ij}(r_{i}-m_{j}-C\normx{x_{i}-x_{j}}) +
    \Lambda^{1}_{ij}(r_{j}-m_{i}-C\normx{x_{i}-x_{j}})],
  \end{multline*}
  where $\nu_{0}, \nu_{1}$ are vectors of Lagrange multipliers associated with
  the constraints~\eqref{eq:constraint-eq-blp}, and $\Lambda^{0}, \Lambda^{1}$
  are matrices of Lagrange multipliers with dimension $n_{1}\times n_{0}$
  associated with the constraints~\eqref{eq:constraint-3-blp}. The first-order
  conditions are given by
  \begin{align}
    \Sigma^{-1}_{d}m_{d}& =\frac{1}{\lambda}(w(d)\iota_{d}
           +{\Lambda^{d}}'\iota_{1-d}-\Sigma_{d}^{-1}Z_{d}\nu_{d})\label{eq:foc-1}\\
    w(d)\iota_{d}
         &=\Lambda^{1-d}\iota_{1-d}\label{eq:foc-2}\\
    \beta_{d}
         & =\frac{2d-1}{\lambda}
           (Z_{d}\Sigma_{d}^{-1} Z_{d})^{-1}(w(1) Z_{1}'\iota_{1}+w(0)Z_{0}'\iota_{0}),
  \end{align}
  for $d\in\{0,1\}$. Furthermore, the constraint $Z_{d}'\Sigma^{-1}_{d}m_{d}=0$,
  combined with~\eqref{eq:foc-1} implies that
  \begin{equation*}
    \nu_{d}= w(d)(Z_{d}'\Sigma_{d}^{-1}Z_{d})^{-1}Z_{d}'\iota_{d}
    +(Z_{d}'\Sigma_{d}^{-1}Z_{d})^{-1}Z_{d}'{\Lambda^{d}}'\iota_{1-d}
  \end{equation*}
  so that
  \begin{equation*}
    \Sigma_{d}^{-1} m_{d}=
    \frac{1}{\lambda}(I-\Sigma_{d}^{-1}Z_{d}(Z_{d}'\Sigma_{d}^{-1}Z_{d})^{-1}Z_{d}')(w(d)\iota_{d}
    +{\Lambda^{d}}'\iota_{1-d}).
  \end{equation*}
  Now, since by assumption, $(w(1)n_{1}+w(0)n_{0})=1$, we have
  $\iota_{1}'\Sigma_{1}^{-1}m_{1}+\iota_{1}'\Sigma_{1}^{-1}Z_{1}\beta_{1}=
  \frac{1}{\lambda}$. Hence, by \cref{eq:optimal-weights}, the optimal weights
  for the treated and untreated units, respectively, take the form
  \begin{multline*}
      k_{d}=(2d-1)\lambda\Sigma_{d}^{-1}m_{d}+\lambda\Sigma_{d}^{-1}Z_{d}\beta_{d}
       =(2d-1) [w(d)\iota_{d}\\
      +(I-\Sigma_{d}^{-1}Z_{d}(Z_{d}'\Sigma_{d}^{-1}Z_{d})^{-1}Z_{d}'){\Lambda^{d}}'\iota_{1-d}
       + \Sigma_{d}^{-1}Z_{d} (Z_{d}\Sigma_{d}^{-1}
      Z_{d})^{-1}w(1-d)Z_{1-d}'\iota_{1-d}],
  \end{multline*}
  with the optimal estimator given by
  $\hat{L}_{\delta}=\sum_{i=1}^{n}(2d_{i}-1)[w(d_{i})(Y_{i}-z_{i}'\hat{\beta}_{1-d_{i}})
  -\sum_{j}\Lambda^{1-d_{i}}_{ij}(Y_{j}-z_{i}'\hat{\beta}_{d_{i}})]$. The result
  then follows from the fact that if $w(d)=0$, $\Lambda^{1-d}=0$ at optimum
  by~\cref{eq:foc-2} since $\Lambda^{1-d}\geq 0$ by the complementary slackness
  constraint. Otherwise, if $w(d)>0$, for each $i$, $\Lambda_{ij}^{1-d}$ is
  non-zero for the set corresponding to the $\argmin$
  of~\cref{eq:constraint-3-blp}. However, for $\delta/C$ and hence
  $\sum_{i}m_{i}^{2}$ small enough, since each observation has a unique closest
  match, the set of minimizers is a singleton given by the closest match of $i$.
  By~\cref{eq:foc-2}, for this closest match $j$, $\Lambda_{ij}^{1-d}=w(d)$,
  which yields the result.
\end{proof}

\subsection{Proof of \texorpdfstring{\Cref{lipschitz_optimal_thm}}{Theorem~\ref{lipschitz_optimal_thm}}}\label{sec:opt_proof}

The dual problem to~\eqref{eq:modulus-lipchitz} is to minimize
$\sum_{i=1}^{n}f(x_{i}, d_{i})^{2}/\sigma^{2}(x_{i}, d_{i})$ subject to a lower
bound on $Lf/C$. When $\sigma^{2}(x, d)=\sigma^{2}(d)$, the Lagrangian for this
problem has the form
\begin{equation}\label{eq:dual-problem}
  \min_{f\in\FLipn(C)}\frac{1}{2}\sum_{i=1}^{n}\frac{f(x_{i}, d_{i})^{2}}{\sigma^{2}(x_{i}, d_{i})}-
  \mu Lf/C= \min_{f\in\FLipn(1)}\frac{C^{2}}{2}\sum_{i=1}^{n}\frac{f(x_{i}, d_{i})^{2}}{\sigma^{2}(x_{i}, d_{i})}-
  \mu Lf,
\end{equation}
where we use the observation that if $f\in\FLipn(C)$, then $f/C\in\FLipn(1)$.
Let $g^{*}_{\mu}$ denote the solution to the minimization problem on the
right-hand side of~\eqref{eq:dual-problem}. Because for each $\delta>0$, the
program~\eqref{eq:modulus-lipchitz} is strictly feasible at $f=0$, Slater's
condition holds, and the solution path $\{f^{*}_{\delta}\}_{\delta>0}$ can be
identified with the solution path $\{Cg^{*}_{\mu}\}_{\mu>0}$.

It will be convenient to state the algorithm using the notation
$m_{i}=(2d_{i}-1)g(x_{i}, d_{i})$, and $r_{i}=(1-2d_{i})g(x_{i},1-d_{i})$, in
analogy to the notation in the proof of \Cref{theorem:reg-adjusted-matching}.
Then $Lf=\sum_{i=1}^{n}w_{i}(m_{i}+r_{i})$. Next, we claim that the constraint
$g\in\FLipn(1)$ can be replaced with the constraint
\begin{equation}\label{eq:cij}
    r_{j}\leq m_{i}+\normx{x_{i}-x_{j}}, \quad d_{i}\neq d_{j},
\end{equation}
This follows by observing that at optimum, if $w(d)=0$, then
$m_{i}=\mu w(d)\sigma^{2}(d)$ and $r_{j}=\mu w(d)\sigma^{2}(d)$ achieves the
optimum, so the constraint holds trivially. If $w(d)>0$, at least one of the
constraints in \cref{eq:cij} must bind, for each $i$, otherwise increasing
$r_{j}$ would increase the value of the objective function. Thus, by part (i)
\Cref{lemma:linear-constraints}, we can replace the constraint $g\in\FLipn(1)$
with \cref{eq:cij} and
\begin{equation}\label{eq:cii}
  m_{i}\leq m_{i'}+\normx{x_{i}-x_{i'}}, \quad d_{i}= d_{i'}.
\end{equation}
If we only impose the constraints in~\eqref{eq:cij}, the Lagrangian for the
program~\eqref{eq:dual-problem} can be written as
\begin{multline}\label{eq:lagrangian-mu}
  \frac{1}{2}\sum_{i=1}^{n}\frac{m_{i}^{2}}{\sigma^{2}(d_{i})}
  -\mu\left(\sum_{i=1}^{n}w(d_{i})(m_{i}+r_{i})
  \right)\\
  +\sum_{i, j\colon d_{i}=1-d_{j}=1}\left[\Lambda^{0}_{ij}(r_{i}-m_{j}
    - \normx{x_{i}-x_{j}}) +\Lambda^{1}_{ij}(r_{j}- m_{i}-
    \normx{x_{i}-x_{j}})\right].
\end{multline}
This Lagrangian implies that~\eqref{eq:cii} must hold automatically at the
optimum. Otherwise, if for some $i, i'$ with $d_{i}=d_{i'}=1$,
$m_{i}>m_{i'}+\normx{x_{i}-x_{i'}}$, then for all
$j$ with $d_{i}=0$,
\begin{equation*}
  r_{j}\leq m_{i'}+\normx{x_{i'}-x_{j}}
  \leq m_{i'}+\normx{x_{i'}-x_{i}}+\normx{x_{i}-x_{j}}
  < m_{i}+\normx{x_{i}-x_{j}},
\end{equation*}
The complementary slackness condition
$\Lambda^{1}_{ij}(r_{j}-m_{i}-\normx{x_{i}-x_{j}})=0$ then implies that
$\sum_{j}\Lambda^{1}_{ij}=0$, and it follows from the first-order condition that
$m_{i}/\sigma^{2}(1)=\mu w(1)\leq m_{i'}/\sigma^{2}(1)$, which contradicts the
assertion that $m_{i}>m_{i'}+\normx{x_{i}-x_{i'}}$.

To describe the algorithm, we need additional notation. Let $m(\mu)$, $r(\mu)$,
$\Lambda^{0}(\mu)$, and $\Lambda^{1}(\mu)$ denote the values of $m, r$, and of
the Lagrange multipliers at the optimum of~\eqref{eq:lagrangian-mu}. For
$d\in\{0,1\}$, let $N^{d}(\mu)\in \mathbb{R}^{n_{1}\times n_{0}}$ denote a
matrix with elements $N_{ij}^{d}(\mu)=1$ if the constraint associated with
$\Lambda^{d}_{ij}(\mu)$ is active, and $N^{d}_{ij}(\mu)=0$ otherwise. Let
$G^{0}\in\mathbb{R}^{n_{0}\times n_{0}}$ and
$G^{1}\in\mathbb{R}^{n_{1}\times n_{1}}$ denote matrices with elements
$G^{0}_{jj'}=\1{\sum_{i}N^{0}_{ij}(\mu)N^{0}_{ij'}(\mu)>0}$, and
$G^{1}_{ii'}=\1{\sum_{j}N^{1}_{ij}(\mu)N^{1}_{i'j}(\mu)>0}$. Then $G^{0}$
defines a graph (adjacency matrix) of a network in which $j$ and $j'$ are linked
if the constraints associated with $\Lambda^{0}_{ij}$ and $\Lambda^{0}_{ij'}$ are
both active for some $i$. Similarly, $G^{1}$ defines a graph of a network in
which $i$ and $i'$ are linked if the constraints associated with
$\Lambda^{1}_{i'j}$ and $\Lambda^{1}_{ij}$ are both active for some $j$. Let
$\{\mathcal{M}^{0}_{1}, \dotsc, \mathcal{M}^{0}_{K_{0}}\}$ denote a partition of
$\{1,\dotsc, n_{0}\}$ according to the connected components of $G^{0}$, so that
if $j, j'\in\mathcal{M}_{k}^{0}$ then there exists a path from $j$ to $j'$. Let
$\{\mathcal{R}^{0}_{1}, \dotsc, \mathcal{R}^{0}_{k}\}$ be a corresponding
partition of $\{1,\dotsc, n_{1}\}$, defined by
$\mathcal{R}_{k}^{0}=\{i\in\{1,\dotsc, n_{1}\}\colon\text{$N_{ij}^{0}(\mu)=1$ for
  some $j\in\mathcal{M}^{0}_{k}$}\}$. Similarly, let
$\{\mathcal{M}^{1}_{1}, \dotsc, \mathcal{M}^{1}_{K_{1}}\}$ denote a partition of
$\{1,\dotsc, n_{1}\}$ according to the connected components of $G^{1}$, and let
$\mathcal{R}^{1}_{k}=\{j\in\{1,\dotsc, n_{0}\}\colon\text{$N_{ij}^{1}(\mu)=1$ for
  some $i\in\mathcal{M}^{1}_{k}$}\}$.

In the supplemental materials, we show that the solution path for $m(\mu)$ is
piecewise linear in $\mu$, with points of non-differentiability when either a
new constraint becomes active, or else the Lagrange multiplies
$\Lambda_{ij}^{d}(\mu)$ associated with an active constraint decreases to zero.
We also derive the formulas for the slope of $m(\mu)$, $r(\mu)$, and
$\Lambda^{d}(\mu)$ at points of differentiability. This leads to the following
algorithm that is similar to the LAR algorithm in \citet{rosset_piecewise_2007}
and \citet{ehjt04} for computing the LASSO path.

\begin{enumerate}
\item Initialize $\mu=0$, $m=0$, $\Lambda^{0}=0$, and $\Lambda^{1}=0$. Let
  $D^{0}, D^{1}\in\mathbb{R}^{n_{1}\times n_{0}}$ be matrices with elements
  $D^{d}_{ij}=\normx{x_{i}-x_{j}}$, $d\in\{0,1\}$, $d_{i}=1-d_{j}=1$. Let $r$ be
  a vector with elements $r_{j}=\min_{i=1,\dotsc, n_{1}}\{D^{1}_{ij}\}$, if
  $d_{j}=0$, and $r_{i}=\min_{j=1,\dotsc, n_{0}}\{D_{ij}^{0}\}$, if $d_{i}=1$.
  Let $N^{0}, N^{1}\in\mathbb{R}^{n_{1}\times n_{0}}$ be matrices with elements
  $N^{0}_{ij}=\1{D^{0}_{ij}=r_{i}}$ and $N^{1}_{ij}=\1{D^{1}_{ij}=r_{j}}$.
\item While $\mu<\infty$:
  \begin{enumerate}
  \item Calculate the partitions $\mathcal{M}_{k}^{d}$ and $\mathcal{R}^{d}_{k}$
    associated with $N^{d}$, $d\in\{0,1\}$. Calculate directions $\delta$ for
    $m$ and a direction $\delta_{r}$ for $r$ as
    $\delta_{r, i}=\delta_{j}=\sigma^{2}(0)
    (w(0)+(\#\mathcal{R}_{k}^{0}/\#\mathcal{M}^{0}_{k})w(1))$ for
    $i\in\mathcal{R}^{0}_{k}$ and $j\in\mathcal{M}_{k}^{0}$, and
    $\delta_{r, j}=\delta_{i}=\sigma^{2}(1)
    (w(1)+(\#\mathcal{R}_{k}^{1}/\#\mathcal{M}^{1}_{k})w(0))$ for
    $i\in\mathcal{R}^{0}_{k}$ and $j\in\mathcal{M}_{k}^{0}$.
  \item\label{item:Lam-direction} Calculate directions $\Delta^{d}$ for
    $\Lambda^{d}$ by setting $\Delta^{d}_{ij}=0$ if $N^{d}_{ij}=0$, with the
    remaining elements given by a solution to the systems of $n$ equations (i)
    $\sum_{i=1}^{n_{0}} \Delta_{ij}^{1}=\delta_{j}/\sigma^{2}(0)-w(0)$,
    $j=1,\dotsc, n_{0}$ and $\sum_{j=1}^{n_{0}}\Delta_{ij}^{0}=w(1)$,
    $i=1,\dotsc, n_{1}$ and (ii)
    $\sum_{j=1}^{n_{0}}\Delta_{ij}^{1}=\delta_{i}/\sigma^{2}(1)-w(1)$,
    $i=1,\dotsc, n_{1}$ and $\sum_{i=1}^{n_{1}}\Delta_{ij}^{0}=w(0)$,
    $j=1,\dotsc, n_{0}$.
  \item Calculate step size $s$ as
    $s=\min\{s^{0}_{1}, s^{0}_{2}, s_{1}^{1}, s_{2}^{1}\}$, where
    \begin{align*}
      s^{0}_{1}
      &=\min\{s\geq 0\colon
        r_{i}+\delta_{r, i}s=\delta_{j}s+D_{ij}^{0}
        \;\text{some $(i, j)$ s.t.\ $N^{0}_{ij}=0$, $\delta_{j}>
        \delta_{r, i}$}\}\\
      s^{1}_{1}
      &=\min\{s\geq 0\colon
        r_{j}+\delta_{r, j}s=\delta_{i}s+D_{ij}^{1}
        \;\text{some $(i, j)$ s.t.\ $N^{1}_{ij}=0$, $\delta_{i}>
        \delta_{rj}$} \}\\
      s^{d}_{2}
      &=\min\{s\geq 0\colon \Lambda^{d}_{ij}+
        s\Delta^{d}_{ij}=0\;\text{among $(i, j)$ with
        $N^{d}_{ij}=1$ and $\Delta^{d}_{ij}<0$}\}
    \end{align*}
  \item Update $\mu\mapsto \mu+s$, $m\mapsto m+s\delta$,
    $r\mapsto r+s\delta_{r}$ $\Lambda^{d}\mapsto \Lambda^{d}+s\Delta^{d}$,
    $D_{ij}^{0}\mapsto D_{ij}^{0}+s\delta_{j}$,
    $D_{ij}^{1}\mapsto D_{ij}^{1}+s\delta_{i}$ If $s=s^{d}_{1}$, then
    update $N^{d}_{ij}=1$, where $(i, j)$ is the index defining $s^{d}_{1}$. If
    $s=s^{d}_{2}$, update $N^{d}_{ij}=0$, where $(i, j)$ is the index defining
    $s^{d}_{2}$.
  \end{enumerate}
\end{enumerate}
Given the solution path $\{m(\mu)\}_{\mu>0}$, the optimal estimator
$\hat{L}_{\delta}$ and its worst-case bias can then be easily computed. For
simplicity, we specialize to the \ac{CATE} case, $w(1)=w(0)=1/n$. Let
$\delta(\mu)=2C\sqrt{m(\mu)'m(\mu)}$. It then follows from the formulas in
\Cref{sec:gener-setup-results} and the first-order conditions associated with
the Lagrangian~\eqref{eq:lagrangian-mu} (see the supplemental materials) that
the optimal estimator takes the form
$\hat{L}_{\delta(\mu)}
=\frac{1}{n}\sum_{i=1}^{n}(\hat{f}_{\mu}(x_{i},1)-\hat{f}_{\mu}(x_{i},0))$,
where $\hat{f}_{\mu}(x_{j},1)=\sum_{i} n\Lambda_{ij}^{1}(\mu)/\mu Y_{i}$ if
$d_{j}=0$; $\hat{f}_{\mu}(x_{i},1)=Y_{i}$ if $d_{i}=1$;
$\hat{f}_{\mu}(x_{j},0)=Y_{j}$ if $d_{j}=0$; and
$\hat{f}_{\mu}(x_{i},0)=\sum_{j}n\Lambda_{i, j}^{0}(\mu)/\mu Y_{j}$ if
$d_{i}=1$. The worst-case bias of the estimator is given by
$C(\sum_{i=1}^{n}(m_{i}(\mu)+r_{i}(\mu))/n-\sum_{i=1}^{n}m_{i}(\mu)^{2}/\mu)$.

For the interpretation of $\hat{L}_{\delta}$ as a matching estimator with a
variable number of matches, observe that
$\sum_{i} n\Lambda_{ij}^{1}(\mu)/\mu=\sum_{j}
n\Lambda_{ij}^{0}(\mu)/\mu= 1$. Also, $N_{ij}^{0}(\mu)=0$ and hence
$\Lambda_{ij}^{0}(\mu)=0$ unless
$D_{ij}^{0}(\mu)=\min_{\ell}D_{i \ell}^{0}(\mu)$. Similarly,
$\Lambda_{ij}^{1}(\mu)=0$ unless
$D_{ij}^{1}(\mu)=\min_{\ell}D_{\ell j}^{1}(\mu)$. Thus, the counterfactual
outcome for each observation $i$ is given by a weighted average of outcomes for
observations with opposite treatment status that are closest to it in terms of
the ``effective distance'' matrices $D_{ik}^{0}(\mu)$ (if $d_{i}=1$)
or $D^{1}_{ki}(\mu)$ (if $d_{i}=0$). Since
$D^{0}_{ik}(\mu)=m_{k}(\mu)+\normx{x_{i}-x_{k}}$
$D^{1}_{ki}(\mu)=m_{i}(\mu)+\normx{x_{i}-x_{k}}$, and $m_{k}(\mu)$
is increasing in the number of times $k$ has been used as a match, observations
that have been used more often as a match are considered to be further away
according to these effective distance matrices.

\section{Proofs for asymptotic results}\label{sec:proofs-asympt-result}

This appendix gives additional details and proofs for the results
in~\Cref{sec:pate,sec:asymptotic-results}.

\subsection{Proofs and details for Section~\ref{sec:pate}}\label{pate_sec_append}

We now analyze the asymptotic coverage properties of the \ac{CI}
in~\eqref{eq:pate_ci} and its one-sided analog $\hor{\hat{c}, \infty}$, where
\begin{equation}\label{eq:pate_onesided_ci}
  \hat{c}=\hat{L}_{k}-\maxbias_{\mathcal{F}}(\hat{L}_k)-z_{1-\alpha} \sepate(\hat{L}_{k}).
\end{equation}

We give coverage results that are uniform over a certain class $\mathcal{P}$ of
underlying distributions. To this end, we index the population from which the
data is drawn by $P$, and use $P$ and $E_{P}$ to denote probability statements
and expectations taken under $n$ i.i.d.\ draws from the population. We make the
dependence of conditional distributions on $P$ explicit by writing
$f_P(x, d)=E_P[Y_{i}\mid X_i=x, D_i=d]$ and
$\sigma^2_P(x, d)=\var_P(Y_{i}\mid X_i=x, D_i=d)$, with $Lf_{P}$ and
$\tau(P)=E_{P}Lf_{P}$ denoting the \ac{CATE} and \ac{PATE}, respectively. The
asymptotic coverage requirement for a sequence of \acp{CI} $\mathcal{C}$ based
on data $\{Y_{i}, X_{i}, D_i\}_{i=1}^n$ is
\begin{equation}\label{pate_asym_coverage_requirement_eq}
  \liminf_n\inf_{P\in\mathcal{P}} P\left(\tau(P)\in\mathcal{C} \right)\ge 1-\alpha.
\end{equation}
To construct $\mathcal{P}$, we assume that $f_P\in\mathcal{F}$ for all
$P\in\mathcal{P}$, and that $\mathcal{F}$ is convex and centrosymmetric. We
allow the distribution of $u_i=Y_{i}-f_P(X_i)$ and $X_i$ to vary over a class
that places uniform integrability or moment conditions on $u_i$ and conditions
on the support of $X_{i}$. As before, we require $\mathcal{F}$ to be fully
specified. However, we do not need to know the constants governing the
regularity of $u_{i}$ and $X_{i}$. The weights $k$
in~\eqref{linear_estimator_eq} may depend on the entire set
$\{X_{i}, D_{i}\}_{i=1}^{n}$, but not on the outcomes $Y_{i}$. Let
$V_{2,n}(P)=E_P((f_P(X_{i},1)-f_P(X_{i},0)-\tau(P))^2)/n$ denote the variance
of the \ac{CATE}.

We now give asymptotic coverage results for the \acp{CI}
in~\cref{eq:pate_ci,eq:pate_onesided_ci} under the following high-level
assumption.
\begin{assumption}\label{pate_high_level_assump}
  For some sequence of constants $V_{1,n}(P)$, (i)
  $\frac{\sum_{i=1}^{n}
    k(X_i,D_i)^2\sigma^2_P(X_i,D_i)}{V_{1,n}(P)}\overset{p}{\to} 1$ uniformly
  over $P\in\mathcal{P}$ for all $\varepsilon>0$ and (ii) uniformly over
  $P\in\mathcal{P}$,
  \begin{equation*}
    \frac{\sum_{i=1}^{n} E_P[k(X_i,D_i)^2u_i^2\1{k(X_i,D_i)^2u_i^2>
        \varepsilon V_{1,n}(P)}]}{V_{1,n}(P)}\to 0.
  \end{equation*}
\end{assumption}
The supplemental materials verify \Cref{pate_high_level_assump} for the matching
estimator, and discuss consistency of standard errors.

\begin{theorem}\label{theorem:pate}
  Suppose that \Cref{pate_high_level_assump} holds,
  $[V_{1,n}(P)+V_{2,n}(P)]/\sepate(\hat{L}_{k})^2\stackrel{p}{\to} 1$ uniformly
  over $P\in\mathcal{P}$, and that for some $\eta>0$ and $d\in\{0,1\}$
  $E_P\abs{f_P(X_i, d)}^{2+\eta}<1/\eta$ for all $P\in\mathcal{P}$. Then the
  confidence sets in~\cref{eq:pate_ci,eq:pate_onesided_ci}
  satisfy~\eqref{pate_asym_coverage_requirement_eq}.
\end{theorem}
\begin{proof}
  Given a sequence $P_n\in\mathcal{P}$, we need to show that
  $\liminf_n P_n(\tau(P_n)\in\hor{\hat{c}, \infty})\ge 1-\alpha$. Since the
  \ac{CI} in~\eqref{eq:pate_ci} is given by the intersection of two one-sided
  \acp{CI}, its asymptotic validity then follows immediately from Bonferroni's
  inequality.

  To this end, use a martingale representation, analogous to the martingale
  representation for matching estimators noted by
  \citet{abadie_martingale_2012}. For $i=1, \dotsc, n$, let
  $\xi_{n, i}=\frac{1}{n}(f_{P_{n}}(X_i,1)-f_{P_{n}}(X_i,0)-\tau(P_n))$ and let
  $\mathcal{H}_{n, i}$ be the $\sigma$-algebra generated by $D_1, \dotsc, D_n$
  and $X_1, \dotsc, X_i$. For $i=n+1, \dotsc, 2n$, let
  $\xi_{n, i}=k(X_{i-n}, D_{i-n})u_{i-n}$, and let $\mathcal{H}_{n, i}$ be the
  $\sigma$-algebra generated by $\mathcal{H}_{n, n}$ and
  $u_{1}, \dotsc, u_{i-n}$. Then $\sum_{i=1}^j \xi_{n, i}$ is a martingale with
  respect to the filtration $\mathcal{H}_{n, j}$. Let
  $s_n(P_n)^2=\sum_{i=1}^{2n}E_{P_n}(\xi_{n, i}^2\mid \mathcal{H}_{n,i-1})
  =V_{2,n}(P_n) + \sum_{i=1}^{n} k(X_i,D_i)^2\sigma^2_P(X_i, D_i)$ and let
  $\tilde s_n(P_n)^2=V_{2,n}(P_n)+V_{1,n}(P_n)$. We apply the martingale central
  limit theorem, Theorem 35.12 in \citet{billingsley_probability_1995}, to the
  martingale $\sum_{i=1}^j\tilde \xi_{n, i}$ where
  $\tilde \xi_{n, i}=\xi_{n, i}/\tilde s_n(P_n)$. By
  \Cref{pate_high_level_assump}, $s_n(P_n)/\tilde s_n(P_n)$ converges in
  probability to one under $P_n$. Thus,
  $\sum_{i=1}^{2n}E_{P_n}(\tilde\xi_{n, i}^2 \mid
  \mathcal{H}_{n,i-1})=s_n({P_n})^{2} / \tilde{s}_n({P_n})^2$ converges to one
  under $P_n$, which gives condition (35.35) in
  \citet{billingsley_probability_1995}. To verify the Lindeberg condition
  (35.36) in \citet{billingsley_probability_1995}, note that, for
  $i=1, \dotsc, n$,
  $\tilde{\xi}_{n, i}^{2} = (f_{P_n}(X_{i}, 1)-f_{P_{n}}(X_{i},
  0)-\tau({P_n}))^2/(n^2\tilde{s}_n({P_n})^2) =W_i^2/[nE_{P_n}(W_i^2)+n^{2}
  V_{1,n}({P_n})]\le W_i^2/[nE_{P_n}(W_i^2)]= \tilde{W}_i^2/n$ where
  $W_i=f_{P_n}(X_i,1)-f_{P_n}(X_i,0)-\tau({P_n})$ and
  $\tilde W_i=W_i/\sqrt{E_{P_n}(W_i^2)}$. Thus, for any $\varepsilon>0$,
  $\sum_{i=1}^n\tilde\xi^2\1{\tilde \xi^2>\varepsilon}\le E_{P_n}
  \tilde{W}_i^2\1{\tilde W_i^2> n}$ which converges to zero by the uniform
  $2+\eta$ moment bounds. For the remaining terms, we have
\begin{equation*}
  \sum_{i=n+1}^{2n} \tilde\xi_{n, i}^2\1{\tilde \xi_{n, i}^2>\varepsilon}
  = \frac{\sum_{i=1}^{n} E_{P_n}[k(X_{i},D_i)^2u_i^2
    \1{k(X_i,D_i)^2u_i^2>\varepsilon(V_{2,n}({P_n})+V_{1,n}({P_n}))}]}{
    V_{2,n}({P_n})+V_{1,n}({P_n})}
\end{equation*}
which converges to $0$ by \Cref{pate_high_level_assump}. Thus,
$\sum_{i=1}^{2n}\tilde \xi_{n, i}\stackrel{d}{\to} N(0,1)$ under $P_n$.

To complete the proof, note that
\begin{multline*}
    P_n (\tau({P_n})\notin \hor{\underline{\hat c}_\alpha, \infty}) = P_{n}
    \left(\tau({P_n})<\hat L_k-\maxbias_{\mathcal{F}}(\hat L_k)
      -z_{1-\alpha} \sepate(\hat{L}_{k})\right) \\
    \le P_{n}\left(\tau({P_n})<\sum_{i=1}^{n} k(X_i,D_i)u_i +Lf_{P_{n}}
      -z_{1-\alpha} \sepate(\hat{L}_{k})\right)
    = P_{n}\left(z_{1-\alpha} < \frac{\tilde s_n({P_n})}{\sepate(\hat{L}_{k})}
      \sum_{i=1}^{2n}\tilde \xi_{n, i} \right),
  \end{multline*}
which converges to $\alpha$ since $\tilde s_n(P)/\sepate(\hat{L}_{k})$ converges
to one uniformly over $P\in\mathcal{P}$,
\end{proof}

\subsection{Proof of
  \texorpdfstring{\Cref{sate_rate_thm}}{Theorem~\ref{sate_rate_thm}}}\label{efficiency_bound_sec_append}

The fact that $X_i$ has a bounded density conditional on $D_i$ means that there
exists some $a<b$ such that $X_i$ has a density bounded away from zero and
infinity on $[a, b]^{\dimx}$ conditional on $D_i=1$. Let
$\mathcal{N}_{d, n}=\{i\colon D_i=d, i\in\{1, \dotsc, n\}\}$,
$\norm{x}^{2}=\sum_{j=1}^{\dimx}x_{j}^{2}$, and let
  \begin{equation*}
    \mathcal{I}_n(h)
    =\{i\in \mathcal{N}_{1,n}\colon X_i\in[a, b]^{\dimx}\text{ and for all }j\in\mathcal{N}_{0,n},
    \norm{X_i-X_j}> 2h\}
  \end{equation*}
  Let $\mathcal{E}$ denote the $\sigma$-algebra generated by
  $\{D_i\}_{i=1}^\infty$ and $\{X_i\colon D_i=0,i\in\mathbb{N}\}$. Note that,
  conditional on $\mathcal{E}$, the observations
  $\{X_i\colon i\in\mathcal{N}_{1,n}\}$ are i.i.d.\ with density bounded away
  from zero and infinity on $[a, b]^{\dimx}$. The following lemma is proved in
  the supplemental materials.
\begin{lemma}\label{In_lemma}
  There exists $\eta>0$ such that, if $\limsup_{n} h_{n}n^{1/\dimx}\le \eta$,
  then almost surely, $\liminf_{n}\#\mathcal{I}_n(h_n)/n\ge \eta$.
\end{lemma}

Let $\tilde{\mathcal{X}}_n(h, \eta)$ be the set of elements $\tilde x$ in the
grid
\begin{equation*}
  \{a+j h\eta\colon j=(j_1, \dotsc, j_{\dimx})\in \{1, \dotsc, \lfloor
  h^{-1}\rfloor(b-a)\}^{\dimx}\}
\end{equation*}
such that there exists $i\in\mathcal{I}_n(h)$
with $\max_{1\le k\le \dimx}\abs{\tilde x_k-X_{i, k}}\le h\eta$. Note that, for any
$\tilde x\in\tilde{\mathcal{X}}_n(h, \eta)$, the closest element $X_i$ with
$i\in\mathcal{I}_n(h)$ satisfies
$\norm{\tilde x-X_i}\le \dimx h\eta$. Thus, for any $X_j$ with
$D_j=0$, we have
\begin{equation*}
  \norm{\tilde x-X_j}
  \ge \norm{X_j-X_i}-\norm{\tilde x-X_i}
  \ge 2h-\dimx\eta h
  > h
\end{equation*}
for $\eta$ small enough, where the first inequality follows from rearranging the
triangle inequality. Let $k\in\Sigma(1,\gamma)$ be a nonnegative function with
support contained in $\{x\colon \norm{x}\le 1\}$, with
$k(x)\ge \underline k$ on $\{x\colon \max_{1\le k\le \dimx} \abs{x_k}\le \eta\}$ for
some $\underline k>0$. By the above display, the function $f_n(x, d)
  =\sum_{\tilde x\in \tilde{\mathcal{X}}_n(h, \eta)} (1-d)k((x-\tilde x)/h)$
is equal to zero for $(x, d)=(X_i,D_i)$ for all $i=1, \dotsc, n$. Thus, it is
observationally equivalent to the zero function conditional on
$\{X_i,D_i\}_{i=1}^n$:
$P_{f_{n}}(\cdot\mid\{X_i,D_i\}_{i=1}^n)=P_{0}(\cdot
\mid\{X_i,D_i\}_{i=1}^n)$. Furthermore, we have
\begin{equation}\label{fn_sate_bound_eq}
  Lf_{n}
  = -\frac{1}{n}\sum_{i=1}^n \sum_{\tilde x\in\tilde{\mathcal{X}}_n(h, \eta)}k((X_i-\tilde x)/h)
    \le -\underline k \frac{\#\mathcal{I}_n(h)}{n},
\end{equation}
where the last step follows since, for each $i\in \mathcal{I}_n(h)$, there is a
$\tilde x\in\tilde{\mathcal{X}}_n(h, \eta)$ such that
$\max_{1\le k\le \dimx}\abs{\tilde x_k-X_{i, k}}/h\le \eta$.

Now let us consider the Hölder condition on $f_{n}$. Let
$\ell$ be the greatest integer strictly less than $\gamma$ and let $D^r$ denote
the derivative with respect to the multi-index $r=r_1, \dotsc, r_{\dimx}$ for some
$r$ with $\sum_{i=1}^{\dimx}r_i=\ell$. Let $x, x'\in \mathbb{R}^{\dimx}$. Let
$\mathcal{A}(x, x')\subseteq \tilde{\mathcal{X}}_n(h, \eta)$ denote the set of
$\tilde x\in\tilde{\mathcal{X}}_n(h, \eta)$ such that
$\max\{k((x-\tilde x)/h), k((x'-\tilde x)/h)\}>0$. By the support conditions on
$k$, there exists a constant $K$ depending only on $\dimx$ such that
$\#\mathcal{A}(x, x')\le K/\eta^{\dimx}$. Thus,
\begin{multline*}
  \abs{D^{r}f_{n}(x, d)
    -D^{r} f_{n}(x', d)} \\
  \le h^{-\ell}(K/\eta^{\dimx}) \sup_{\tilde x\in \mathcal{A}(x, x')}
    \abs{D^{r} k((x-\tilde x)/h)-D^{r} k((x'-\tilde x)/h)} \\
  \le h^{-\ell}(K/\eta^{\dimx}) \norm{(x-x')/h}^{\gamma-\ell}
    =h^{-\gamma}(K/\eta^{\dimx}) \norm{x-x'}^{\gamma},
\end{multline*}
which implies that $\tilde f_{n}\in \Sigma(C, \gamma)$ where
$\tilde f_{n}(x, d)=\frac{h^\gamma C}{K/\eta^{\dimx}}
f_{n}(x, d)$. By~\eqref{fn_sate_bound_eq}, the \ac{CATE}
under $\tilde f_{n}$ is bounded from above by
$-\underline k\frac{h^\gamma C}{K/\eta^{\dimx}}\frac{\#\mathcal{I}_n(h)}{n}$,
which, by \Cref{In_lemma}, is bounded from above by a constant times
$h_n^\gamma$ for large enough $n$ on a probability one event for $h_n$ a small
enough multiple of $n^{-1/\dimx}$. Thus, there exists $\varepsilon>0$ such that
the \ac{CATE} under $\tilde f_{n}$ is bounded from above by
$-\varepsilon n^{-1/\dimx}$ for large enough $n$ with probability one. On this
probability one event,
\begin{multline*}
  \liminf_n P_0\left(\hat c_n \le -\varepsilon n^{-\gamma}
    \mid \{X_i,D_i\}_{i=1}^n\right)
    =\liminf_n P_{\tilde f_{n}}\left(\hat c_n \le \varepsilon n^{-\gamma}
    \mid \{X_i,D_i\}_{i=1}^n\right) \\
  \ge \liminf_n \inf_{f(\cdot,0), f(\cdot,1)\in \Sigma(C, \gamma)}
    P_f\left(Lf\in \hor{\hat c_n, \infty}
    \mid \{X_i,D_i\}_{i=1}^n\right)
    \ge 1-\alpha,
\end{multline*}
which gives the result.

\subsection{Proofs of \texorpdfstring{\Cref{asymptotic_coverage_theorem,match_asymptotic_coverage_theorem}}{%
Theorems~\ref{asymptotic_coverage_theorem} and~\ref{match_asymptotic_coverage_theorem}
}}\label{nonnormal_sec_append}

We use the following result, which shows that it suffices to verify
\cref{lindeberg_eq} for the weights $\tilde{k}^{*}_{\delta}$ and
$k_{\text{match}, M}$. The proof follows by Theorem F.1 in \citet{ArKo18optimal}
and standard arguments, and is deferred to the supplemental materials.
\begin{lemma}\label{asymptotic_coverage_theorem_general}
  Suppose condition (a) of~\Cref{asymptotic_coverage_theorem} holds, and suppose
  that for $J$ fixed, $\max_{i}C_{n}\normx{x_{\ell_{J}(i)}-x_{i}}\to 0$, where
  $\ell_{J}(i)$ is the $J$th nearest neighbor of $i$. Let $\mathcal{C}$ be the
  \ac{CI} in \Cref{remark:other_feasible_cis} based on a linear estimator such that
  \cref{lindeberg_eq} holds, with $\hat{u}_{i}^{2}$ given by the nearest
  neighbor estimator with $J$ neighbors. Then
  $\liminf_{n\to\infty}\inf_{f\in\FLip(C_n)}P_f(Lf\in \mathcal{C})\ge 1-\alpha$.
\end{lemma}

\begin{proof}[Proof of \Cref{asymptotic_coverage_theorem}]
  By boundedness of $\tilde\sigma(x_i,d_i)$ away from zero and infinity,
  \cref{lindeberg_eq} for $k=\tilde{k}^*_\delta$ is equivalent to
  \begin{equation*}
    \frac{\max_{1\le i\le n}\tilde{f}^*_\delta(x_i,d_i)^2}{
      \sum_{i=1}^n \tilde f^*_\delta(x_i,d_i)^2} \to 0,
  \end{equation*}
  where $\tilde f^*_\delta$ is the solution to the optimization problem defined
  by~\eqref{eq:single-class-modulus-problem} and
  (\ref{eq:lipschitz-constraints-finite}) with $\tilde\sigma(x, d)$ in place of
  $\sigma(x, d)$. Since the constraint on
  $\sum_{i=1}^n\frac{\tilde f^*_\delta(x_i,d_i)^2}{\tilde\sigma^2(x_i,d_i)}$
  in~\eqref{eq:single-class-modulus-problem} binds, the denominator is bounded
  from above and below by constants that depend only on $\delta$ and the upper
  and lower bounds on $\tilde \sigma^2(x_i,d_i)$. Thus, it suffices to show
  that $\max_{1\le i\le n}\tilde f^*_\delta(x_i,d_i)^2\to 0$.

  To get a contradiction, suppose that there exists $\eta>0$ and a sequence
  $i_n^*$ such that $\tilde f^*_\delta(x_{i_n^*}, d_{i_n^*})^2> \eta^2$
  infinitely often. Then, by the Lipschitz condition,
  $\abs{\tilde f^*_\delta(x, d_{i_n^*})}\ge \eta-C_n\normx{x-x_{i_n^*}}$ so
  that, for $\normx{x-x_{i_n^*}}\le \eta/(2C_n)$, we have
  $\abs{\tilde f^*_\delta(x, d_{i_n^*})}\ge \eta/2$. Thus, we have
  \begin{equation*}
    \sum_{i=1}^n \tilde f^*_\delta(x_i,d_i)^2
    \ge \sum_{i: d_i=d_{i_n^*}} \tilde f^*_\delta(x_i,d_i)^2
    \ge (\eta/2)^2\#\{i\colon \normx{x_i-x_{i_n^*}}\le \eta/(2C_n), d_i=d_{i^*_n}\}
  \end{equation*}
  infinitely often. This gives a contradiction so long
  as~\eqref{x_density_condition} holds.
\end{proof}

\begin{proof}[Proof of \Cref{match_asymptotic_coverage_theorem}]
  To show that~\eqref{lindeberg_eq} holds for
  $k_{\text{match}, M}(x_i, d_i)=(1+K_M(i))/n$, it is sufficient to show that
  $\max_{1\le i\le n}K_M(i)^2/n\to 0$. To this end, let
  $U_M(x, d)=\normx{x_j-x}$ where $x_j$ is the $M$th closest
  observation to $x$ among observations $i$ with $d_i=d$, so that
  $K_M(i)=\#\{j\colon d_j\ne d_i, \normx{x_j-x_i}\le U_M(x_j,d_i)\}$.
  When~\eqref{overlap_event_eq} holds and $n$ is large enough so that
  $n \underline G(a_n)\ge M$, we will have $U_M(x, d)\le a_n$ for all
  $x\in\mathcal{X}$. By definition of $K_M(i)$, the upper bound
  in~\eqref{overlap_event_eq} then implies $K_M(i)\le n \overline G(a_n)$. Thus,
  it suffices to show that $[n \overline G(a_n)]^2/n=n \overline G(a_n)^2\to 0$.

  Let $c_n= n\underline G(a_n)/\log n$ and
  $b(t)=\overline G(\underline G^{-1}(t))^2/[t/\log t^{-1}]$ (so that
  $\lim_{t\to 0}b(t)=0$ under the conditions of
  \Cref{match_asymptotic_coverage_theorem}). Then
  $a_n= \underline G^{-1}(c_n(\log n)/n)$ so that
  \begin{equation*}
    n \overline G(a_n)^2=n \overline G(\underline G^{-1}(c_n(\log n)/n))^2
    =b(c_n(\log n)/n)\frac{c_n \log n}{\log n-\log c_n-\log\log n}.
  \end{equation*}
  This converges to zero so long as $c_n$ increases slowly enough (it suffices
  to take $c_n$ to be the minimum of $\log n$ and $1/\sqrt{b((\log n)^2/n)}$).
\end{proof}

\subsection{Proof of Lemma~\ref{x_density_sufficient_condition_lemma}}

It suffices to show that, for i.i.d.\ variables $w_i$ taking values in Euclidean
space with finite support $\mathcal{W}$,
$\inf_{w\in\mathcal{W}} \#\{i\in\{1, \dotsc, n\}\colon \norm{w-w_i}\le
\varepsilon\} \to \infty$ with probability one. To this end, for any $w$ and
$r$, let $B_r(w)=\{\tilde w\colon \norm{w-\tilde w}< r\}$ denote the open ball centered
at $w$ with radius $r$. Given $\delta>0$, let $\widetilde{\mathcal{W}}_\delta$
be a grid of meshwidth $\delta$ on $\mathcal{W}$. If $\delta$ is chosen to be
small enough, then, for every $w\in\mathcal{W}$, there exists
$\tilde w\in \widetilde{\mathcal{W}}_\delta$ such that
$B_\delta(\tilde w)\subseteq B_\varepsilon(w)$, and hence the quantity of
interest is bounded from below by
\begin{equation}\label{eq:finite_Wdelta}
  \min_{w\in\widetilde{\mathcal{W}}_\delta} \#\{i\in\{1, \dotsc, n\}\colon
  \norm{w-w_i} < \delta\},
\end{equation}
where we note that the infimum is now a minimum over a finite set. Since each
$w\in\widetilde{\mathcal{W}}_\delta$ is contained in the support of $w_i$, we
have $\min_{w\in\widetilde{\mathcal{W}}_\delta} P(\norm{w-w_i}< \delta)>0$, so it
follows from the strong law of large numbers~\eqref{eq:finite_Wdelta} converges
to infinity almost surely.

\end{appendices}

\onehalfspacing
\phantomsection%
\addcontentsline{toc}{section}{References}
\bibliography{../../np-testing-library}

\newpage

\end{document}


\maketitle

\renewcommand{\theequation}{S\arabic{equation}}
\renewcommand{\thetable}{S\arabic{table}}
\renewcommand{\thefigure}{S\arabic{figure}}
\renewcommand{\thepage}{S\arabic{page}}

\begin{appendices}
  \crefalias{section}{sappsec}%
  \crefalias{subsection}{sappsubsec}%
  \crefalias{subsubsection}{sappsubsubsec}%
  \setcounter{section}{2}
  \allowdisplaybreaks%

  These supplemental materials are organized as follows.
  \Cref{other_distance_sec} gives additional empirical results.
  \Cref{sec:proofs-auxil-lemm} proves \Cref*{lemma:linear-constraints}, gives
  the derivation of the solution path in the proof of
  \Cref*{lipschitz_optimal_thm}, completes the proof of
  \Cref*{match_optimality_thm}, proves \Cref*{In_lemma} and
  \Cref*{asymptotic_coverage_theorem_general}, gives conditions for asymptotic
  efficiency of the matching estimator with a single match, and finally verifies
  \Cref*{pate_high_level_assump} for the matching estimator.

\section{Additional empirical results: Other choices of distance}\label{other_distance_sec}

\begin{table}[tp]
  \centering
  \begin{threeparttable}
    \caption{Diagonal elements of the weight matrix $A$ in definition of the
      norm in~\cref*{eq:norm-nsw} for the main
      specification, $A_{\text{main}}$, and alternative specification,
      $A_{\text{ne}}$.}\label{Ane_table_supp}
  \begin{tabular}{@{}lrrrrrcccc@{}}
    &&&&&& \multicolumn{2}{c}{Earnings}
    & \multicolumn{2}{c@{}}{Employed}\\
    \cmidrule(rl){7-8}\cmidrule(l){9-10}
    & Age & Educ. & Black & Hispanic & Married& 1974 & 1975 & 1974 & 1975 \\
    \midrule
    $A_{\text{main}}$ &0.15& 0.60& 2.50& 2.50& 2.50& 0.50&0.50&0.10& 0.10\\
    $A_{\text{ne}}$   &0.10& 0.33& 2.20& 5.49& 2.60& 0.07&0.07&2.98& 2.93
  \end{tabular}
\end{threeparttable}
\end{table}

A disadvantage of the distance based on $A=A_{\text{main}}$ is that it requires
prior knowledge of the relative importance of different pretreatment variables
in explaining the outcome variable. An alternative is to specify the distance
using moments of the pretreatment variables in a way that ensures invariance to
scale transformations. For example, \citet{abadie_bias-corrected_2011} form
matching estimators using the weighted Euclidean norm (so $q=2$) with
$A=A_{\text{ne}}\equiv \text{diag}(1/\text{std}(x_1), \dotsc,
1/\text{std}(x_{\dimx}))$, where $\text{std}$ denotes sample standard deviation.
\Cref{Ane_table_supp} shows the diagonal elements of $A_\text{ne}$. It can be seen
that this distance is most likely not the best way of encoding a researcher's
prior beliefs about Lipschitz constraints. For example, the bound on the
difference in average earnings between blacks and non-black non-Hispanics is
substantially smaller than the bound on the difference in average earnings
between Hispanics and non-black non-Hispanics.

If the constant $C$ is to be chosen conservatively, the derivative of $f(x, d)$
with respect to each of these variables must be bounded by $C$ times the
corresponding element in this table. If one allows for somewhat persistent
earnings, then $C$ should be chosen in the range of $10$ or above: to allow
previous years' earnings to have a one-to-one effect, we would need to take
$C=1/\sqrt{.07^2+.07^2}=10.1$. For this $C$, when $\delta$ is chosen to optimize
\ac{CI} length, the resulting \ac{CI} is given by $1.72\pm 7.63$, which is much
wider than the \acp{CI} reported in~\Cref*{nsw_results_table_Amain_C1}.

In \Cref*{match_optimality_thm}, we showed that the matching estimator with a
single match is optimal for $C$ large enough. For this result, it is important
that the norm used to construct the matches is the same as the norm defining the
Lipschitz class. To illustrate this point, consider a matching estimator
considered in \citet{abadie_bias-corrected_2011}, that uses $q=2$ and
$A=A_{\text{ne}}$. The \ac{RMSE} efficiency of this estimator under our main
specification ($A_{\text{main}}$, $q=1$ and $C=1$) is 77.5\%; for
\ac{CI} length, its efficiency is 74.6\%. This is considerably lower than the
efficiencies of the matching estimator that matched on the norm defining the
Lipschitz class reported in \Cref*{sec:results}. Furthermore, the efficiency is
never higher than 80.1\%, even for large values of $C$.

\section{Proofs of auxiliary Lemmas and additional details}\label{sec:proofs-auxil-lemm}

\subsection{Proof of~\texorpdfstring{\Cref*{lemma:linear-constraints}}{Lemma
    \ref*{lemma:linear-constraints}}}

We will show that~\cref*{eq:lip-blp-n} holds for (a) all $i, j$ with
$d_{i}=d_{j}=1-d$, (b) all $i, j$ with $d_{i}=1-d_{j}=d$, and for part (ii) that
it also holds (c) for all $i, j$ with $d_{i}=d_{j}=d$. Let $g_{i}$ denote the
$i$th element of the vector $(g(x_{1}, d), \dotsc, g(x_{n}, d))'$.

  For (a), if~\cref*{eq:lip-blp-n} didn't hold for some $i, j$ with
  $d_{i}=d_{j}=1-d$, then by the triangle inequality, for all $j'$ with
  $d_{j'}=d$,
    \begin{equation*}
      g_{j}+C\normx{x_{i}-x_{j}}<g_{i}\leq g_{j'}+C\normx{x_{i}-x_{j'}}
      \leq g_{j'}+C\normx{x_{i}-x_{j}}+C\normx{x_{j}-x_{j'}},
    \end{equation*}
    contradicting the assertion in both part (i) and part (ii)
    that~\cref*{eq:lip-blp-n} holds with equality for at least one $j'$ with
    $d_{j'}=d$. Similarly, for (c), if it didn't hold for some $i, j$, then for
    all $i'$ with $d_{i'}=1-d$, by the triangle inequality,
    \begin{equation*}
      g_{i'}\leq
      g_{j}+C\normx{x_{i'}-x_{j}}
      < g_{i}+C\normx{x_{i'}-x_{j}}-C\normx{x_{i}-x_{j}}
      \leq g_{i}+C\normx{x_{i'}-x_{i}},
    \end{equation*}
    contradicting the assertion that~\cref*{eq:lip-blp-n} holds with equality
    for at least one $i'$ with $d_{i'}=1-d$. Finally, for (b),
    if~\cref*{eq:lip-blp-n} didn't hold for some $i', j'$ with
    $d_{i'}=1-d_{j'}=d$, then by the triangle inequality, denoting by
    $j^*(j')$ an element with $d_{j^*}=d$ such that \cref*{eq:lip-blp-n} holds
    with equality when $i=j'$ and $j=j^*$,
    \begin{equation*}
      g_{i'}-g_{j^{*}(j')}=g_{i'}+C\normx{x_{j^{*}(j')}-x_{j'}}-g_{j'}>
      C\normx{x_{j^{*}(j')}-x_{j'}}+C\normx{x_{i'}-x_{j'}}
      \geq C\normx{x_{j^{*}(j')}-x_{i'}},
    \end{equation*}
    which violates (c).

\subsection{Derivation of algorithm for solution path}
Observe that $\Lambda^{0}_{ij}=0$ unless for some $k$, $i\in\mathcal{R}^{0}_{k}$
and $j\in\mathcal{M}^{0}_{k}$, and similarly $\Lambda^{1}_{ij}=0$ unless for
some $k$,
$j\in\mathcal{R}^{1}_{k}$ and $i\in\mathcal{M}^{1}_{k}$. Therefore, the
first-order conditions for the Lagrangian can be written as
\begin{align}\label{eq:foc0a}
  m_{j}/\sigma^{2}(0)&=\mu w(0)+\sum_{i\in\mathcal{R}_{k}^{0}}\Lambda^{0}_{ij}
  &j&\in\mathcal{M}_{k}^{0},
  & \mu w(1)&=\sum_{j\in\mathcal{M}_{k}^{0}}\Lambda^{0}_{ij}&i&\in\mathcal{R}_{k}^{0}, \\
  m_{i}/\sigma^{2}(1)&=\mu w(1)+\sum_{j\in\mathcal{R}^{1}_{k}}\Lambda^{1}_{ij}&i&\in\mathcal{M}_{k}^{1},
  &\mu w(0)&=\sum_{i\in\mathcal{M}_{k}^{1}}\Lambda^{1}_{ij}&j&\in\mathcal{R}_{k}^{1}.\label{eq:foc1a}
\end{align}
Summing up these conditions then yields
\begin{align*}
  \sum_{j\in\mathcal{M}_{k}^{0}}m_{j}/\sigma^{2}(0)
  &=\mu w(0)\cdot
    \#{\mathcal{M}_{k}^{0}}+\sum_{j\in\mathcal{M}_{k}^{0}}\sum_{i\in\mathcal{R}_{k}^{0}}\Lambda^{0}_{ij}
    =\#{\mathcal{M}_{k}^{0}}\cdot \mu w(0)+ \#\mathcal{R}_{k}^{0}\cdot \mu w(1), \\
  \sum_{i\in\mathcal{M}_{k}^{1}}m_{i}/\sigma^{2}(1)
  &=\mu w(1)\cdot\#\mathcal{M}_{k}^{1}
    +\sum_{i\in\mathcal{M}_{k}^{1}}\sum_{j\in\mathcal{R}^{1}_{k}}\Lambda^{1}_{ij}=
    \#\mathcal{M}_{k}^{1}\cdot\mu w(1)+\#\mathcal{R}_{k}^{1}\cdot\mu w(0).
\end{align*}
Following the argument in \citet[Section 4]{osborne00}, by continuity of the
solution path, for a small enough perturbation $s$,
${N}^{d}(\mu+s)={N}^{d}(\mu)$, so long as the elements of
$\Lambda^{d}(\mu)$ associated with the active constraints are strictly positive. In
other words, the set of active constraints doesn't change for small enough
changes in $\mu$. Hence, the partition $\mathcal{M}_{k}^{d}$ remains the same
for small enough changes in $\mu$ and the solution path is differentiable.
Differentiating the preceding display yields
\begin{align*}
\frac{1}{\sigma^{2}(0)} \sum_{j\in\mathcal{M}_{k}^{0}}\frac{\partial
  m_{j}(\mu)}{\partial \mu}
  &=\#{\mathcal{M}_{k}^{0}}\cdot w(0)+ \#\mathcal{R}_{k}^{0}\cdot w(1), \\
\frac{1}{\sigma^{2}(1)} \sum_{i\in\mathcal{M}_{k}^{1}}\frac{\partial
  m_{i}(\mu)}{\partial \mu}
  &= \#\mathcal{M}_{k}^{1}\cdot w(1)+\#\mathcal{R}_{k}^{1}\cdot w(0).
\end{align*}

If $j\in\mathcal{M}^{0}_{k}$, then there exists a $j'$ and $i$ such that the
constraints associated with $\Lambda_{ij}^{0}$ and $\Lambda_{ij'}^{0}$ are both
active, so that
$m_{j}+\normx{x_{i}-x_{j}}=r_{i}=m_{j'}+\normx{x_{i}-x_{j'}}$,
which implies that
$\partial m_{j}(\mu)/\partial \mu=\partial m_{j'}(\mu)/\partial \mu$. Since all
elements in $\mathcal{M}^{0}_{k}$ are connected, it follows that the
derivative $\partial m_{j}(\mu)/\partial \mu$ is the same for all $j$ in
$\mathcal{M}_{k}^{0}$. Similarly, $\partial m_{j}(\mu)/\partial \mu$ is the same for
all $j$ in $\mathcal{M}_{k}^{1}$. Combining these observations with the
preceding display implies
\begin{align*}
\frac{1}{\sigma^{2}(0)} \frac{\partial
  m_{j}(\mu)}{\partial \mu}
  &= w(0)+ \frac{\#\mathcal{R}_{k(j)}^{0}}{\#{\mathcal{M}_{k(j)}^{0}}} w(1), &
\frac{1}{\sigma^{2}(1)} \frac{\partial
  m_{i}(\mu)}{\partial \mu}
  &= w(1)+\frac{\#\mathcal{R}_{k(i)}^{1}}{\#\mathcal{M}_{k(i)}^{1}} w(0),
\end{align*}
where $k(i)$ and $k(j)$ are the partitions that $i$ and $j$ belong to.
Differentiating the first-order
conditions~\eqref{eq:foc0a} and \eqref{eq:foc1a} and combining them
with the restriction that $\partial \Lambda_{ij}^{d}(\mu)/\partial \mu=0$ if
$N^{d}_{ij}(\mu)=0$ then yields the following set of linear equations for
$\partial\Lambda^{d}(\mu)/\partial\mu$:
\begin{align*}
  \frac{\#\mathcal{R}^{0}_{k}}{\#{\mathcal{M}_{k}^{0}}}w(1)
  & =\sum_{i\in\mathcal{R}_{k}^{0}}\frac{\partial \Lambda^{0}_{ij}(\mu)}{\partial\mu},
  & w(1)&=\sum_{j\in\mathcal{M}_{k}^{0}}\frac{\partial\Lambda^{0}_{ij}(\mu)}{\partial\mu},
\\
  \frac{\#\mathcal{R}^{1}_{k}}{\#{\mathcal{M}^{1}_{k}}}w(0)
  & = \sum_{j\in\mathcal{R}_{k}^{1}}\frac{\partial\Lambda^{1}_{ij}(\mu)}{\partial\mu},
  & w(0)&=\sum_{i\in\mathcal{M}_{k}^{1}}\frac{\partial\Lambda^{1}_{ij}(\mu)}{\partial\mu},
  &\frac{\partial\Lambda^{d}_{ij}(\mu)}{\partial\mu}=0\qquad \text{if $N_{ij}^{d}(\mu)=0$}.
\end{align*}
Therefore, $m(\mu)$, $\Lambda^{0}(\mu)$, and $\Lambda^{1}(\mu)$ are all
piecewise linear in $\mu$. Furthermore, since for $i\in\mathcal{R}^{0}_{k}$,
$r_{i}(\mu)=m_{j}(\mu)+\normx{x_{i}-x_{j}}$ where $j\in\mathcal{M}_{k}^{0}$,
it follows that
\begin{equation*}
    \frac{\partial r_{i}(\mu)}{\partial \mu}
  = \frac{\partial m_{j}(\mu)}{\partial \mu}
  =\sigma^{2}(0)\left[w(0)+ \frac{\#\mathcal{R}_{k}^{0}}{\#{\mathcal{M}_{k}^{0}}}
    w(1)\right].
\end{equation*}
Similarly, since for $j\in\mathcal{R}^{1}_{k}$, and $i\in\mathcal{M}^{1}_{k}$
$r_{j}(\mu) =m_{i}(\mu)+\normx{x_{i}-x_{j}}$, where
$j\in\mathcal{M}^{0}_{k}$, we have
\begin{equation*}
\qquad
  \frac{\partial r_{j}(\mu)}{\partial \mu}
  = \frac{\partial m_{i}(\mu)}{\partial \mu}
  =\sigma^{2}(1)\left[
    w(1)+\frac{\#\mathcal{R}_{k}^{1}}{\#\mathcal{M}_{k}^{1}} w(0)\right].
\end{equation*}
Thus, $r(\mu)$ is also piecewise linear in $\mu$.

Differentiability of $m$ and $\Lambda^{d}$ is violated if the condition that the
elements of $\Lambda^{d}$ associated with the active constraints are all
strictly positive is violated. This happens if one of the non-zero elements of
$\Lambda^{d}(\mu)$ decreases to zero, or else if a non-active constraint becomes
active, so that for some $i$ and $j$ with $N^{0}_{ij}(\mu)=0$,
$r_{i}(\mu)=m_{j}(\mu)+\normx{x_{i}-x_{j}}$, or for some $i$ and
$j$ with $N^{1}_{ij}(\mu)=0$,
$r_{j}(\mu)=m_{i}(\mu)+\normx{x_{i}-x_{j}}$. This determines the
step size $s$ in the algorithm.

\subsection{Bounds on optimal \texorpdfstring{$\delta$}{delta} for
  \texorpdfstring{\Cref*{match_optimality_thm}}{Theorem~\ref{match_optimality_thm}}}\label{sec:optimal_delta_bound}

\Cref*{match_optimality_thm} follows from \Cref*{theorem:reg-adjusted-matching}
so long as the optimal $\delta$ for the \ac{FLCI} and \ac{RMSE} criteria do not
increase without bound as $C$ increases. This section shows that this is indeed the
case.

Let $S(\delta, C)=\sd(\hat L_\delta)$ and let
$B(\delta, C)=\maxbias_{\mathcal{F}}(\hat L_\delta)$ denote standard deviation
and worst-case bias when $\mathcal{F}$ is given by the Lipschitz class with
constant $C$, and $\hat L_\delta$ is computed with this class. Let
$\mathcal{A}(C)$ denote the feasible set of worst-case bias and standard
deviation pairs for this problem. Note that the set $\mathcal{A}(C)$ is convex.
In particular, given estimators $\hat L_1$ and $\hat L_2$ with worst-case bias
$B_{1}, B_{2}$ and standard deviation $S_{1}, S_{2}$, the estimator
$\lambda \hat L_1+(1-\lambda)\hat L_2$ has worst-case bias bounded by
$\lambda B_1+(1-\lambda) B_2$ and standard deviation bounded by
$\lambda S_1+(1-\lambda) S_2$, which then allows for the construction of an
affine estimator with worst-case bias and standard deviation exactly equal to
these quantities by adding a nonrandom constant and a multiple of a
$\mathcal{N}(0,1)$ variable independent of the observed data (adding a
$\mathcal{N}(0,1)$ variable to the sample will not change the calculations for
the optimal estimator for \ac{RMSE} or \ac{FLCI} length).

Let $R(B, S)$ be the \ac{RMSE} criterion ($R(B, S)=\sqrt{B^2+S^2}$) or the
\ac{FLCI} length criterion ($R(B, S)=\cv_\alpha(B/S)S$). Let
$\delta^*=\delta^*(C)$ minimize $R(B(\delta, C), S(\delta, C))$. Then
$B(\delta^*, C), S(\delta^*, C)$ optimizes $R(B, S)$ over the feasible set
$\mathcal{A}(C)$. Let $\delta\ne \delta^*$ be given. By convexity of the
feasible set $\mathcal{A}(C)$, we have, for all $t\in [0,1]$,
\begin{equation*}
  R((B(\delta, C)-B(\delta^*, C))t+B(\delta^*, C), (S(\delta, C)-S(\delta^*, C))t+S(\delta^*, C))
  - R(B(\delta^*, C), S(\delta^*, C)) \ge 0.
\end{equation*}
Dividing both sides by $t$ and taking the limit as $t\to 0$, we obtain
\begin{equation*}
  R_1^*(C) [B(\delta, C)-B(\delta^*, C)] + R_2^*(C) [S(\delta, C) - S(\delta^*, C)] \ge 0,
\end{equation*}
where $(R_1^*(C), R_2^*(C))$ is the derivative of $R(B, S)$ at
$(B(\delta^*, C), S(\delta^*, C))$.  It now follows that $\delta^*$ minimizes
\begin{equation*}
  2 B(\delta) + [2 R_2^*(C)/R_1^*(C)] S(\delta)
\end{equation*}
over $\delta>0$. Note, however, that this is simply the worst-case $\beta$
quantile of excess length of a one-sided $1-\alpha$ CI when
$z_{1-\alpha}+z_\beta=2 R_2^*(C)/R_1^*(C)$, so this means that $\delta^*(C)$ is
also optimal for this criterion. By \Cref*{theorem:Ldelta-optimality}, the
estimator $\hat L_{\tilde\delta}$ where $\tilde{\delta}=2 R_2^*(C)/R_1^*(C)$ is
also optimal for this criterion. Furthermore, the estimator that optimizes this
criterion is unique in this setting, so it follows that the estimator that
optimizes the criterion $R(B, S)$ is equal to the estimator
$\hat L_{\tilde\delta}$.

To show that this estimator is equal to the matching estimator with a single
match once $C$ is large enough, it now suffices to show that
$R_2^*(C)/(2R_1^*(C))$ is bounded as $C\to \infty$ so that $C> K
R_2^*(C)/(2R_1^*(C))$ once $C$ is large enough.  This can be checked by noting
that, for the \ac{FLCI} length and RMSE criteria, $R_1^*(C)$ is bounded from below
and $R_2^*(C)$ is bounded from above, over the set $(B(\delta, C), S(\delta, C))$
with $C>0$, using the fact that $S(\delta, C)$ is bounded from above and below
away from zero over this set.

\subsection{Proof of \texorpdfstring{\Cref*{In_lemma}}{Lemma~\ref{In_lemma}}}
Let $A_n=\{x\in [a, b]^{\dimx}\colon \text{there exists $j$ such that $D_j=0$ and
  $\norm{x-X_j}\leq 2h$}\}$. Then
  $\# \mathcal{I}_n(h)=\sum_{i\in \mathcal{N}_{1,n}}[\1{X_i\in
    [a, b]^{\dimx}}-\1{X_i\in A_n}]$. Note that, conditional on $\mathcal{E}$,
  the random variables $\1{X_i\in A_n}$ with $i\in\mathcal{N}_{1,n}$ are i.i.d.\
  $\text{Bernoulli}(\nu_n)$ with
  $\nu_n=P(X_i\in A_n\mid \mathcal{E})=\int \1{x\in A_n}f_{X\mid D}(x\mid 1)\, dx\le
  K\lambda(A_n)$ where $f_{X\mid D}(x\mid 1)$ is the conditional density of $X_i$ given
  $D_i=1$, $\lambda$ is the Lebesgue measure and $K$ is an upper bound on this
  density. Under the assumption that $\limsup_{n} h_{n} n^{1/\dimx}\le \eta$, we
  have $\lambda(A_n)\le (4h_n)^{\dimx}n\le 8^{\dimx}\eta^{\dimx}$ where the last
  inequality holds for large enough $n$. Thus, letting
  $\overline\nu=8^{\dimx}\eta^{\dimx}K$, we can construct random variables $Z_i$
  for each $i\in \mathcal{N}_{1,n}$ that are i.i.d.\
  $\text{Bernoulli}(\overline\nu)$ conditional on $\mathcal{E}$ such that
  $\1{X_i\in A_n}\le Z_i$. Applying the strong law of large numbers, it follows
  that
  \begin{equation*}
    \begin{split}
      \liminf_n \#\mathcal{I}_n(h)/n &\ge \liminf_n
      \frac{\#\mathcal{N}_{1,n}}{n}
      \frac{1}{\#\mathcal{N}_{1,n}}\sum_{i\in\mathcal{N}_{1,n}} (\1{X_i\in [a, b]^{\dimx}}-Z_i) \\
      &\ge P(D_i=1)(P(X_i\in [a, b]^{\dimx}\mid D_i=1)-8^{\dimx}\eta^{\dimx}K)
    \end{split}
  \end{equation*}
  almost surely. This will be greater than $\eta$ for $\eta$ small enough.

\subsection{Proof of \texorpdfstring{\Cref*{asymptotic_coverage_theorem_general}}{Lemma~\ref{asymptotic_coverage_theorem_general}}}

The result follows from verifying the conditions of Theorem F.1 in
\citet{ArKo18optimal}. In particular, we need to show that the weights $k$ are
such that $\sum_{i=1}^{n}k(x_{i}, d_{i})u_{i}/\sd_{k}$ converges in distribution
to $N(0,1)$ (condition (S13) in \citealp{ArKo18optimal}) and
$\sum_{i}\hat{u}_{i}^{2}k(x_{i}, d_{i})^{2}/\sd^{2}_{k}$ converges in probability
to 1, uniformly over $f\in\FLip(C_n)$ (S14), where
$\sd_{k}^{2}=\sum_{i=1}^{n}\sigma^{2}(x_{i}, d_{i})k(x_{i}, d_{i})^{2}$.

Under the moment bounds on $u_i$, \cref*{lindeberg_eq} directly implies the
Lindeberg condition that is needed for condition (S13) to hold. To show that it
also implies (S14), note that (S14) is equivalent to the requirement that
$\sum_{i=1}^n\hat{u}_{i}^{2} a_{ni}-\sum_{i=1}^n \sigma^2(x_i,n_i)a_{ni}$
converges to zero uniformly over $f\in\FLip(C_n)$, where
\begin{equation*}
  a_{ni}=k(x_i, d_i)^2/\sum_{j=1}^{n}[\sigma^2(x_j,d_j)k(x_j,d_j)^2].
\end{equation*}
By an inequality of \citet{von_bahr_inequalities_1965},
\begin{multline*}
  E\abs*{\sum_{i=1}^n (u_i^2-\sigma^{2}(x_{i}, d_{i}))a_{ni}}^{1+1/(2K)}\leq 2
  \sum_{i=1}^{n}a_{ni}^{1+1/(2K)}E\abs{u_i^2-\sigma^2(x_i, d_i)}^{1+1/(2K)}\\
  \le \max_{1\le i\le n}a_{ni}^{1/(2K)}E\abs{u_i^2 - \sigma^2(x_i,
    d_i)}^{1+1/(2K)}\cdot \sum_{i=1}^n a_{ni}.
\end{multline*}
Note that, by boundedness of $\sigma(x, d)$ away from zero and infinity,
$\sum_{i=1}^n a_{ni}$ is uniformly bounded. Furthermore, it follows
from~\cref*{lindeberg_eq} that $\max_{1\le i\le n} a_{ni}\to 0$. From this and
the moment bounds on $u_i$, it follows that the above display converges to zero.
It therefore suffices to show that
$\sum_{i=1}^{n}(\hat{u}_{i}^{2}-u_{i}^{2})a_{ni}$ converges to zero. This
follows from the following result.
\begin{lemma}\label{lemma:nn-consistency}
  Consider the model in \cref*{fixed_design_eq}. Suppose that
  $1/K\le Eu_i^2\le K$ and $E\abs{u_{i}}^{2+1/K}\le K$ for some constant $K$,
  and that $\sigma^{2}(x, d)$ is uniformly continuous in $x$ for $d\in\{0,1\}$.
  Let $\ell_{j}(i)$ be the $j$th closest unit to $i$, with respect to some norm
  $\norm{\cdot}$, among units with the same value of the treatment. Let
  $\hat{u}^{2}_{i}=\frac{J}{J+1}(Y_{i}-\sum_{j=1}^{J}Y_{\ell_{j}(i)}/J)^{2}$,
  and let $a_{ni}\geq 0$ be a non-random sequence such that
  $\max_{i}{a_{ni}}\to 0$, and that $\sum_{i=1}^{n}a_{ni}$ is uniformly bounded.
  If $\max_{i}C_{n}\norm{x_{\ell_{J}(i)}-x_{i}}\to 0$, then
  $\sum_{i}a_{ni}(\hat{u}_{i}^{2}-u_{i}^{2})$ converges in probability to zero,
  uniformly over $\FLip(C_{n})$.
\end{lemma}

\begin{proof}
  The proof is based on the arguments in \citet{AbIm08cv}. For ease of notation,
  let $f_{i}=f(x_{i}, d_{i})$, $\sigma^{2}_{i}=\sigma^{2}(x_{i}, d_{i})$, and
  let $\overline{f}_{i}=J^{-1}\sum_{j=1}^{J}f_{\ell_{j}(i)}$ and
  $\overline{u}_{i}=J^{-1}\sum_{j=1}^{J}u_{\ell_{j}(i)}$. Then we can decompose
  \begin{multline*}
    \frac{J+1}{J}(\hat{u}_{i}^{2}-u_{i}^{2})=[
    f_{i}-\overline{f}_{i}+u_{i}-\overline{u}_{i}
    ]^{2}-\frac{J+1}{J}u_{i}^{2}\\
    =[(f_{i}-\overline{f}_{i})^{2}
    +2(u_{i}-\overline{u}_{i})(f_{i}-\overline{f}_{i})] -2\overline{u}_{i}u_{i}
    + \frac{2}{J^{2}} \sum_{j=1}^{J} \sum_{k=1}^{j-1}
    u_{\ell_{j}(i)}u_{\ell_{k}(i)}
    +\frac{1}{J^{2}}\sum_{j=1}^{J}(u_{\ell_{j}(i)}^{2}-u_{i}^{2})\\
    =T_{1i}+2T_{2i}+2T_{3i}+T_{4i}+T_{5i}+
    \frac{1}{J^{2}}\sum_{j=1}^{J}(\sigma_{\ell_{j}(i)}^{2}-\sigma_{i}^{2}),
  \end{multline*}
  where
  \begin{align*}
    T_{1i}
    &=[(f_{i}-\overline{f}_{i})^{2}
      +2(u_{i}-\overline{u}_{i})(f_{i}-\overline{f}_{i})],
    & T_{2i}&=\overline{u}_{i}u_{i}\\
    T_{3i}&=\frac{1}{J^{2}} \sum_{j=1}^{J}
            \sum_{k=1}^{j-1} u_{\ell_{j}(i)}u_{\ell_{k}(i)},
    & T_{4i} &=\frac{1}{J^{2}}\sum_{j=1}^{J}(u_{\ell_{j}(i)}^{2}-\sigma_{\ell_{j}(i)}^{2}),
          & T_{5i}&= \sigma_{i}^{2}-u_{i}^{2}.
  \end{align*}
  Since $\max_{i}\norm{x_{\ell_{J}(i)}-x_{i}}\to 0$ and since
  $\sigma^{2}(\cdot, d)$ is uniformly continuous, it follows that
  \begin{equation*}
    \max_{i}\max_{1\leq j\leq J} \abs{\sigma^{2}_{\ell_{j}(i)}-\sigma^{2}_{i}}\to
    0,
  \end{equation*}
  and hence that
  $\abs{\sum_{i=1}^{n}a_{ni}J^{-1}\sum_{j=1}^{J}
    (\sigma_{\ell_{j}(i)}^{2}-\sigma_{i}^{2})}\leq \max_{i} \max_{j=1,\dotsc, J}
  (\sigma_{\ell_{j}(i)}^{2}-\sigma_{i}^{2})\sum_{i=1}^{n}a_{ni}\to 0$. To prove
  the lemma, it therefore suffices to show that the sums
  $\sum_{i=1}^{n}a_{ni}T_{qi}$ all converge to zero.

  To that end,
  \begin{equation*}
    E\abs{\sum_{i}a_{ni}T_{1i}}\leq \max_{i}(f_{i}-\overline{f}_{i})^{2}
    \sum_{i}a_{ni}
    +2\max_{i}\abs{f_{i}-\overline{f}_{i}}
    \sum_{i}a_{ni}E\abs{u_{i}-\overline{u}_{i}},
  \end{equation*}
  which converges to zero since
  $ \max_{i}\abs{f_{i}-\overline{f}_{i}} \leq \max_{i}\max_{j=1,\dotsc,
    J}(f_{i}-f_{\ell_{j}(i)})\leq C_{n}\max_{i}\normx{x_{i}-x_{\ell_{J}(i)}}\to
  0$. Next, by the von Bahr-Esseen inequality,
  \begin{equation*}
    E\abs{\sum_{i=1}^{n}a_{ni}T_{5i}}^{1+1/2K}\leq
    2\sum_{i=1}^{n}a_{ni}^{1+1/2K}E\abs{T_{5i}}^{1+1/2K}\leq
    2\max_{i}a_{ni}^{1/2K}\max_{j}E
    \abs{T_{5j}}^{1+1/2K} \sum_{k=1}^{n}a_{nk}\to 0.
  \end{equation*}
  Let $\mathcal{I}_{j}$ denote the set of observations for which an observation
  $j$ is used as a match. To show that the remaining terms converge to zero, let
  we use the fact $\#\mathcal{I}_{j}$ is bounded by $J\overline{L}$, where
  $\overline{L}$ is the kissing number, defined as the maximum number of
  non-overlapping unit balls that can be arranged such that they each touch a
  common unit ball (\citealp[Lemma 3.2.1]{mttv97}; see also \citealp{AbIm08cv}).
  $\overline{L}$ is a finite constant that depends only on the dimension of the
  covariates (for example, $\overline{L}=2$ if $\dim(x_{i})=1$). Now,
  \begin{equation*}
    \sum_{i}a_{ni}T_{4i}=\frac{1}{J^{2}}
    \sum_{j=1}^{n}(u_{j}-\sigma^{2}_{j})\sum_{i\in \mathcal{I}_{j}}a_{ni},
  \end{equation*}
  and so by the von Bahr-Esseen inequality,
  \begin{multline*}
    E\abs{\sum_{i}a_{ni}T_{4i}}^{1+1/2K}\leq \frac{2}{J^{2+1/K}}
    \sum_{j=1}^{n}E\abs{u_{j}-\sigma^{2}_{j}}^{1+1/2K}
    \left(\sum_{i\in \mathcal{I}_{j}}a_{ni}\right)^{1+1/2K}\\
    \leq \frac{(J\overline{L})^{1/2K}}{J^{2+1/K}}
    \max_{k}E\abs{u_{k}-\sigma^{2}_{k}}^{1+1/2K} \max_{i}a_{ni}^{1+1/2K}
    \sum_{j=1}^{n}\sum_{i\in \mathcal{I}_{j}}a_{ni},
  \end{multline*}
  which is bounded by a constant times
  $\max_{i}a_{ni}^{1+1/2K} \sum_{j=1}^{n}\sum_{i\in \mathcal{I}_{j}}a_{ni}=
  \max_{i}a_{ni}^{1+1/2K} J\sum_{i}a_{ni}\to 0$. Next, since
  $E[u_{i}u_{i'}u_{\ell_{j}(i)}u_{\ell_{k}(i')}]$ is non-zero only if either
  $i=i'$ and $\ell_{j}(i)=\ell_{k}(i')$, or else if $i=\ell_{k}(i')$ and
  $i'=\ell_{j}(i)$, we have
  $\sum_{i'=1}^{n}a_{ni'}E[u_{i}u_{i'}u_{\ell_{j}(i)}u_{\ell_{k}(i')}]\leq\max_{i'}a_{ni'}\left(
    \sigma^{2}_{i}\sigma^{2}_{\ell_{j}(i)}+
    \sigma_{\ell_{j}(i)}^{2}\sigma^{2}_{i}\right)$, so that
  \begin{equation*}
    \var(\sum_{i}a_{ni}T_{2i})=
    \frac{1}{J^{2}}
    \sum_{i, j, k, i'}a_{ni}a_{ni'}E[u_{i}u_{\ell_{k}(i')}
    u_{i'}u_{\ell_{j}(i)}]\leq 2K^{2}\max_{i'}a_{ni'}
    \sum_{i}a_{ni}
    \to 0.
  \end{equation*}
  Similarly for $j\neq k$ and $j'\neq k$,
  $\sum_{i'=1}^{n}a_{ni'}
  E[u_{\ell_{j}(i)}u_{\ell_{k}(i)}u_{\ell_{j'}(i')}u_{\ell_{k'}(i')}]\leq
  \max_{i'} 2\sigma^{2}_{\ell_{j}(i)}\sigma^{2}_{\ell_{k}(i)}$, so that
  \begin{multline*}
    \var\Big(\sum_{i}a_{ni}T_{3i}\Big)\\
    =\frac{1}{J^{4}}\sum_{i, i', j, j'}\sum_{k=1}^{j-1}\sum_{k'=1}^{j'-1}
    a_{ni}a_{ni'}E[u_{\ell_{j}(i)}u_{\ell_{k}(i)}
    u_{\ell_{j'}(i')}u_{\ell_{k'}(i')}]\leq
    2K^{2}\max_{i'}a_{ni'}\sum_{i}a_{ni}\to 0.
  \end{multline*}
\end{proof}

\subsection{Asymptotic efficiency of the matching estimator}

By \Cref*{lipschitz_optimal_thm}, the matching estimator with $M=1$ is efficient
in finite samples if the Lipschitz constant $C$ is large enough. We now give
conditions for its asymptotic optimality.

\begin{theorem}\label{match_asym_efficiency_thm}
  Suppose that the assumptions of \Cref*{sate_rate_thm} hold, and that
  $\sigma^2(x, d)$ is bounded away from zero and infinity. Suppose that, for
  some functions $\overline{G}\colon \mathbb{R}^+\to\mathbb{R}^+$ and
  $\underline{G}\colon\mathbb{R}^+\to\mathbb{R}^+$ with
  $\lim_{t\to 0}\overline G(\underline G^{-1}(t))^2/[t/\log
  t^{-1}]^{2/\dimx+1}=0$,
  \begin{equation*}
    \underline G(a)
    \le P(\normx{X_i-x}\le a, \, D_i=d)
    \le \overline G(a).
  \end{equation*}
  Let $R^*_{n, \text{match}, \text{RMSE}}$ denote the worst-case \ac{RMSE} of
  the matching estimator with $M=1$, and let $R^*_{n, \text{opt}, \text{RMSE}}$
  denote the minimax \ac{RMSE} among linear estimators, conditional on
  $\{X_{i}, D_{i}\}_{i=1}^n$, for the class $\FLip(C)$. Then
  $R^*_{n, \text{match}, \text{RMSE}}/R^*_{n, \text{opt}, \text{RMSE}}\to 1$
  almost surely. The same holds with ``RMSE'' replaced by ``\ac{CI} length'' or
  ``$\beta$ quantile of excess length of a one-sided \ac{CI}.''
\end{theorem}

If $X_i$ has sufficiently regular support and the conditional density of $X_i$
given $D_i$ is bounded away from zero on the support of $X_i$ for both $D_i=0$
and $D_i=1$, then the conditions of \Cref{match_asym_efficiency_thm} hold with
$\underline G(a)$ and $\overline G(a)$ both given by constants times $a^\dimx$,
so that $\overline G(\underline G(a))$ decreases like $a$ as $a\to 0$. Thus, the
conditions of \Cref{match_asym_efficiency_thm} hold so long as $\dimx>2$ and
there is sufficient overlap.

\begin{proof}
  Let $\sd_{\delta_{\RMSE}, n}$ and $\maxbias_{\delta_{\RMSE}, n}$ denote the
  standard deviation and worst-case bias of the minimax linear estimator and let
  $\sd_{\text{match},1}$ and $\maxbias_{\text{match},1}$ denote the standard
  deviation and worst-case bias of the estimator with a single match
  (conditional on $\{(X_i,D_i)_{i=1}^n\}$). Since worst-case bias is increasing
  in $\delta$ and variance is decreasing in $\delta$, and since the matching
  estimator with $M=1$ solves the modulus problem for small enough $\delta$ by
  \Cref*{match_optimality_thm}, we have
  $\maxbias_{\delta_{\RMSE}, n}\ge \maxbias_{\text{match},1}$. Thus,
\begin{equation*}
1\le \frac{\maxbias_{\text{match},1}^2+\sd_{\text{match},1}^2}{\maxbias_{\delta{\RMSE}, n}^2+\sd_{\delta{\RMSE}, n}^2}
  \le \frac{\maxbias_{\delta{\RMSE}, n}^2+\sd_{\text{match},1}^2}{\maxbias_{\delta{\RMSE}, n}^2+\sd_{\delta{\RMSE}, n}^2}
  \le 1 + \frac{\sd_{\text{match},1}^2}{\maxbias_{\delta{\RMSE}, n}^2+\sd_{\delta{\RMSE}, n}^2}.
\end{equation*}
By the arguments in the proof of \Cref*{sate_rate_thm}, there exists
$\varepsilon>0$ such that $\maxbias_{\delta{\RMSE}, n}\ge \varepsilon n^{-2/p}$
almost surely. In addition, by Theorem 37 in Chapter 2 of
\citet{pollard_convergence_1984}, the conditions of
\Cref*{match_asymptotic_coverage_theorem} hold almost surely (with
$\underline G(a)$ and $\overline G(a)$ multiplied by some positive constants).
Arguing as in the proof of \Cref*{match_asymptotic_coverage_theorem} then gives
the bound
$\sd_{\text{match},1}^2 \le [2\max_{1\le i\le n} K_1(i)]^2/n \le [2n\overline
G(a_n)]^2/n$ for any sequence $a_n= \underline G^{-1}(c_n(\log n)/n)$ with
$c_n=n\overline G(a_n)/\log n\to\infty$. Plugging these bounds into the above
display gives a bound proportional to
\begin{equation*}
  \overline G(\underline G^{-1}(c_n(\log n)/n))^2 n^{2/p+1}
  =b(c_n(\log n)/n) \left[\frac{c_n(\log n)/n}{\log n-\log c_n-\log\log n} \right]^{2/p+1}
  n^{2/p+1},
\end{equation*}
where $b(t)=\overline G(\underline G^{-1}(t))^2/[t/\log t^{-1}]^{2/p+1}$. If
$\lim_{t\to 0} b(t)=0$, then this can be made to converge to zero by choosing
$c_n$ to increase slowly enough. Similar arguments apply to the other
performance criteria.
\end{proof}

\subsection{Verification of the conditions in
\texorpdfstring{\Cref*{theorem:pate}}{Theorem \ref{theorem:pate}}
for the matching estimator}\label{pate_standard_errors_sec}

For matching estimators with a fixed number of matches we use results from
\citet{AbIm06match} and \citet{abadie_matching_2016} to verify
\Cref*{pate_high_level_assump}. Since such results appear to be available only
for the case where $X_i$ is scalar, we restrict ourselves to this case, and we
leave the question of verifying \Cref*{pate_high_level_assump} when $X_i$ is
multivariate for future research. Since these results are stated for a single
underlying distribution, we restrict attention to the case where the
distribution of $(X_i,D_i)$ is fixed over $P\in\mathcal{P}$ (but where the
conditional expectation function $f_P$ is allowed to vary over the given class
$\mathcal{F}$).

\begin{theorem}\label{theorem:prim-cond-match}
  Suppose that the class $\mathcal{P}$ is such that the marginal distribution of
  $(X_i,D_i)$ and the conditional variance function $\sigma^2_P(x, d)$ is
  the same for all $P\in\mathcal{P}$,
  and such that the following conditions hold:
  (i) $X_i$ is scalar, and is supported on a compact interval $[a, b]$ with
  continuous density (ii) $\sigma^2_P(x, d)$ is continuous and uniformly bounded
  away from zero and infinity (iii) $0<P(D_i=1)<1$ and letting $g(x\mid d)$ denote
  the density of $X_i$ given $D_i$, $g(x\mid 1)/g(x\mid 0)$ is uniformly bounded from
  above and below away from zero on $[a, b]$. Suppose, in addition, that, for
  some $\eta$, $E_P(u_i^{2+\eta}\mid X_i=x, D_i=d)\le 1/\eta$ for $d\in\{0,1\}$, all
  $x$ and all $P\in\mathcal{P}$. Then \Cref*{pate_high_level_assump} holds for
  the weights
  $k(X_i, D_i)=\frac{1}{n}(2D_i-1)\left(1+\frac{K_M(i)}{M} \right)$ for the
  matching estimator with $M$ matches.
\end{theorem}
\begin{proof}
  Part (i) of \Cref*{pate_high_level_assump} follows from Lemma S.11 in
  \citet{abadie_matching_2016}. The formula for $V_{1,n}(P)$ follows from this
  lemma as well, and is given by a constant times $1/n$ (where, under our
  assumptions, the constant is strictly positive and does not depend on $P$).
  Thus, to verify part (ii) of \Cref*{pate_high_level_assump}, it suffices to
  show this condition with $V_{1,n}(P)$ replaced by $1/n$. To this end, note
  that replacing $V_{1,n}(P)$ with $1/n$ in this condition gives
  \begin{equation*}
    n^2 E_P[k(X_i,D_i)^2u_i^2\1{k(X_i,D_i)^2u_i^2>\varepsilon/n}]
    =E_P[(1+K_M(i)/M)^2u_i^2\1{(1+K_M(i))^2u_i^2>\varepsilon\cdot n}].
  \end{equation*}
  This will converge to zero by the standard arguments showing that the
  Lyapunov condition implies the Lindeberg condition, so long as
  $E_P[(1+K_M(i)/M)^{2+\eta}u_i^{2+\eta}]$ is uniformly bounded. Indeed, the
  bound on the conditional $2+\eta$ moment of $u_i$ implies that this is bounded
  by a constant times $E_P[(1+K_M(i)/M)^{2+\eta}]$, which is bounded uniformly
  in $i$ and $n$ by Lemma S.8 in \citet{abadie_matching_2016}.
\end{proof}

We now consider construction of the standard error $\sepate(\hat{L}_{k})$.
For matching estimators with a fixed number of matches, standard errors for the
PATE are available, for example, in \citet{AbIm06match}.  For completeness, we
provide a generic formulation and consistency result that applies to arbitrary
estimators $\hat{L}_k$ in our setting.

In \Cref*{asymptotic_coverage_theorem,match_asymptotic_coverage_theorem}, we gave
conditions under which the conditional standard error $\se(\hat{L}_{k})$ is
consistent in the sense that
$\se(\hat L_k)^2/\sum_{i=1}^{n} k(X_i,D_i)^2\sigma^2_P(X_i,D_i)$ converges in
probability to one conditional on $\{X_i,D_i\}_{i=1}^n$, along with conditions
on the marginal distribution of $(X_i, D_i)$ such that this holds for
$\{X_i,D_i\}_{i=1}^\infty$ in a probability one set. This implies that
$\se(\hat L_k)^2/\sum_{i=1}^{n} k(X_i,D_i)^2\sigma^2_P(X_i,D_i)$ converges in
probability to one unconditionally under these conditions. Thus, if
\Cref*{pate_high_level_assump} holds as well, $\se(\hat L_k)^2/V_{1,n}(P)$ will
converge in probability to one. %

Thus, it suffices to estimate
$n V_{2,n}(P)=E_P((f_P(X_i,1)-f(X_i,0)-\tau(P))^2)$. \citet[Theorem
7]{AbIm06match} give consistency conditions for the matching estimator described
in the text. We therefore focus on the estimator
$n \hat{V}_2=\frac{1}{n}\sum_{i=1}^n (\hat f(X_i,1)-\hat f(X_i,0))^2-\hat L_k^2$.

\begin{theorem}
  Suppose that
  $\max_{1\le i\le n, d\in\{0,1\}}\abs{\hat f(X_i, d)-f_P(X_i, d)}\overset{p}{\to} 0$
  and $\hat L_k\overset{p}{\to} \tau(P)$ uniformly over $P\in\mathcal{P}$, and
  that \Cref*{pate_high_level_assump} holds, with $n[V_{1,n}(P)+V_{2,n}(P)]$
  bounded away from zero uniformly over $P\in\mathcal{P}$. Let $\hat V_{2,n}$ be
  given above. Then $[\hat V_{2,n}-V_{2,n}(P)]/[V_{1,n}(P)+V_{2,n}(P)]$
  converges in probability to zero uniformly over $P\in\mathcal{P}$.
  Furthermore, if $\sepate(\hat{L}_{k})^2=\se(\hat L_k)^2+\hat V_{2,n}$ where
  $\se(\hat L_k)^2/V_{1,n}(P)$ converges in probability to one uniformly over
  $P\in\mathcal{P}$, then
  $[V_{1,n}(P)+V_{2,n}(P)]/\sepate(\hat{L}_{k})^{2} \overset{p}{\to} 1$
  uniformly over $P\in\mathcal{P}$.
\end{theorem}
\begin{proof}
We have
\begin{multline*}
  \abs{\hat V_{2,n}/n-V_{2,n}(P)/n}\\
  =\abs*{\frac{1}{n}\sum_{i=1}^{n}
    \{[\hat f(X_i,1)-\hat f(X_i,0)]^2-[f_P(X_i,1)-f_P(X_i,0)]^2\} + \tau(P)^2-\hat L_k^2} \\
  \le 2\max_{1\le i\le n, d\in\{0,1\}}\abs{\hat f(X_i, d)-f_P(X_i, d)}^2 + \abs{
    \hat L_k^2 - \tau(P)^{2}},
\end{multline*}
which converges in probability to zero uniformly over $P\in\mathcal{P}$.  By the
$\mathcal{O}(1/n)$ lower bound on $V_{1,n}(P)+V_{2,n}(P)$, it then follows that
$[\hat V_{2,n}-V_{2,n}(P)]/[V_{1,n}(P)+V_{2,n}(P)]$ converges in probability to
zero uniformly over $P\in\mathcal{P}$.
\end{proof}

\end{appendices}

\bibliography{../../np-testing-library}